\documentstyle[11pt,epsfig,a4]{article}

\renewcommand{\theequation}{\thesection.\arabic{equation}}
\newcommand{\beq}{\begin{equation}}
\newcommand{\eeq}{\end{equation}}
\newcommand{\bea}{\begin{eqnarray}}
\newcommand{\eea}{\end{eqnarray}}
\newcommand{\nn}{\nonumber}
\newcommand{\gsim}{\raisebox{-0.07cm}{$\:\stackrel{>}{{\scriptstyle
 \sim}}\: $} }

\newcommand{\ra}{\rightarrow}
\newcommand\MSb{$\overline{\mbox{MS}}$}
\newcommand\Nf{N_{\rm f}}

\newcommand\cV{\mbox{\boldmath $c$}}
\newcommand\eV{\mbox{\boldmath $e$}}
\newcommand\qV{\mbox{\boldmath $q$}}
\newcommand\CV{\mbox{\boldmath $C$}}
\newcommand\PV{\mbox{\boldmath $P$}}
\newcommand\RV{\mbox{\boldmath $R$}}
\newcommand\UV{\mbox{\boldmath $U$}}
\newcommand\ZV{\mbox{\boldmath $Z$}}
\newcommand\uV{\mbox{\boldmath $1$}}

\def\a{\alpha} \def\b{\beta}  \def\d{\delta} 
\def\e{\epsilon} \def\f{\phi} \def\g{\gamma}  
 \def\m{\mu} \def\n{\nu}    
    \def\z{\zeta}  
 \def\G{\Gamma}  \def\S{\Sigma}

\begin{document}
\setlength{\parskip}{0.2cm}
\begin{titlepage}

\noindent
\large
{\tt hep-ph/0110331} \hfill TTP01-23 \\
\hspace*{\fill} NIKHEF 01-015 \\
\hspace*{\fill} October 2001 \\
\vspace{1.5cm}
\begin{center}
\LARGE
{\bf Next-to-Next-to-Leading Order QCD Corrections} \\
\vspace{0.15cm}
{\bf to the Photon's Parton Structure} \\
\vspace{2.2cm}
\Large
S. Moch$^{\, a}$, J.A.M. Vermaseren$^{\, b}$ and A. Vogt$^{\, b}$\\
\vspace{0.8cm}
\large
{\it $^a$Institut f\"{u}r Theoretische Teilchenphysik \\
\vspace{0.1cm}
Universit\"{a}t Karlsruhe, D--76128 Karlsruhe, Germany} \\
\vspace{0.5cm}
{\it $^b$NIKHEF Theory Group \\
\vspace{0.1cm}
Kruislaan 409, 1098 SJ Amsterdam, The Netherlands} \\
\vspace{4.5cm}
\Large
{\bf Abstract}
\vspace{-0.3cm}
\end{center}
\normalsize
%
The next-to-next-to-leading order (NNLO) corrections in massless 
perturbative QCD are derived for the parton distributions of the photon 
and the deep-inelastic structure functions $F_1^{\,\g}$ and 
$F_2^{\,\g}$. 
We present the full photonic coefficient functions at order 
$\a \a_{\rm s}$ and calculate the first six even-integer moments of the 
corresponding ${\cal O}(\a \a_{\rm s}^2)$ photon-quark and photon-gluon 
splitting functions together with the moments of the $\a \a_{\rm s}^2$ 
coefficient functions which enter only beyond NNLO. 
These results are employed to construct parametrizations of the 
splitting functions which prove to be sufficiently accurate at 
least for momentum fractions $x \,\gsim\, 0.05$.
We also present explicit expressions for the transformation from the 
\MSb\ to the DIS$_{\gamma}$ factorization scheme and write down the
solution of the evolution equations. The numerical impact of the NNLO
corrections is discussed in both~schemes. 
\vfill
\end{titlepage} 
\normalsize
%
%
\section{Introduction}
%
%
The hadronic structure of the photon, in particular the deep-inelastic
structure function $F_2^{\,\g}$, has attracted interest since the early 
days of perturbative QCD. Indeed, the leading-order (LO) corrections to 
the `pointlike' parton-model result~\cite{QPM} were derived twenty-five 
years ago in ref.~\cite{Witt}, and the next-to-leading order (NLO) 
contributions followed a few years later \cite{BaBu}. These results, 
obtained in the framework of the operator product expansion (OPE) 
\cite{CHM}, were recast in the language of evolution equations for the 
photon's quark and gluon momentum distributions in refs.~\cite{WJSWW} 
and~\cite{GR83}. An error in the NLO photon-gluon anomalous dimension 
was corrected ten years ago \cite{FP92,GRVg1}. Unlike the case of 
lepton-nucleon deep-inelastic scattering (DIS) 
\mbox{[9$\,$--$\,$15]}, the next-to-next-to-leading order (NNLO) QCD 
corrections have not been addressed for the photon structure up to now. 

So far the measurements of $F_2^{\,\g}$ have been performed using the 
process $e^+ e^- \!\ra\! e^+ e^- +{\it hadrons}\,$ at electron-positron 
colliders; see refs.~\cite{exrev} for recent overviews. While data from 
LEP have greatly improved the situation, an accuracy comparable to that 
achieved in lepton-hadron DIS can only be envisaged if $e\g$ collisions 
will be realized via laser back-scattering \cite{lback} of one of the 
electron beams of a future linear collider \cite{LColl,ph99}. 
Important information on the photon structure can also be expected from 
photoproduction of jets at HERA, which has been treated at NLO thus far 
\mbox{[20$\,$--$\,$23]}. 
The extension to NNLO is under way, e.g., the 
two-loop matrix element required also for hadronic collisions have been 
derived in refs.~\cite{MatEl} using the pioneering results~\cite{dbox} 
for the scalar double box diagrams; the corresponding results with
one external photon will be available soon~\cite{NGprp}.

In this article we present, with one qualification concerning the 
splitting functions, the NNLO corrections for electron-photon DIS and 
the evolution of the parton densities of the photon in massless 
perturbative QCD. 
In Sect.~2 we extend the OPE analysis of the photon structure 
\cite{Witt,BaBu} to the required accuracy: The partonic forward 
amplitudes for the scattering of a virtual photon (and of a fictitious 
scalar directly coupled only to gluons) off a real photon are expressed 
in terms of the anomalous dimensions and coefficient functions up to 
second order in the strong coupling constant. Our calculation of these 
amplitudes for the lowest six even-integer values of the Mellin 
variable, $N = 2,\, \ldots ,\, 12$, then facilitates the extraction of 
the corresponding anomalous dimensions (up to NNLO) and coefficient 
functions (up to the next-to-next-to-next-to-leading order, N$^3$LO). 
The results for these quantities in the \MSb\ scheme are presented in 
Sect.~3 in numerical form, together with a brief discussion of the 
actual computation which closely followed the lines of 
refs.~\cite{moms,RV00}. 
The analytic expressions for these results can be found in Appendix A.

In Sect.~4 we switch to the parton language and specify the dependence 
of the photon-parton splitting functions and the photonic coefficient 
functions on the renormalization and factorization scales. After
recalling the general factorization-scheme transformation, we then
derive the NNLO corrections in the DIS$_\g$ scheme~\cite{GRVg1} and 
discuss the `physical' kernel for the non-singlet evolution of the 
structure functions at large values of the Bjorken variable $x$. The 
NNLO solution of the evolution equations is also given in this section.
Explicit $x$-space expressions for the photonic coefficient functions 
and the photon-parton splitting functions up to NNLO are presented in 
Sect.~5. For the splitting functions we have to rely on our finite-$N$ 
results of Sect.~3, thus we can only provide approximations analogous 
to those derived in refs.~\cite{NV1,NV2,NV3} for the three-loop QCD 
splitting functions. For the NNLO coefficient functions and the 
corresponding transformation of the splitting functions to the DIS$_\g$ 
scheme we present, besides the exact results (deferred to Appendix~B 
for the latter quantities), also compact approximate expressions.
In Sect.~6 we finally illustrate the numerical effect of the NNLO 
corrections on the evolution kernels and on the solution of the 
evolution equations. Our conclusions are presented in Sect.~7.

\newpage
%
%
\section{Moments: formalism and method}
%
%
The subject of our calculation is inclusive hadron production in 
unpolarized electromagnetic (e.m.) deep-inelastic electron-photon 
scattering, 
\beq
\label{egeX}
  e(k) + \g (p) \:\ra\: e(k^{\prime}) +  X \:\: , 
\eeq
where `$X$' stands for all hadronic states allowed by quantum number
conservation. The hadronic part of the corresponding amplitude is given 
by the (spin-averaged) tensor
\bea
\label{Wmunu}
  W^{\,\g\G}_{\m\n}(p,q) & = &  
  \frac{1}{4\pi} \int\! d^{\, 4} z\, e^{iq \cdot z} \, 
  \langle \G, p \vert \, J_{\m}(z)J_{\n}(0) \, 
  \vert \G, p \rangle \nn \\
  &=& e_{\m\n}\, \frac{1}{2x}F_L^{\,\g}(x,Q^2) + 
      d_{\m\n}\, \frac{1}{2x}F_2^{\,\g}(x,Q^2)
\eea
with
\bea
\label{emunu}
  e_{\m\n} & = &  g_{\m \n}-\frac{q_{\m} q_{\n}}{q^2} \nn \\
  d_{\m\n} & = & -g_{\m \n}-p_{\m}p_{\n}\,\frac{4x^2}{q^2}
              -(p_{\m}q_{\n}+p_{\n}q_{\m})\,\frac{2x}{q^2} \:\: . 
\eea
Here $\vert \G, p \rangle$ denotes the physical photon state (including
a non-perturbative hadronic component) with momentum $p$, and $J_{\m}$ 
represents the e.m.~quark current. $q = k - k^{\prime}$ is the momentum 
transferred by the electron, $Q^2=-q^2$, and $x = Q^2 / (2p \cdot q)$ 
is the Bjorken variable ($0 < x < 1$).
The longitudinal structure function $F_L$ is related to the structure 
function $F_1$ by $ F_L = F_2 -2xF_1$.

The optical theorem relates the tensor $W_{\m\n}$ in Eq.~(\ref{Wmunu})
to the forward amplitude $T_{\m\n}$ for the scattering of a virtual 
photon off a real photon,
\beq 
\label{Tmunu}
  T^{\,\g\G}_{\m\n}(p,q) \: = \: i \! \int \! d^{\, 4}z \, e^{iqz} \, 
  \langle \G, p \vert \, T \big(J_{\m}(z)J_{\n}(0) \big) \,
  \vert \G, p \rangle \:\: .
\eeq
This quantity represents a convenient starting point for practical 
calculations, due to the presence of the time-ordered product of 
currents to which standard perturbation theory applies. In fact, the
operator product expansion for this product and the subsequent 
application of a dispersion relation to Eq.~(\ref{Tmunu}) very closely 
follow the procedure for standard lepton-hadron deep-inelastic 
scattering discussed, for example, in refs.~\cite{MV1,moms,revs} to 
which we refer the reader for details.
In the leading-twist sector addressed in this article, the only, but 
crucial difference to the lepton-hadron case is the presence of the 
spin-$N$ twist-2 photon operators~\cite{Witt,BaBu}
\beq
\label{Ogam}
  O_{\g}^{\{\m^{\,}_1,\cdots ,\m^{\,}_N\}} \: = \: F^{\n\{\m^{\,}_1} 
  D^{\m^{\,}_2}\cdots D^{\m^{\,}_{N-1}} F^{\m^{\,}_N \} \n }
\eeq
and their coefficient functions $C_{i,\g}^{N}$ in addition to the 
usual quark (flavour non-singlet and singlet) and gluon operators, 
$O_{\rm ns}$, $O_{\rm q}$ and $O_{\rm g}$, and their respective 
coefficient functions. $D^{\m}$ in Eq.~(\ref{Ogam}) denotes the 
covariant derivative, and $F^{\m\n}$ represents the e.m.\ field 
strength tensor. The spin-averaged matrix elements of these 
(renormalized) operators are given by
\bea
\label{OME}
  \langle \G, p \vert \, O_{\a}^{\{\m^{\,}_1,...,\m^{\,}_N\}} 
  \,\vert \G, p \rangle \: = \: p^{\{\m^{\,}_1}...p^{\m^{\,}_N\}} 
  \, A_{\,\G,\a}^N (\m^2) \:\:  , \quad\quad \a \, = \, 
  \mbox{ns, q, g, }\g \:\: , 
\eea
where $\m$ stands for the renormalization scale. It is understood in 
Eqs.~(\ref{Ogam}) and (\ref{OME}) that the symmetric and traceless part 
is taken with respect to the indices in curved brackets.

Following the procedure for lepton-hadron DIS \cite{MV1,moms,revs}, the 
even-integer Mellin-$N$ moments of the structure functions $F_2^{\,\g}$ 
and $F_L^{\,\g}$ in Eq.~(\ref{Wmunu}) 
\beq
\label{Mtrf}
  F_i^{\,\g,N}(Q^2) \: = \: 
  \int_0^1 \! dx\, x^{N-1} {\cal F}_i^{\,\g}(x,Q^2) \:\: , \quad\quad
  {\cal F}_i(x) = \frac{1}{x}\, F_i(x)
\eeq
can then be expressed in terms of the parameters of the OPE,
\beq
\label{OPE}
  F_i^{\,\g,N}(Q^2) = \sum_{\a = {\rm ns, q, g,} \g}
  C_{i,\a}^{N} \left (\frac{Q^2}{\m^2},a_{\rm s},a_{\rm em} \right) 
  A_{\,\G,\a}^N (\m^2) \:\: , 
  \quad\quad i = 2,L \:\: .
\eeq
Here and throughout the whole article we use the notation 
\beq
\label{coupl}
  a_{\rm s}  \: = \: \frac{\a_{\rm s}} {4 \pi} \:\: , \quad
  a_{\rm em} \: = \: \frac{\a}{4 \pi}
\eeq
for the strong and electromagnetic coupling constants. The present
study addresses the higher-order QCD corrections to the photon 
structure functions $F_{2,L}^{\,\g}$ at the leading order of QED, 
$a_{\rm em}^1$. Consequently the quantities entering the r.h.s.\ of
Eq.~(\ref{OPE}) are only needed at their respective lowest e.m.\ 
orders, i.e., $A_{\,\G,\g}$ and $C_{i,\rm{p}}$ ($\rm{p = ns,q,g}$) at 
$a_{\rm em}^0$, and $A_{\,\G,\rm{p}}$ and $C_{i,\g}$ at $a_{\rm em}^1$.

The operators $O_{\a}$ in Eq.~(\ref{OME}) mix under renormalization.
Expressing the renormalized operators in terms of their bare 
counterparts, this mixing can be written as
\beq
\label{Oren}
  O_\a \: = \: Z_{\a\b}\,O_\b^{\rm bare} \:\: .
\eeq
Here and in the next two equations the summation convention is used,
and the range of all indices is as specified in Eq.~(\ref{OME}) above.
The anomalous dimensions $\g_{\a\b}$ governing the scale dependence of 
the operators $O_{\a}$,
\beq
\label{gamma}
  \frac{d}{d \ln \m^2 }\, O_\a \: = \: - \,\g_{\a\b}\, O_\b \:\: , 
\eeq
are connected to the mixing matrix $Z_{\a\b}$ in Eq.~(\ref{Oren}) by
\beq
\label{gamZ}
 \g_{\a\b} \: = \: -\,\left( \frac{d }{d\ln\m^2 }\, Z_{\a\a^{\prime}} 
 \right) (Z^{-1})_{\a^{\prime}\b} \:\: . 
\eeq 
Keeping only those terms which are relevant for the structure functions
$F_{2,L}^{\,\g}$ at order $a_{\rm em}^{1}$, the matrices $Z$ and $\g$ 
take the form 
\beq
\label{Zmat}
  Z \: = \: \left( 
  \begin{array}{cccc} 
  Z_{\rm ns } & 0 & 0 & Z_{\rm ns\g} \\
  0 & Z_{\rm qq } & Z_{\rm qg } & Z_{\rm q \g} \\
  0 & Z_{\rm gq } & Z_{\rm gg } & Z_{\rm g \g} \\
  0 & 0 & 0 & 1 
  \end{array}                                
  \right) 
\eeq
and
\beq
\label{gmat}
  \gamma \: = \: \left( 
  \begin{array}{cccc} 
  \gamma_{\rm ns} & 0 & 0 & k_{\rm ns} \\
  0 & \gamma_{\rm qq} & \gamma_{\rm qg} & k_{\rm q} \\
  0 & \gamma_{\rm gq} & \gamma_{\rm gg} & k_{\rm g} \\
  0 & 0 & 0 & 0 
  \end{array}                                
  \right) 
\eeq
with the perturbative expansions (where $k_{\rm g}^{(0)} = 0$)
\beq
\label{adims}
  \g_{\rm p(p^{\prime})} \: = \: \sum_{l=0}^{\infty}\, 
  a_{\rm s}^{l+1}\, \gamma_{\rm p(p^{\prime})}^{(l)} \:\: , \quad\quad 
  k_{\rm{p}} \: = \: \sum_{l=0}^{\infty}\, 
  a_{\rm em}\, a_{\rm s}^l\, k_{\rm{p}}^{(l)} \:\: .
\eeq 

In order to make practical use of Eq.~(\ref{gamZ}) a regularization 
procedure and a renormalization scheme need to be selected. We choose
dimensional regularization~\cite{dreg} and the modified~\cite{BBDM}
minimal subtraction \cite{MS} scheme, \MSb\ --- the standard choice for
modern multi-loop calculations in QCD. For this choice the running
couplings in $ D = 4 - 2\e $ dimensions evolve according to
\beq
\label{arun}
  \frac{d\, a_j}{d \ln \m^2} \: = \: - \e\, a_j + \b_j (a_j) \:\: ,
  \quad\quad j = \mbox{em, s} \:\: ,
\eeq
where $\b_{\rm em}$ and $\b_{\rm s}$ denote the usual four-dimensional 
beta functions of QED and QCD, respectively. $\b_{\rm em}$ does 
actually not enter the present calculation; for $\b_{\rm s}$ we employ 
the standard notation
\beq
\label{bQCD}
  \b_{\rm s} \: = \: - \b_0\, a_{\rm s}^2 - \b_1\, a_{\rm s}^3 
                     - \b_2\, a_{\rm s}^4 - \ldots
  \:\: 
\eeq
with $\b_{0} = 11 - 2/3\, \Nf$, where $\Nf$ stands for the number of 
effectively massless quark flavours.

Inserting Eqs.~(\ref{Zmat})--(\ref{bQCD}) into Eq.~(\ref{gamZ}) and 
solving for the photon-quark and photon-gluon renormalization factors
$Z_{\rm ns\g}$, $Z_{\rm q\g}$ and $Z_{\rm g\g}$ we obtain, up 
to the desired order $a_{\rm em}^{} a_{\rm s}^2$,
\bea
\label{Zns}
  a_{\rm em}^{-1}\, Z_{\rm ns\g}\! & = & 
  \frac{1}{\e}\, k^{(0)}_{\rm ns} 
  \, + \, a_{\rm s}\, \left[ 
    \frac{1}{\e^2} \left\{ \frac{1}{2} k^{(0)}_{\rm ns} 
      \g^{(0)}_{\rm ns} \right\}  
    + \frac{1}{2\e}\, k^{(1)}_{\rm ns} \right]
  \nn \\ & & \mbox{}
  + \, a_{\rm s}^2\, \left[ \,
    \frac{1}{\e^3} \left\{ 
      \frac{1}{6} k^{(0)}_{\rm ns} \left(\g^{(0)}_{\rm ns}\right)^2 
      - \frac{1}{6} \beta_0 k^{(0)}_{\rm ns} \g^{(0)}_{\rm ns} 
      \right\} \right. 
  \\ & & \quad\quad\: \mbox{} \left.
    + \frac{1}{\e^2} \left\{ \frac{1}{3} k^{(0)}_{\rm ns} 
      \g^{(1)}_{\rm ns} + \frac{1}{6} k^{(1)}_{\rm ns} 
      \g^{(0)}_{\rm ns} - \frac{1}{6} \beta_0 k^{(1)}_{\rm ns} 
      \right\}
    + \frac{1}{3\e}\, k^{(2)}_{\rm ns} \right] 
  \:\: ,\nn \\[2ex]
\label{Zq}
  a_{\rm em}^{-1}\, Z_{\rm q\g} & = & 
  \frac{1}{\e}\, k^{(0)}_{\rm{q}} 
  \, + \, a_{\rm s}\, \left[ \,  
    \frac{1}{\e^2} \left\{ \frac{1}{2} k^{(0)}_{\rm q} 
      \g^{(0)}_{\rm qq} \right\} 
    + \frac{1}{2\e}\, k^{(1)}_{\rm q} \right]
  \nn \\ & & \mbox{}
  + \, a_{\rm s}^2\, \left[ \,
    \frac{1}{\e^3} \left\{ \frac{1}{6} k^{(0)}_{\rm q} \left( 
      \g^{(0)}_{\rm qq}\right)^2 + \frac{1}{6} k^{(0)}_{\rm q} 
      \g^{(0)}_{\rm qg} \g^{(0)}_{\rm gq} - \frac{1}{6} \beta_0 
      k^{(0)}_{\rm q} \g^{(0)}_{\rm qq} \right\} \right. 
  \\ & & \quad\quad\: \mbox{} \left.
    + \frac{1}{\e^2}  \left\{ \frac{1}{3} k^{(0)}_{\rm q} 
      \g^{(1)}_{\rm qq} + \frac{1}{6} k^{(1)}_{\rm q} 
      \g^{(0)}_{\rm qq} + \frac{1}{6} k^{(1)}_{\rm g} 
      \g^{(0)}_{\rm qg} - \frac{1}{6} \beta_0 k^{(1)}_{\rm q} \right\}
    + \frac{1}{3\e}\,  k^{(2)}_{\rm q} \right]
  \:\: ,\nn \\[2ex]
\label{Zg}
  a_{\rm em}^{-1}\, Z_{\rm g\g} & = &
  \: + \: a_{\rm s}\: \left[ \,  
    \frac{1}{\e^2} \left\{ \frac{1}{2} k^{(0)}_{\rm q} 
      \g^{(0)}_{\rm gq} \right\} 
    + \frac{1}{2\e}\, k^{(1)}_{\rm g} \right]
  \nn \\ & & \mbox{}
  + \, a_{\rm s}^2\, \left[ \,
    \frac{1}{\e^3} \left\{ \frac{1}{6} k^{(0)}_{\rm{q}} 
       \g^{(0)}_{\rm gq} \left( \g^{(0)}_{\rm qq} + \g^{(0)}_{\rm gg} 
       \right) - \frac{1}{6} \beta_0 k^{(0)}_{\rm{q}} \g^{(0)}_{\rm gq} 
       \right\} \right. 
  \\ & & \quad\quad\: \mbox{} \left.
    + \frac{1}{\e^2}  \left\{ \frac{1}{3} k^{(0)}_{\rm q} 
      \g^{(1)}_{\rm gq} + \frac{1}{6} k^{(1)}_{\rm q} 
      \g^{(0)}_{\rm gq} + \frac{1}{6} k^{(1)}_{\rm g} 
      \g^{(0)}_{\rm gg} - \frac{1}{6} \beta_0 k^{(1)}_{\rm g} \right\}
    + \frac{1}{3\e}\,  k^{(2)}_{\rm g} \right] \nn \:\: .
\eea   
The photon-parton anomalous dimensions $k_{\rm ns}$, $k_{\rm q}$ and 
$k_{\rm g}$ can thus be read off order-by-order from the $\e^{-1}$ 
terms of the corresponding renormalization factors, while the higher
poles in $1/\e$ in Eqs.\ (\ref{Zns})--(\ref{Zg}) can serve as checks 
for the calculation. The coefficient functions in Eq.~(\ref{OPE}), on 
the other hand, have an expansion in positive powers of $\e$, viz
\bea
\label{cfcts}
  C_{i,\rm p} & = & \d_{i2}\, (1 - \d_{\rm pg}) + \sum_{l=1}^{\infty} 
    \, a_{\rm s}^{l} \left( c_{i,\rm p}^{(l)} + \e a_{i,\rm p}^{(l)} + 
    \e^2 b_{i,\rm p}^{(l)} + \ldots \right)
    \nn \\
  C_{i,\g}    & = & \sum_{l=1}^{\infty} \, a_{\rm em}\, a_{\rm s}^{l-1}
    \left( c_{i,\g}^{(l)} + \e a_{i,\g}^{(l)} + \e^2 b_{i,\g}^{(l)} + 
    \ldots \right) \:\: ,
\eea
where $i = 2,\, L$ and p = ns, q, g. Also here only the lowest-order 
terms in $a_{\rm em}$ have been retained as discussed below 
Eq.~(\ref{OPE}), thus, like the anomalous dimensions 
$\g_{\rm p(p^{\prime})}$ in Eq.~(\ref{adims}), the coefficient 
functions $C_{i,\rm p}$ in Eq.~(\ref{cfcts}) are just the standard QCD 
quantities entering lepton-hadron DIS. 

Due to the partly non-perturbative physical photon state $\vert \G, p 
\rangle $, Eqs.~(\ref{Tmunu}) and (\ref{OPE}) are not accessible to a
perturbative computation. However, as the OPE represents an operator 
relation, the anomalous dimensions (\ref{adims}) and the coefficient 
functions (\ref{cfcts}) do not depend on this state. Hence, again 
closely following the procedure \cite{MV1,moms,RV00} for lepton-nucleon 
DIS, the calculation can be performed using a partonic photon state 
$\vert \g, p \rangle $. Instead of Eq.~(\ref{Tmunu}) we thus consider
\beq
\label{Tpart}
  T^{\,\g\g}_{\m\n}(p,q) \: = \: i \! \int \! d^{\, 4}z \, e^{iqz} \,
  \langle \g, p \vert \, T \big(J_{\m}(z)J_{\n}(0) \big) \,
  \vert \g, p \rangle \:\: .
\eeq
At leading-twist accuracy the decomposition of $T^{\,\g\g}_{\m\n}$ into 
$T_{2,\g}$ and $T_{L,\g}$ analogous to Eq.~(\ref{Wmunu}) is 
provided by 
\bea
\label{Tproj}
  T_{L,\g}(x,Q^2) & = & - \,\frac{q^2}{(p\cdot q)^2}\, p^\m p^\n \: 
    T^{\,\g\g}_{\m\n}(p,q) 
  \nn \\
  T_{2,\g}(x,Q^2) & = & - \left( \frac{3-2\e}{2-2\e}\: \frac{q^2}
    {(p\cdot q)^2}\, p^\m p^\n + \frac{1}{2-2\e}\: g^{\m\n} \right) 
    T^{\,\g\g}_{\mu\nu}(p,q) 
  \:\: .
\eea
The $N^{\rm th}$ moments are obtained from Eqs.~(\ref{Tproj}) by 
applying the projection operator ${\cal P}_N$ \cite{GLT}, 
\bea
\label{PNop}
  T^N_{i,\g} \left(\frac{Q^2}{\m^2},a_{\rm s},a_{\rm em},\e\right)
  & = & {\cal P}_N\, T_{i,\g}(x,Q^2,a_{\rm s},a_{\rm em},\e ) \\
  & \equiv & \left[ \frac{q^{ \{\m_1}\cdots q^{\m_N \}}}{N !}\,
  \frac{\partial ^N}{\partial p^{\m_1} \ldots  \partial p^{\m_N}} 
  \right] \, T_{i,\g}(x,Q^2,a_{\rm s},a_{\rm em},\e ) \Bigg|_{p=0} 
  \:\: ,
  \nn
\eea
where $q^{ \{\m_1}\cdots q^{\m_N \}}$ is the harmonic, i.e., the 
symmetric and traceless part of the tensor $q^{\m_1}\cdots q^{\m_N}$.

This operator does not act on the coefficient functions $C_{i,\a}$ and
the renormalization constants $Z_{\a\b}$ in Eq.~(\ref{Oren}), which are 
functions only of $N$, $\a_{\rm s}$, $\a$ and $\e$. It does 
act, however, on the bare matrix elements $A_{\g,\a}^N$ (defined 
analogously to Eq.~(\ref{OME})$\,\!$) and eliminates all diagrams 
containing loops, as the nullification of $p$ transform these diagrams 
to massless tadpole diagrams which are put to zero in dimensional 
regularization. This removes the operator matrix elements 
$A_{\g,\rm p}^N$, p = ns, q, g, which only start at the one-loop level. 
Hence only the matrix elements $A_{\g\g}^{N,\,\rm tree}$ of the photon 
operators (\ref{Ogam}) remain. These matrix elements are given by
\beq
  A_{\g\g}^{N,\,\rm tree}(\e) \: = \: (1- \e)\cdot {\rm const}_N \:\: ,
\eeq
where the factor $(1- \e)$ arises from the number of photon 
polarizations in $D = 4 - 2\e$ dimensions.
We thus arrive at 
\beq
\label{T2LN}
  T^N_{i,\g} \left (\frac{Q^2}{\m^2}, a_{\rm s},a_{\rm em},\e \right) 
  \: = \sum_{\a = {\rm ns, q, g,} \g} C_{i,\a}^{N} 
    \left (\frac{Q^2}{\m^2},a_{\rm s},a_{\rm em},\e \right)
    Z_{\a\g}^N \left(a_{\rm s},a_{\rm em},\frac{1}{\e} \right)
    A_{\g\g}^{N,\,\rm tree}(\e) 
\eeq
with $i=2,L$. This relation, after expanding in powers of $a_{\rm em}$, 
$a_{\rm s}$ and $\e$, provides a system of coupled equations which can 
be solved for the photon-parton anomalous dimensions and the photon 
coefficient functions. In particular, by computing $T^N_{i,\g}$ to the 
order $a_{\rm em}\,a_{\rm s}^2$, we can derive the desired coefficients 
$k_{\rm ns} ^{(2)N}$, $k_{\rm q} ^{(2)N}$, and $c_{i,\g}^{(2)N}$ in 
Eqs.~(\ref{adims}) and (\ref{cfcts}). 
The quantities $k_{\rm g}^{(2)N}$, on the other hand, cannot be 
determined in this manner since the gluonic coefficient function
$C_{i,\rm g}$ in Eq.~(\ref{cfcts}), unlike its quark counterparts, 
starts at order $a_{\rm s}$ only. This problem is overcome by 
considering, in addition to Eq.~(\ref{Tpart}), another unphysical Green
function $T^{\f\g}$ where the virtual-photon probe is replaced by an
external scalar field $\f$ coupling directly only to gluons, see below.

The expansion of Eq.~(\ref{T2LN}) to order $a_{\rm em}^{}\,a_{\rm s}^2$
can be written as 
\beq
\label{Texp} 
  T^{N}_{i,\gamma} \: = \: \sum_{l=0}^{2}\, a_{\rm em}\, a_{\rm s}^l
  \, S_\e^{\, l+1} \bigg( \frac{\mu^2}{Q^2} \bigg)^{(l+1) \e} 
  T^{(l)N}_{i,\gamma} A_{\g\g}^{N,\,\rm tree} \:\: .
\eeq
The factor $S_\e =\exp\left[ \e\{\ln(4\pi-\gamma^{}_{\rm E}\} \right]$,
where $\gamma^{}_{\rm E}$ denotes the Euler-Mascheroni constant, is an 
artefact of dimensional regularization kept out of the coefficient 
functions and anomalous dimensions in the \MSb\ scheme. 
The expansion coefficients $T^{(l)}_{i,\gamma}$ can be decomposed into 
flavour non-singlet (ns) and singlet (s) pieces, 
\bea
\label{Tsns}
  T^{(l)}_{i,\g} & = & \d_{\rm ns}\, T^{(l),\rm ns}_{i,\g} \, + \, 
  \langle e^2 \rangle \,\d_{\rm s}\, T^{(l),\rm s}_{i,\g} 
  \\[1ex]
  & \equiv & \d_{\rm ns} \, T^{(l),\rm ns}_{i,\g} \, + \,  
 \langle e^2 \rangle \,\d_{\rm s}\, \Big( T^{(l),\rm ns}_{i,\g}
  + T^{(l),\rm ps}_{i,\g} \Big) \nn \:\: ,
\eea
which collect the contributions proportional to the respective 
combinations of quark charges,
\beq
\label{delta}
  \d_{\rm ns} \: = \: 3 \Nf \left( \langle e^4 \rangle 
  - \langle e^4 \rangle ^2 \right)
  \:\: , \quad\quad
  \d_{\rm s}  \:\equiv\: \d_{\rm q} \: = \: 3 \Nf\, \langle e^2 \rangle 
  \:\equiv\: 3\, {\textstyle \sum_{j=1}^{\Nf}\, e_{q_j}^2} \:\: .
\eeq
The pure-singlet (ps) contribution defined in the second line of 
Eq.~(\ref{Tsns}) starts only at order~$a_{\rm s}^2$. Consequently the
anomalous dimensions $k^{(l)}_{\rm ns}$ and $k^{(l)}_{\rm q}$ in Eq.\
(\ref{adims}) are identical for $l \leq 1$ except for their obvious 
charge factors. The same applies, for $l \leq 2$, to the non-singlet
and singlet photon coefficient functions in  
\beq
\label{cpdec}
  c_{i,\g}^{(l)} \:  = \:  c_{i,\g}^{(l),\rm ns} 
  + \langle e^2 \rangle\, c_{i,\g}^{(l),\rm s} \:\: .
\eeq
The corresponding decomposition for the hadronic quantities reads
\beq
\label{gcps}
  \g_{\rm qq}^{(l)} \: = \:\g_{\rm ns}^{(l)} \: + \,\g_{\rm ps}^{(l)}
  \:\: , \quad\quad
  c_{i,\rm q}^{(l)} \: = \: c_{i,\rm ns}^{(l)} + \,c_{i,\rm ps}^{(l)}
  \:\: ,
\eeq
where the pure-singlet contributions are non-vanishing for $l \geq 1\:
$(2) in the first (second) relation. 

The explicit expressions for the first two singlet expansion 
coefficients in Eq.~(\ref{Tsns}) in terms of the anomalous dimensions 
and coefficient functions are given by
\bea
  \label{T2L0}
  \d_{\rm s} T^{(0),\rm s}_{L,\g}   & = &    
     c^{(1),\rm s}_{L,\g} 
  \nn \\
  \d_{\rm s}\, T^{(0),\rm s}_{2,\g} & = &   
    \frac{1}{\e}\, k^{(0)}_{\rm q} + c^{(1),\rm s}_{2,\g} 
\eea
and
\bea
\label{T2L1}
  \d_{\rm s}\, T^{(1),\rm s}_{L,\g} & = & 
  \,\frac{1}{\e}\, \left\{ k^{(0)}_{\rm{q}} c^{(1)}_{L,\rm q} \right\} 
  + c^{(2),\rm s}_{L,\g} + k^{(0)}_{\rm q} a^{(1)}_{L,\rm q} 
  \nn \\[1ex]
  \d_{\rm s}\, T^{(1),\rm s}_{2,\g} & = & 
  \frac{1}{\e^2} \left\{ \frac{1}{2} k^{(0)}_{\rm q} 
    \gamma^{(0)}_{\rm qq} \right\} +
  \frac{1}{\e} \left\{ \frac{1}{2} k^{(1)}_{\rm q} + k^{(0)}_{\rm q} 
    c^{(1)}_{2, \rm q} \right\} + 
    c^{(2),\rm s}_{2,\gamma} + k^{(0)}_{\rm{q}} a^{(1)}_{2,\rm q}
  \:\: .
\eea
The corresponding non-singlet relations are obvious as discussed below
Eq.~(\ref{delta}), hence they are not written out for brevity. The 
contributions at the order $a_{\rm em}^{}\, a_{\rm s}^2$ read
\bea
\label{T2L2ns}
  \d_{\rm ns}\, T^{(2),\rm ns}_{L,\g} & = & 
  \frac{1}{\e^2} \left\{ \frac{1}{2} k^{(0)}_{\rm ns} \g^{(0)}_{\rm qq} 
     c^{(1)}_{L,\rm q} \right\} + 
  \frac{1}{\e} \left\{ 
     k^{(0)}_{\rm ns} c^{(2)}_{L, \rm ns} 
     + \frac{1}{2} k^{(1)}_{\rm ns} c^{(1)}_{L, \rm q} 
     + \frac{1}{2} k^{(0)}_{\rm ns} \g^{(0)}_{\rm qq} a^{(1)}_{L,\rm q} 
     \right\}
 \nn \\[0.5ex] & & \mbox{}
  + c^{(3),\rm ns}_{L,\gamma}  + k^{(0)}_{\rm ns} a^{(2)}_{L,\rm ns} 
     + \frac{1}{2} k^{(1)}_{\rm ns} a^{(1)}_{L, \rm q}
     + \frac{1}{2} k^{(0)}_{\rm ns}\g^{(0)}_{\rm qq} b^{(1)}_{L, \rm q} 
 \nn \:\: , \\[1ex]
  \d_{\rm ns}\, T^{(2),\rm{ns}}_{2,\g} & = & \quad\!
  \frac{1}{\e^3} \left\{ 
    \frac{1}{6} k^{(0)}_{\rm ns} \left( \g^{(0)}_{\rm qq} \right)^2 
    - \frac{1}{6} \beta_0 k^{(0)}_{\rm ns} \g^{(0)}_{\rm qq} \right\} 
  \nn \\ & & \mbox{} \!\! +
  \frac{1}{\e^2} \left\{ \frac{1}{2} k^{(0)}_{\rm ns}
    \g^{(0)}_{\rm qq} c^{(1)}_{2,\rm q} 
    + \frac{1}{3} k^{(0)}_{\rm ns} \g^{(1)}_{\rm ns} 
    + \frac{1}{6} k^{(1)}_{\rm ns} \g^{(0)}_{\rm qq}
    - \frac{1}{6} \beta_0 k^{(1)}_{\rm ns} \right\}  
  \nn \\ & & \mbox{} \!\! + \,
  \frac{1}{\e}\: \left\{ \frac{1}{3} k^{(2)}_{\rm ns}
    + k^{(0)}_{\rm ns} c^{(2)}_{2,\rm ns} 
    + \frac{1}{2} k^{(1)}_{\rm ns} c^{(1)}_{2,\rm q} 
    + \frac{1}{2} k^{(0)}_{\rm ns} \g^{(0)}_{\rm qq} a^{(1)}_{2,\rm q}
    \right\}
  \nn \\ & & \mbox{} \!\! + \:
    c^{(3),\rm ns}_{2,\g} + k^{(0)}_{\rm ns} a^{(2)}_{2,\rm ns} 
    + \frac{1}{2} k^{(1)}_{\rm ns} a^{(1)}_{2,\rm q}
    + \frac{1}{2} k^{(0)}_{\rm ns} \g^{(0)}_{\rm qq} b^{(1)}_{2,\rm q} 
\eea
and
\bea
\label{T2L2ps}
  \d_{\rm s}\, T^{(2),\rm ps}_{L,\g} & = & 
  \frac{1}{\e^2}\left\{ \frac{1}{2} k^{(0)}_{\rm q} \g^{(0)}_{\rm{gq}} 
     c^{(1)}_{L,\rm g} \right\} + 
  \frac{1}{\e} \left\{ 
     k^{(0)}_{\rm q} c^{(2)}_{L,\rm ps} 
     + \frac{1}{2} k^{(1)}_{\rm g} c^{(1)}_{L,{\rm g}} 
     + \frac{1}{2} k^{(0)}_{\rm q} \g^{(0)}_{\rm gq} a^{(1)}_{L,\rm g}
     \right\}
  \nn \\ & & \mbox{}
  + c^{(3),{\rm ps}}_{L,\g} + k^{(0)}_{\rm q} a^{(2)}_{L,{\rm ps}} 
    + \frac{1}{2} k^{(1)}_{\rm g} a^{(1)}_{L,{\rm g}} 
    + \frac{1}{2} k^{(0)}_{\rm q} \g^{(0)}_{\rm gq} b^{(1)}_{L,\rm g}
  \:\: , \nn \\[1ex] 
  \d_{\rm s}\, T^{(2),\rm ps}_{2,\g} & = & 
  \frac{1}{\e^3} \left\{ \frac{1}{6} k^{(0)}_{\rm q} \g^{(0)}_{\rm qg} 
    \g^{(0)}_{\rm gq} \right\} + 
  \frac{1}{\e^2} \left\{ 
    \frac{1}{2} k^{(0)}_{\rm{q}} \g^{(0)}_{\rm gq} c^{(1)}_{2,\rm g} 
    + \frac{1}{6} k^{(1)}_{\rm g} \g^{(0)}_{\rm qg}
    + \frac{1}{3} k^{(0)}_{\rm q} \g^{(1)}_{\rm ps} \right\} 
  \nn \\ & & \mbox{} +
  \frac{1}{\e} \left\{ 
    \frac{1}{3} k^{(2)}_{\rm ps} 
    + k^{(0)}_{\rm q} c^{(2)}_{2,{\rm ps}} 
    + \frac{1}{2} k^{(1)}_{\rm g} c^{(1)}_{2,{\rm g}} 
    + \frac{1}{2} k^{(0)}_{\rm q} \g^{(0)}_{\rm gq} a^{(1)}_{2,\rm g}
    \right\}
  \nn \\ & & \mbox{} + \, 
  c^{(3),\rm ps}_{2,\g} + k^{(0)}_{\rm{q}} a^{(2)}_{2,{\rm ps}} 
    + \frac{1}{2} k^{(1)}_{\rm g} a^{(1)}_{2,{\rm g}} 
    + \frac{1}{2} k^{(0)}_{\rm q} \g^{(0)}_{\rm gq} b^{(1)}_{2,\rm g}
    \:\: .
\eea

Finally we derive the corresponding expressions for the unphysical
flavour-singlet Green function $T^{\f\g}(p,q)$ mentioned below 
Eq.~(\ref{T2LN}). After application of the projection operator 
${\cal P}_N$ in Eq.~(\ref{PNop}) the moments of $T^{\f\g}$ can be 
written as
\beq
\label{TphN}
  T^N_{\f,\g} \left (\frac{Q^2}{\m^2}, a_{\rm s},a_{\rm em},\e \right) 
  \: = \sum_{\a = {\rm q, g,} \g} C_{\phi,\a}^{N} 
    \left (\frac{Q^2}{\m^2},a_{\rm s},a_{\rm em},\e \right)
    Z_{\a\g}^N \left(a_{\rm s},a_{\rm em},\frac{1}{\e} \right)
    A_{\g\g}^{N,\,\rm tree}(\e) \:\: .
\eeq
The coefficient functions $C_{\phi,\a}$ have an expansion analogous to
Eq.~(\ref{cfcts}), but with $C_{\phi,\rm g} = {\cal O}(1)$ and 
$C_{\phi,\rm q} = {\cal O}(a_{\rm s})$. It is understood in 
Eq.~(\ref{TphN}) that the external gluon operator $G_{\m\n}G^{\m\n}$
employed for the coupling to the scalar field $\f$ has been 
renormalized according to \cite{ZG2} 
\beq
\label{ZG2}
  G^{\mu\nu} G_{\mu\nu} \: = \: Z_{G^2} \, \left( G^{\mu\nu} G_{\mu\nu} 
  \right)^{\rm bare}\, + \,\ldots \:\: , \quad\quad
  Z_{G^2} \: = \: \frac{1}{1-\beta(\alpha_{\rm s})/(\e\a_{\rm s})} 
  \:\: ,
\eeq
where the dots indicate mixing with unphysical operators which give 
vanishing contributions under the conditions of the present calculation.
Expanding Eq.~(\ref{TphN}) analogously to Eq.~(\ref{Texp}), the 
coefficients $T_{\phi,\g}^{(l)}$, $l = 1, 2$ read
\beq
\label{Tph1}
  T^{(1)}_{\phi, \g} \: = \:
  \frac{1}{\e^2} \left\{ \frac{1}{2} k^{(0)}_{\rm{q}} 
     \g^{(0)}_{\rm gq} \right\}
  + \frac{1}{\e} \left\{ \frac{1}{2} k^{(1)}_{\rm g} 
    + k^{(0)}_{\rm q} c^{(1)}_{\phi,\rm q} \right\}
  + c^{(2)}_{\phi,\g} + k^{(0)}_{\rm q} a^{(1)}_{\phi,\rm q} 
\eeq
and
\bea
\label{Tph2}
  T^{(2)}_{\phi,\g} & \! = \! & \:
  \frac{1}{\e^3} \left\{ \frac{1}{6} k^{(0)}_{\rm q} \g^{(0)}_{\rm gq} 
    \left( \g^{(0)}_{\rm qq} + \g^{(0)}_{\rm gg}  \right)
    - \frac{1}{6} \beta_0 k^{(0)}_{\rm q} \g^{(0)}_{\rm gq} \right\}
  \nn \\[0.5ex] & & \mbox{} \!\!\!\!\!\! +
  \frac{1}{\e^2} \left\{ 
    \frac{1}{2} k^{(0)}_{\rm q} \g^{(0)}_{\rm gq} c^{(1)}_{\phi,\rm g} 
    + \frac{1}{2} k^{(0)}_{\rm q}\g^{(0)}_{\rm qq} c^{(1)}_{\phi,\rm q} 
    + \frac{1}{3} k^{(0)}_{\rm q} \g^{(1)}_{\rm gq} 
    + \frac{1}{6} k^{(1)}_{\rm q} \g^{(0)}_{\rm gq} 
    + \frac{1}{6} k^{(1)}_{\rm g} \g^{(0)}_{\rm gg} 
    - \frac{1}{6} \beta_0 k^{(1)}_{\rm g} \right\}
  \nn \\[0.5ex] & & \mbox{} \!\!\!\!\!\! + \:
  \frac{1}{\e}\, \left\{ 
    \frac{1}{3} k^{(2)}_{\rm{g}} 
    + k^{(0)}_{\rm q} c^{(2)}_{\phi,{\rm q}} 
    + \frac{1}{2} k^{(1)}_{\rm g} c^{(1)}_{\phi,{\rm g}} 
    + \frac{1}{2} k^{(1)}_{\rm q} c^{(1)}_{\phi,{\rm q}} 
    + \frac{1}{2} k^{(0)}_{\rm q}\g^{(0)}_{\rm gq} a^{(1)}_{\phi,\rm g} 
    + \frac{1}{2} k^{(0)}_{\rm q}\g^{(0)}_{\rm qq} a^{(1)}_{\phi,\rm q} 
   \right\} 
  \: . \quad\quad
\end{eqnarray}
In Eq.~(\ref{Tph2}) we have not written out the $\e^0$ term fixing 
the unphysical coefficient function $c^{(3)}_{\phi,\g}$.
%
%
\setcounter{equation}{0}
\section{Moments: calculation and results}
%
%
The calculation of the moments (\ref{T2LN}) and (\ref{TphN}) of the 
Green functions $T^{\,\g\g}$ and $T^{\f\g}$ can be performed quite
analogously to that of $T^{\,\g \rm g}$ and $T^{\,\f \rm g}$ in refs.\
\cite{moms,RV00} (where these quantities are denoted as $T^{\rm g \g
 g \g}$ and $T^{\rm g \f g \f}$), to which the reader is referred for a 
more detailed discussion. Indeed, the Feynman diagrams contributing to 
$T^{\,\g\g}$ ($T^{\f\g}$) derive from a subset of those for $T^{\,\g 
\rm g}$ ($T^{\f \rm g}$): at three loops 117 (57) out of 366 (7162) 
diagrams contribute, respectively, using the counting of refs.~\cite
{moms,RV00}. Despite this reduction we are not able to compute higher
moments than obtained for the hadronic case in ref.~\cite{RV00}, as
some of the most storage-consuming diagrams remain.

The moments $N = 2,\, \ldots ,\, 8$ have been calculated from scratch.
The diagrams are generated using a special version of QGRAF 
\cite{QGRAF}. The actual computation is done using optimized FORM
\cite{FORM} programs using the MINCER package \cite{MINC} for the 
scalar three-loop integrals. The moments $N = 2, 4$ of $T^{\,\g\g}$ and 
$N = 2$ of $T^{\,\f\g}$ have been computed in an arbitrary covariant 
gauge, i.e., keeping the gauge parameter $\xi$ in the gluon propagator, 
$ i\, [-g^{\m\n}+(1-\xi) q^{\m}q^{\n}/(q^2+i\e)]/(q^2+i\e)$ 
as a free parameter.  The explicit cancellation of the gauge dependence 
in the anomalous dimensions and coefficient functions provides an 
important check of the results.
The direct calculation of the moments $N\! =\! 10$ and $N\! =\! 12$ is 
rather time-consuming and requires running FORM on a computer with 
64-bit architecture instead on a standard PC~\cite{RV00}. 
Therefore we have for these moments (and as a further check also for 
the lower moments) made use of the diagram database of ref.~\cite{RV00} 
by replacing the colour factors of $T^{\,\g\rm g}$ and $T^{\,\f \rm g}$ 
by those for our cases and then re-assembling the integrated results of 
all diagrams, thus sidestepping the involved parts of the computation. 

{}From the results of these calculations the ${\cal O} (a_{\rm em}^{} 
a_{\rm s}^2)$ photon-parton anomalous dimensions and photon coefficient 
functions at $N = 2,\, \ldots ,\, 12$ can be obtained by means of 
Eqs.~(\ref{Texp}), (\ref{Tsns}), (\ref{T2L0})--(\ref{T2L2ps}) and 
(\ref{Tph1})--(\ref{Tph2}). Here we present the results in numerical 
form; the full analytic expressions can be found in Appendix A.
The non-singlet ($\rm p = ns$) and singlet ($\rm p = s$) photon-quark 
anomalous dimensions read
\bea
\label{kqmom}
  (\d_{\rm p}\, a_{\rm em})^{-1}\, k_{\rm p}^{N=2} & = &
    \mbox{} - 1.333333333 \:\:
    - a_{\rm s}\: 7.308641975 
    \nn \\ & & \mbox{}
    + a_{\rm s}^2\,(\, - 3.368998628\,f_{\rm ps}\, \Nf 
    - 86.97527479 + 1.470507545\,\Nf \,)
  \nn \\[0.5ex]
  (\d_{\rm p}\, a_{\rm em})^{-1}\, k_{\rm p}^{N=4} & = &
    \mbox{} - 0.7333333333
    - a_{\rm s}\: 8.343259259
    \nn \\ & & \mbox{}
    + a_{\rm s}^2\,(\, - 0.4223537997\,f_{\rm ps}\, \Nf 
    - 102.8310333 + 1.477370645\,\Nf \,)
  \nn \\[0.5ex]
  (\d_{\rm p}\, a_{\rm em})^{-1}\, k_{\rm p}^{N=6} & = &
    \mbox{} - 0.5238095238
    - a_{\rm s}\: 8.525587590 
    \nn \\ & & \mbox{}
    + a_{\rm s}^2\,(\, - 0.1483655726\,f_{\rm ps}\, \Nf 
    - 109.2777760 + 1.656530217\,\Nf \,)
  \nn \\[0.5ex]
  (\d_{\rm p}\, a_{\rm em})^{-1}\, k_{\rm p}^{N=8} & = &
    \mbox{} - 0.4111111111
    - a_{\rm s}\: 8.343215919
    \nn \\ & & \mbox{}
    + a_{\rm s}^2\,(\, - 0.06961022336\,f_{\rm ps}\, \Nf 
    - 111.1669036 + 1.695497529\,\Nf \,)
  \nn \\[0.5ex]
  (\d_{\rm p}\, a_{\rm em})^{-1}\, k_{\rm p}^{N=10} \!\! & = &
    \mbox{} - 0.3393939394
    - a_{\rm s}\: 8.045400445
    \nn \\ & & \mbox{}
    + a_{\rm s}^2\,(\, - 0.03823222708\,f_{\rm ps}\, \Nf 
    - 111.0346700 + 1.670607074\,\Nf \,)
  \nn \\[0.5ex]
  (\d_{\rm p}\, a_{\rm em})^{-1}\, k_{\rm p}^{N=12} \!\! & = &
    \mbox{} - 0.2893772894
    - a_{\rm s}\: 7.720564967
    \\ & & \mbox{}
    + a_{\rm s}^2\,(\, - 0.02324832852\,f_{\rm ps}\, \Nf 
    - 109.9430144 + 1.619078961 \,\Nf \,) \:\: , \quad
    \nn
\eea
where the factors $\d_{\rm ns}$ and $\d_{\rm s}$ have been defined in
Eq.~(\ref{delta}), and
\beq
\label{fps}
  f_{\rm ps} \: \equiv \: \left\{ \begin{array}{l}
  0 \:\: , \quad\quad {\rm p = ns} \\
  1 \:\: , \quad\quad {\rm p = s} \:\: . \end{array} \right.
\eeq
The corresponding results for the photon-gluon anomalous dimensions are
given by
\bea
\label{kgmom}
  (\d_{\rm s}\, a_{\rm em})^{-1}\, k_{\rm g}^{N=2} & = &
    a_{\rm s}\: 1.975308642 \:\:\:  + \, 
    a_{\rm s}^2\,(\, 31.41971923  + 5.157750343\,\Nf \,)
  \nn \\[0.5ex]
  (\d_{\rm s}\, a_{\rm em})^{-1}\, k_{\rm g}^{N=4} & = &
    a_{\rm s}\: 0.8743703704 \, + \, 
    a_{\rm s}^2\,(\, 23.94271102  + 1.108863649\,\Nf \,)
  \nn \\[0.5ex]
  (\d_{\rm s}\, a_{\rm em})^{-1}\, k_{\rm g}^{N=6} & = &
    a_{\rm s}\: 0.4439549365 \, + \, 
    a_{\rm s}^2\,(\, 15.65166504  + 0.6959529917\,\Nf \,)
  \nn \\[0.5ex]
  (\d_{\rm s}\, a_{\rm em})^{-1}\, k_{\rm g}^{N=8} & = &
    a_{\rm s}\: 0.2669916389 \, + \, 
    a_{\rm s}^2\,(\, 10.96606146  + 0.4981961912\,\Nf \,)
  \nn \\[0.5ex]
  (\d_{\rm s}\, a_{\rm em})^{-1}\, k_{\rm g}^{N=10} \!\! & = &
    a_{\rm s}\: 0.1780490673 \, + \, 
    a_{\rm s}^2\,(\, 8.160308004  + 0.3790604508\,\Nf \,)
  \\[0.5ex]
  (\d_{\rm s}\, a_{\rm em})^{-1}\, k_{\rm g}^{N=12} \!\! & = &
    a_{\rm s}\: 0.1271644566 \, + \, 
    a_{\rm s}^2\,(\, 6.348294844  + 0.3002739601\,\Nf \,)
   \nn \:\: . \quad
\eea
For the scale choice $\mu^2 = Q^2$ the photon coefficient functions for
$F_2$ and $F_L$ at $N = 2,\,\ldots,\, 12$ read
\bea
\label{c2mom}
  (\d_{\rm p}\, a_{\rm em})^{-1}\, c_{2,\g}^{\,{\rm p},N=2} & = &
    \mbox{} - 1 \:\:
    - a_{\rm s}\: 5.54162471
    \nn \\ & & \mbox{}
    + a_{\rm s}^2\,(\, - 5.818386905\,f_{\rm ps}\, \Nf 
    - 83.61538634 + 12.00984563\,\Nf \,)
  \nn \\[0.5ex]
  (\d_{\rm p}\, a_{\rm em})^{-1}\, c_{2,\g}^{\,{\rm p},N=4} & = &
    \mbox{} - 1.477777778
    - a_{\rm s}\: 12.20420558
    \nn \\ & & \mbox{}
    + a_{\rm s}^2\,(\, - 12.34431320\,f_{\rm ps}\, \Nf 
    - 270.6890263 + 22.99272397\,\Nf \,)
  \nn \\[0.5ex]
  (\d_{\rm p}\, a_{\rm em})^{-1}\, c_{2,\g}^{\,{\rm p},N=6} & = &
    \mbox{} - 1.410317460
    - a_{\rm s}\: 16.88045086
    \nn \\ & & \mbox{}
    + a_{\rm s}^2\,(\, - 10.77343760\,f_{\rm ps}\, \Nf 
    - 461.1411168 + 31.41856945\,\Nf \,)
  \nn \\[0.5ex]
  (\d_{\rm p}\, a_{\rm em})^{-1}\, c_{2,\g}^{\,{\rm p},N=8} & = &
    \mbox{} - 1.288174603
    - a_{\rm s}\: 19.73136088
    \nn \\ & & \mbox{}
    + a_{\rm s}^2\,(\, - 8.964615998\,f_{\rm ps}\, \Nf 
    - 617.8190552 + 37.24672336\,\Nf \,)
  \nn \\[0.5ex]
  (\d_{\rm p}\, a_{\rm em})^{-1}\, c_{2,\g}^{\,{\rm p},N=10} \!\! & = &
    \mbox{} - 1.172255892
    - a_{\rm s}\: 21.44335073
    \nn \\ & & \mbox{}
    + a_{\rm s}^2\,(\, - 7.480750665\,f_{\rm ps}\, \Nf 
    - 741.1418081 + 41.20962391\,\Nf \,)
  \nn \\[0.5ex]
  (\d_{\rm p}\, a_{\rm em})^{-1}\, c_{2,\g}^{\,{\rm p},N=12} \!\! & = &
    \mbox{} - 1.071686118
    - a_{\rm s}\: 22.46033065
    \\ & & \mbox{}
    + a_{\rm s}^2\,(\, - 6.308511998\,f_{\rm ps}\, \Nf 
    - 837.9184018 + 43.90927770 \,\Nf \,) \quad
    \nn
\eea
and
\bea
\label{clmom}
  (\d_{\rm p}\, a_{\rm em})^{-1}\, c_{L,\g}^{\,{\rm p},N=2} & = &
    \mbox{}  1.333333333 \:\:
    - a_{\rm s}\: 12.54891027
    \nn \\ & & \mbox{}
    + a_{\rm s}^2\,(\, 8.693227731\,f_{\rm ps}\, \Nf 
    - 339.8638216 + 26.87433155\,\Nf \,)
  \nn \\[0.5ex]
  (\d_{\rm p}\, a_{\rm em})^{-1}\, c_{L,\g}^{\,{\rm p},N=4} & = &
    \mbox{} 0.5333333333
    - a_{\rm s}\: 6.551703704
    \nn \\ & & \mbox{}
    + a_{\rm s}^2\,(\, - 0.8773595064\,f_{\rm ps}\, \Nf 
    - 208.9707023 + 16.63839609\,\Nf \,)
  \nn \\[0.5ex]
  (\d_{\rm p}\, a_{\rm em})^{-1}\, c_{L,\g}^{\,{\rm p},N=6} & = &
    \mbox{} 0.2857142857
    - a_{\rm s}\: 3.966771047
    \nn \\ & & \mbox{}
    + a_{\rm s}^2\,(\, - 1.452744800\,f_{\rm ps}\, \Nf 
    - 138.7560249 + 10.96941052\,\Nf \,)
  \nn \\[0.5ex]
  (\d_{\rm p}\, a_{\rm em})^{-1}\, c_{L,\g}^{\,{\rm p},N=8} & = &
    \mbox{} 0.1777777778
    - a_{\rm s}\: 2.670167875
    \nn \\ & & \mbox{}
    + a_{\rm s}^2\,(\, - 1.281927733\,f_{\rm ps}\, \Nf 
    - 99.25459329 + 7.787570553\,\Nf \,)
  \nn \\[0.5ex]
  (\d_{\rm p}\, a_{\rm em})^{-1}\, c_{L,\g}^{\,{\rm p},N=10} \!\! & = &
    \mbox{} 0.1212121212
    - a_{\rm s}\: 1.927770988
    \nn \\ & & \mbox{}
    + a_{\rm s}^2\,(\, - 1.056290000\,f_{\rm ps}\, \Nf 
    - 74.93923167 + 5.838900780\,\Nf \,)
  \nn \\[0.5ex]
  (\d_{\rm p}\, a_{\rm em})^{-1}\, c_{L,\g}^{\,{\rm p},N=12} \!\! & = &
    \mbox{} 0.08791208791
    - a_{\rm s}\: 1.461875421
    \\ & & \mbox{}
    + a_{\rm s}^2\,(\, - 0.8667773331\,f_{\rm ps}\, \Nf 
    - 58.86856783 + 4.557749952 \,\Nf \,) \:\: . \quad
    \nn
\eea
The additional terms for $\m^2 \neq Q^2$ do not contain independent 
information, see the next section.
%
%
\setcounter{equation}{0}
\section{Parton distributions and evolution equations}
%
%
In this section we outline the parton formulation of the photon 
structure (introduced in ref.~\cite{WJSWW}) \linebreak
at the next-to-next-to-leading order (NNLO) of perturbative QCD. The 
number \mbox{distributions} of quarks and gluons in the fractional 
photon momentum $x$ are denoted by $q_j^{\,\g}(x,\mu_f^2, \mu_r^2)$ and 
$g^\g(x,\mu_f^2,\mu_r^2)$, where the subscript $j$ indicates the quark 
flavour.  $\mu_f$ represents the mass-factorization scale which, unlike 
in Sect.~2, is not generally identified with the coupling-constant 
renormalization scale denoted by $\mu_r$ from now on.  The photon's 
quark and antiquark distributions are equal due to charge conjugation 
invariance.

First we focus on the flavour-singlet distributions for which we use 
the notation
\beq
\label{qs}
  \qV^\g \: = \: \left( \begin{array}{c} 
              \!\S^{\,\g}\! \\ \!g^{\,\g}\! \end{array} \right)
  \:\: , \quad\quad
  \S^{\,\g} \:\equiv\:\sum_{j=1}^{\Nf}\, ( q_j^{\,\g}+\bar{q}_j^{\,\g}) 
         \: = \: 2\,\sum_{j=1}^{\Nf} q_j^{\,\g} \:\: ,
\eeq
where $\S^{\,\g}$ is the singlet quark density and $\Nf$, as before, 
stands for the number of effectively massless quark flavours.  As in 
some other equations below, the dependence on $x$ and the scales has 
been suppressed in Eq.~(\ref{qs}). At lowest order in the 
electromagnetic coupling $a_{\rm em}$ these distributions are subject 
to the evolution equations  
\beq
\label{qevol}
  \frac{d\, \qV^\g}{d \ln \mu_f^2} \: = \: 
  \PV^\g + \PV \otimes \qV^\g \:\: ,
\eeq
where $\otimes$ represents the Mellin convolution in the momentum
variable,
\beq
\label{conv}
  [ a \otimes b ](x) \:\equiv\: \int_x^1 \! \frac{dy}{y} \: a(y)\,
  b\bigg(\frac{x}{y}\bigg) \:\: ,
\eeq
and the splitting-function matrices are given by
\beq
\label{Pmat}
  \PV^\g \: = \: \left( \begin{array}{c} 
    {\cal P}_{\rm q\g} \\ {\cal P}_{\rm g\g} \end{array} \right) 
  \:\: , \quad\quad
  \PV \: = \: \left( \begin{array}{cc}
    {\cal P}_{\rm qq} & {\cal P}_{\rm qg} \\
    {\cal P}_{\rm gq} & {\cal P}_{\rm gg} \end{array} \right) \:\: .
\eeq
The NNLO expansions of the photon-parton and parton-parton splitting 
functions read
\bea
\label{Pexp}
  \PV^\g \Big( x, a_{\rm em}, a_{\rm s}, L_R \Big) \!\! & =\!\! & \quad
      a_{\rm em}\, \PV_\g^{(0)}(x) + a_{\rm em} a_{\rm s}\, 
      \PV_\g^{(1)}(x) 
  \nn\\ & & \mbox{}\!\!
    + \, a_{\rm em} a_{\rm s}^2 \, \Big( \PV_\g^{(2)}(x) - \beta_0 L_R 
      \PV_\g^{(1)}(x) \Big)
    + \ldots \nn \\[1ex]
  \PV \Big( x, a_{\rm s}, L_R \Big) \:\: & =\!\! & \quad
      a_{\rm s} \, \PV^{(0)}(x) \,
    + a_{\rm s}^2\, \Big( \PV^{(1)}(x) - \beta_0 L_R \PV^{(0)}(x) \Big)
  \:  \\ & & \mbox{}\!\!
    + a_{\rm s}^3 \, \Big( \PV^{(2)}(x) - 2\beta_0 L_R \PV^{(1)}(x)
    - \Big\{ \beta_1 L_R - \beta_0^2 L_R^2 \Big\} \PV^{(0)}(x) \Big)
    + \ldots \nn \:\: . \quad
\eea
The notation for the coupling constants and $\beta_0$ has been defined
in Eqs.~(\ref{coupl}) and (\ref{bQCD}) above, and we abbreviate the
scale logarithms by
\beq
\label{LMLR}
   L_M \:\equiv\: \ln \frac{Q^2}{\mu_f^2} \:\: , \quad
   L_R \:\equiv\: \ln \frac{\mu_f^2}{\mu_r^2} \:\: .
\eeq
The expansion coefficients in Eq.~(\ref{Pexp}) are related to the 
(even-integer $N$) anomalous dimensions in Eq.~(\ref{adims}) by
\beq
  k^{(l)N}_{\rm p} \: = \: 
    - \int_0^1 \! dx \, x^{N-1} P^{(l)}_{\rm p\g}(x)
  \:\: , \quad\quad 
  \g^{(l)N}_{\rm pp'} \: = \: 
    - \int_0^1 \! dx \, x^{N-1} P^{(l)}_{\rm pp'}(x) \:\: .
\eeq
The $x$-space splitting functions, in turn, are uniquely fixed by the
inverse Mellin transform if the anomalous dimensions are known for all 
even moments $N$.

The evolution equations for the non-singlet combinations of the 
photon's quark densities, viz
\beq
\label{qns}
  q^\g_{{\rm ns},ij} \: = \: q^\g_i - q^\g_j 
  \:\: , \quad\quad 1 \leq i \!\neq\! j \leq \Nf
\eeq
and linear combinations thereof, are obtained from Eqs.~(\ref{qevol}) 
and (\ref{Pexp}) by replacing $\PV$ and $\PV^\g$ by their scalar 
non-singlet counterparts $P_{\rm ns}^+$ and $P_{ij}^\g$. The hadronic
quantities $P_{\rm ns}^{\pm ,V}$ have been discussed, e.g., in 
refs.~\cite{NV1,NV3}; the splitting functions $P_{ij}^\g$ are related 
to the non-singlet anomalous dimensions in Eq.~(\ref{adims}) by
\beq
  3 (e_{q_i^{}}^2 - e_{q_j^{}}^2) \,\d_{\rm ns}^{-1} k^{(l)N}_{\rm ns} 
  \: = \: - \int_0^1 \! dx \, x^{N-1} P^{(l)}_{ij,\g}(x) \:\: .
\eeq
Due to $\bar{q}_j^{\,\g} = q_j^{\,\g}$ there are no combinations
evolving with the splitting functions $P_{\rm ns}^-$ and $P_{\rm ns}^V$ 
here, unlike for the nucleon structure. Hence the calculation of 
e.m.~DIS in Sect.~2 and Sect.~3, if extended to all moments, is 
sufficient to fix all NNLO splitting functions relevant to the photon 
structure. Note also that differences of quark densities of equal 
charge, like $d^{\,\g} - s^\g$, evolve like hadronic non-singlet 
distributions, i.e., without an inhomogeneous term in Eq.~(\ref{qevol}).

The anomalous dimensions $k_{\rm p}^{(l)N}$, p = ns, q, g, for the 
next-to-leading order (NLO, $l \!\leq\! 1$) evolution have first been 
written down in ref.~\cite{BaBu}, their $x$-space counterparts 
$P_{\rm p\g}^{(1)}(x)$ in ref.~\cite{GR83}.\footnote
{Our notation for the anomalous dimensions $k_{\rm p}^{(l)N}$ differs 
from that for $K_{\rm p}^{(l),n}$ in ref.~\cite{BaBu} (where the 
analogue of Eq.~(\ref{Oren}) is written in terms of $d/d \ln \m$ 
instead of $d/d \ln \m^2$) by a factor -1/2. Our splitting functions 
$P_{\rm p\g}^{(l)}(x)$ are larger by a factor $2^{l+1}$ than their
counterparts $k_{\rm p}^{(l)}(x)$ in refs.~\cite{GR83,FP92,GRVg1},
where the expansion parameters are normalized as $\alpha_j/2\pi$, $j$ = 
em, s, instead of Eq.~(\ref{coupl}).}
Both results are correct except for the constant-{\it N} ($\,\!\d (1-x)
\,\!$) term erroneously included in $k_{\rm g}^{(1)N}$ ($\,\! 
P_{\rm g\g}^{(1)}(x)\,\!$), an error which has been corrected in 
refs.~\cite{FP92,GRVg1}. 
The $a_{\rm em}^{} a_{\rm s}^2$ contributions $P_{\rm p\g}^{(2)}(x)$ 
to the photon-parton splitting functions are not completely fixed 
at present, as the moments $N \geq 14$ remain uncalculated. Following 
the lines of refs.~\cite{NV1,NV2,NV3}, we will employ our six moments 
(\ref{kqmom}) and (\ref{kgmom}) to derive approximate results for 
$P_{\rm p\g}^{(2)}(x)$ in Sect.~5 below. 

In terms of the parton distributions (\ref{qs}) and (\ref{qns}), the 
${\cal O}(a_{\rm em}^1)$ electromagnetic structure functions 
${\cal F}_1^{\,\g} = \frac{1}{2}\, F_1^{\,\g}$ and ${\cal F}_2^{\,\g} = 
\frac{1}{x}\, F_2^{\,\g}$ are given by
\bea
\label{F12x}
  {\cal F}_i^{\,\g}(x,Q^2) & = & 
   C_{i,\g}^{\,\rm ns} (x,a_{\rm em}, a_{\rm s}, L_M, L_R) \, + 
  \Big[ C_{i,\rm ns}(x, a_{\rm s}, L_M, L_R) \otimes 
  q_{\rm em}^{\,\g} (\m_f^2, \m_r^2) \Big] (x)
  \nn \\ & & \mbox{\hspace{-1.35cm}} 
  + \langle e^2 \rangle \left( 
  C_{i,\g}^{\,\rm s} (x,a_{\rm em}, a_{\rm s}, L_M, L_R) \, + 
  \Big[ \CV_i\, (x, a_{\rm s}, L_M, L_R) \,\otimes\, 
  \qV^{\g} (\m_f^2, \m_r^2) \Big] (x) \right) \:\: . \quad\quad
\eea
The first line in Eq.~(\ref{F12x}) represents the non-singlet
contribution involving the combination
\beq
  q_{\rm em}^{\,\g} \: = \: 2\, \sum_{j=1}^{\Nf} \left( 
  e_{q_j}^2 - \langle e^2 \rangle \right) q_{j}^{\,\g} 
\eeq
of quark densities. The $x$-space coefficient functions $C_{i,\g}(x)$
and $C_{i,\rm p}(x)$ are the inverse Mellin transforms of the $\e^0$ 
terms in Eq.~(\ref{cfcts}), and in the second line of Eq.~(\ref{F12x}) 
we have employed the matrix notation $\CV_i = (\,C_{i,\rm q},\, 
C_{i,\rm g}\, ) $ for the hadronic singlet coefficient functions. Up to 
order $a_{\rm em}^{} a_{\rm s}^2$ the expansion of the photonic 
coefficient functions takes the form
\bea
\label{Cpexp}
  a_{\rm em}^{-1}\, C_{i,\g}(x, a_{\rm em}, a_{\rm s}, L_M, L_R) 
  & = &
    C_{i,\g}^{(1)}(x, L_M) \, +\, 
    a_{\rm s} C_{i,\g}^{(2)}(x,L_M) \, + \,
    a_{\rm s}^2 C_{i,\g}^{(3)}(x,L_M,L_R) 
  \nn \\
  & \!\!\stackrel{{\mu_r = \mu_f}}{=}\!\! &
    \sum_{l=1}^{3}\, a_{\rm s}^{l-1}\left( c_{i,\g}^{(l)}(x) +
    \sum_{m=1_{}}^{l} c_{i,\g}^{(l,m)}(x) L_M^m \right) \:\: .
\eea
The first-order contributions $c_{i,\g}^{(1)}$ have been obtained in
ref.~\cite{BaBu} by exploiting their simple relation to the NLO gluon
coefficient functions $c_{i,\rm g}^{(1)}$ \cite{BBDM}. Also the 
second-order coefficients $c_{i,\g}^{(2)}$ can readily be inferred from 
their gluonic counterparts $c_{i,\rm g}^{(2)}$ first calculated in 
ref.~\cite{ZvN}; explicit expressions will be presented in Sect.~5 
below. The terms up to order $a_{\rm em} a_{\rm s}$ contribute to 
$F_1^{\,\g}$ and $F_2^{\,\g}$ in the NNLO approximation. The 
$a_{\rm em}^{} a_{\rm s}^2$ parts of Eq.~(\ref{Cpexp}) only enter for 
$dF_{1,2}/d\ln Q^2$ to this accuracy, hence the scale-independent 
quantities $c_{i,\g}^{(3)}$, $i = 1, 2$, are not required.

The coefficients $c_{i,\g}^{(l,m)}(x)$ of the scale logarithms in the 
second line of Eq.~(\ref{Cpexp}) can be derived by identifying the 
results of the following two calculations of $(d / d\, \ln Q^2)^l\,
F_{i}$ for $l = 1,\, 2,\, 3$  at $\mu_r^2 =\mu_f^2 = Q^2\,$: 
{\bf (a)} with the scales identified in the beginning, using
Eq.~(\ref{arun}) for $D=4$ and $j = {\rm s}$ together with 
Eq.~(\ref{qevol}), and 
{\bf (b)} with the scales set equal only at the end, after 
differentiating the logarithms in Eq.~(\ref{Cpexp}). 
For the singlet contributions one finds
\bea
\label{clm}
  c_{i,\g}^{(1,1),\rm s} &\!  = \! & 
    \cV_i^{(0)} \otimes \PV_\g^{(0)} \nn \\
  c_{i,\g}^{(2,1),\rm s} &\!  = \! & 
    \cV_i^{(0)} \otimes \PV_\g^{(1)}
  + \cV_i^{(1)} \otimes \PV_\g^{(0)} \nn \\
  c_{i,\g}^{(2,2),\rm s} &\!  = \! & \frac{1}{2}\,  
    \cV_i^{(0)} \otimes \PV^{(0)} \otimes \PV_\g^{(0)} 
   \nn \\
  c_{i,\g}^{(3,1),\rm s} &\!  = \! & 
    \cV_i^{(0)} \otimes \PV_\g^{(2)} 
  + \cV_i^{(1)} \otimes \PV_\g^{(1)} 
  + \cV_i^{(2)} \otimes \PV_\g^{(0)} 
  - \beta_0\, c_{i,\g}^{(2),\rm s} \nn \\
  c_{i,\g}^{(3,2),\rm s} &\!  = \! & \frac{1}{2} \bigg\{ 
    \cV_i^{(0)} \otimes \Big( ( \PV^{(0)} - \beta_0\uV ) \otimes 
    \PV_\g^{(1)} + \PV^{(1)} \otimes \PV_\g^{(0)} \Big)
  + \cV_i^{(1)} \otimes (\PV^{(0)} - 2\beta_0\uV ) \otimes \PV_\g^{(0)}
    \bigg\} \quad
   \nn \\
  c_{i,\g}^{(3,3),\rm s} &\!  = \! & \frac{1}{6}\, 
    \cV_i^{(0)} \otimes \PV^{(0)} \otimes (\PV^{(0)} - 2 \beta_0 \uV) 
      \otimes \PV_\g^{(0)} \:\: ,
\eea
where
$
  \cV_i^{(0)}(x) \equiv (\:\! c_{i,q}^{(0)}(x)\, , \,
  c_{i,g}^{(0)}(x)\:\! ) = ( \:\!\delta (1-x)\, ,\, 0\:\! )
$
is the parton model result for lepton-hadron DIS, and $\uV$ represents 
the $2\!\times\!2$ unit matrix multiplied by $\d (1-x)$. The 
corresponding non-singlet results are obtained from Eqs.~(\ref{clm}) by 
simply replacing all quantities on the right-hand-sides by their scalar 
non-singlet analogues.  
Finally the coefficients $C_{i,\g}^{(3)}$ for $\mu_f \neq \mu_r$ in 
Eq.~(\ref{Cpexp}) are obtained from these results by expanding 
$a_{\rm s} (\m_f^2)$ in terms of $a_{\rm s}(\m_r^2)$, 
\beq
\label{cmur}
  C_{i,\g}^{(3)}(x,L_M,L_R) \: =\: C_{i,\g}^{(3)}(x,L_M,0) 
  - \beta_0 L_R \,C_{i,\g}^{(2)}(x,L_M)  \:\: .
\eeq
The relations for the hadronic coefficient functions $\CV_i(x)$ 
corresponding to Eqs.~(\ref{Cpexp})--(\ref{cmur}) are given in 
Eqs.~(2.16)--(2.18) of ref.~\cite{NV2}. 

The decomposition (`factorization') of the r.h.s.\ of Eq.~(\ref{F12x}) 
into coefficient functions and parton densities is not unique beyond 
the leading order (LO), where the structure functions 
${\cal F}_i^{\,\g}$, $i = 1, 2$, are simply given by 
$q_{\rm em}^{\, \g,\rm LO} + \,\langle e^2 \rangle\,\S^{\, \g,\rm LO}$.
Working, as always, at lowest order in $a_{\rm em}$, the general 
factorization scheme transformation (corresponding to a finite 
renormalization of the operators in Eq.~(\ref{OME})$\:\!$) of the 
singlet parton distributions (\ref{qs}) can be written as
\beq
  \tilde{\qV}^{\g} \: = \: \ZV^\g + \ZV \otimes \qV^{\g} \:\: .
\eeq
Here $\ZV^\g$, like $\qV^{\g}$, is a two-component vector and $\ZV$ 
a $2\!\times\! 2$ matrix. It is sufficient to discuss the scheme
transformations with all scales identified, hence we put $\m_r^2 \, = 
\mu_f^2 \, = Q^2$ in the following. By virtue of Eq.~(\ref{qevol}), the 
evolution equations for the transformed distributions 
$\tilde{\qV}^{\g}$ then read
\bea
\label{qtrf}
  \frac{d\, \tilde{\qV}^{\g}}{d\ln Q^2} & = & 
  \beta_{\rm s}\, \frac{d\ZV^\g}{da_{\rm s}} + \ZV \otimes \PV^\g 
  + \left( \beta_{\rm s}\, \frac{d\ZV}{da_{\rm s}} + \ZV \otimes \PV 
  \right) \otimes \ZV^{-1} \otimes 
  \Big( \tilde{\qV}^{\,\g} - \ZV^\g \Big)  
  \nn \\[1ex] & \equiv &
  \widetilde{\PV}^\g + \widetilde{\PV} \otimes \tilde{\qV}^{\g} 
  \:\: ,
\eea
where $\beta_{\rm s} = \beta_{\rm QCD}$ has been defined in 
Eq.~(\ref{bQCD}). The transformed coefficient functions read
\beq
\label{Ctrf}
  \widetilde{C}_{i,\g} \: = \: C_{i,\g} - \CV_i \otimes \ZV^{-1}
  \otimes \ZV^\g \:\: , \quad\quad
  \widetilde{\CV}_i \: = \: \CV_i \otimes \ZV^{-1} \:\: .
\eeq
The second term proportional to $\qV^{\g}$ in Eq.~(\ref{qtrf}) and 
the second relation in Eq.~(\ref{Ctrf}) are the standard expressions 
for scheme transformations of hadronic parton densities.

Like their gluonic counterparts $c_{i,\rm g}^{(l)}$, the photonic \MSb\
coefficient functions $c_{i,\g}^{(l)}$ are negative and singular for 
$x\!\ra\! 1$, the leading terms read $A_l\ln^{2l-1}(1-x)$ with $A_l>0$. 
Unlike the gluon case, however, the behaviour of $C_{i,\g}$ enters the 
structure functions $(\ref{F12x})$ directly, not via a regularizing 
convolution with a parton density. Hence, in the \MSb\ scheme, these 
singularities have to be compensated by the quark distributions which 
thus have to be rather different from the leading-order quantities at 
NLO and NNLO. Consequently the perturbative stability of the expansion 
in $a_{\rm s}$ is more manifest in factorization schemes in which the 
photonic coefficient functions are removed as far as possible. The 
minimal modification of the \MSb\ scheme that removes (at $\m_f^2 = 
Q^2$) this function for the most important structure function 
$F_2^{\,\g}$ is the DIS$_\g$ scheme introduced in ref.~\cite{GRVg1} 
(for a further discussion see also ref.~\cite{GRVfr}), where 
$F_2^{\,\g}$ is written as
\beq
\label{F2dsg}
  {\cal F}_2^{\,\g} \: = \: 
    C_{2,\rm ns} \otimes q^{\,\g, \rm DIS_\g}_{\rm em}
  + \,\langle e^2 \rangle\, \CV_2 \otimes \qV^{\,\g, \rm DIS_\g} 
\eeq
with the standard \MSb\ hadronic coefficient functions $C_{2,\rm ns}$ 
and $\CV_2$. Eq.~(\ref{F2dsg}) is obtained for the transformation
\beq
\label{Zdsg}
  \ZV^{\g,\rm DIS_\g} \: = \: \left( \begin{array}{c}
  c_{2,\g}^{(1)} + a_{\rm s}\, \Big( c_{2,\g}^{(2),\rm s} 
  - c_{2,\rm q}^{(1)} \otimes c_{2,\g}^{(1),\rm s} \Big) + \ldots 
  \\ 0 \end{array} \right)
  \:\: , \quad\quad
  \ZV^{\rm DIS_\g} \: = \: \uV \:\: .
\eeq
Inserting Eq.~(\ref{Zdsg}) into Eq.~(\ref{qtrf}) yields the NNLO
singlet photon-parton splitting functions in the DIS$_\g$ scheme in
terms of the \MSb\ splitting functions and coefficient functions,%
\footnote{Note that a minus sign is missing on the r.h.s.~of the second 
 line of Eq.~(3.1) in the journal version of ref.~\cite{GRVg1}.}  
\bea
\label{Pdsg}
  P_{\rm q\g}^{\rm DIS_\g} & \! = \! & P_{\rm q\g} 
  - a_{\rm s}\, P^{(0)}_{\rm qq} \otimes c^{(1),\rm s}_{2,\g}
  - a_{\rm s}^2 \left\{ P^{(1)}_{\rm qq} \otimes c^{(1),\rm s}_{2,\g}
  + \left( P^{(0)}_{\rm qq} + \beta_0 \,1 \right) \otimes
    \left( c^{(2),\rm s}_{2,\g} - c^{(1),\rm s}_{2,\g} \otimes 
    c^{(1)}_{2,{\rm q}} \right) \right\} \quad 
  \nn \\[1ex]
  P_{\rm g\g}^{\rm DIS_\g} & \! = \! & P_{\rm g\g} 
  - a_{\rm s}\, P^{(0)}_{\rm gq} \otimes c^{(1),\rm s}_{2,\g}
  - a_{\rm s}^2 \left\{ P^{(1)}_{\rm gq} \otimes c^{(1),\rm s}_{2,\g}
  + P^{(0)}_{\rm gq} \otimes \left( c^{(2),\rm s}_{2,\g} - 
    c^{(1),\rm s}_{2,\g} \otimes c^{(1)}_{2,{\rm q}} \right) \right\} 
    \:\: .
\eea
The corresponding non-singlet relation is obtained from the first line
of Eq.~(\ref{Pdsg}) by substituting the non-singlet counterparts of all 
quantities. Finally, due to Eq.~(\ref{Ctrf}), the photonic coefficient 
functions for ${\cal F}_1^{\,\g}$ in the DIS$_\g$ scheme coincide with 
those of $-{\cal F}_L^{\,\g}$ in the \MSb\ scheme. 

The structure functions $F_i^{\,\g}(x,Q^2)$ evolve approximately like
non-singlet quantities at large~$x$, since the photon's gluon density 
is much smaller than the quark distributions in this region and the
pure-singlet contributions to both the splitting functions and the 
coefficient functions are strongly suppressed at large $x$ (large $N$),
see, e.g.,  Eqs.~(\ref{kqmom}) and (\ref{c2mom}) for the photonic 
quantities. This non-singlet evolution can conveniently be recast in a 
form in which any dependence of the factorization scheme and the scale 
$\m_f$ is explicitly eliminated, viz
\beq
\label{Fevol}
  \frac{d}{d\ln Q^2}\, {\cal F}_{i,\rm ns}^{\,\g}(x,Q^2) \: = \:
  {\cal I}_i^{\,\g}(x,a_{\rm s},L_R) + \Big[ {\cal K}_i(a_{\rm s},L_R) 
  \otimes {\cal F}_{i,\rm ns}^{\,\g}(Q^2) \Big] (x) \:\: .
\eeq
The perturbative expansions of the `physical' evolution kernels 
${\cal I}_i^{\,\g}$ and ${\cal K}_i$, ($i = 1, 2$) for $\m_r^2 = Q^2$ 
 can be obtained by inserting $Z^\g = C_{i,\g}$ and $Z = C_{i,\rm ns}$ 
into the non-singlet analogue of Eq.~(\ref{qtrf}), yielding
\bea
\label{dFinh}
  a_{\rm em}^{-1}\, I_i^{\,\g} & = & 
     P_{\rm ns\g}^{(0)} \,
   + \, a_{\rm s} \left\{ P_{\rm ns\g}^{(1)} 
     + c_{i,\rm ns}^{(1)} \otimes P_{\rm ns\g}^{(0)} 
     - P_{\rm ns}^{(0)} \otimes c_{i,\g}^{(1),\rm ns} \right\} \,
   + \, a_{\rm s}^2 \left\{ P_{\rm ns\g}^{(2)} 
     + c_{i,\rm ns}^{(1)} \otimes P_{\rm ns\g}^{(1)} \right.
  \nn \\[0.5ex] & & \left. \mbox{} 
     + c_{i,\rm ns}^{(2)} \otimes P_{\rm ns\g}^{(0)}
     - P_{\rm ns}^{(0)} \otimes c_{i,\g}^{(2),\rm s}
     - \beta_0\, c_{i,\g}^{(2)\rm s}
     - \Big( P_{\rm ns}^{(1)} - \beta_0 \, c_{i,\rm ns}^{(1)} \Big) 
        \otimes c_{i,\g}^{(1),\rm s} \right\} \, + \, \ldots \quad
\eea
and
\bea
\label{dFhom}
   K_i & \! = \! & a_{\rm s} P_{\rm ns}^{(0)}  
   + a_{\rm s}^2 \left\{ P_{\rm ns}^{(1)} -\beta_0\, c_{i,\rm ns}^{(1)} 
    \right\}
   + a_{\rm s}^3 \left\{ P_{\rm ns}^{(2)} 
     - \beta_0 \Big( 2 c_{i,\rm ns}^{(2)} - c_{i,\rm ns}^{(1)} 
       \otimes c_{i,\rm ns}^{(1)} \Big)
     - \beta_1 c_{i,\rm ns}^{(1)} \right\}  +  \ldots \:\: . 
  \quad \nn \\ & & \mbox{} 
\eea
The kernels ${\cal I}_i^{\,\g}$ and ${\cal K}_i$ for $\mu_r^2 \neq Q^2$ 
can be reconstructed from these relations by inserting the expansion of
$a_{\rm s}(Q^2)$ as a power series in $a_{\rm s}(\mu_r^2)$. 
Eq.~(\ref{dFinh}) has been taken over from ref.~\cite{ph99}$\,$%
\footnote
{We have taken the opportunity to correct the two misprints (one sign 
 and one superscript in the $a_{\rm s}^2$ coefficient) in Eq.~(14) of 
 ref.~\cite{ph99}. These misprints have also been noticed in 
 ref.~\cite{Chyla}.}; 
the hadronic contribution (\ref{dFhom}) has been discussed up to order
$a_{\rm s}^5$ in ref.~\cite{NV4}.

The LO splitting functions $P_{\rm ns\g}^{(0)}$ and $P_{\rm ns}^{(0)}$, 
and the combinations of splitting functions and coefficient functions
enclosed in curved brackets in Eq.~(\ref{dFinh}) and (\ref{dFhom})
represent factorization scheme invariants. Hence, together with the
hadronic coefficient functions $c_{i,\rm ns}^{(1)}$, the quantities 
$P_{\rm ns}^{(1)}$, $c_{i,\g}^{(1),\rm ns}$ and $P_{\rm ns\g}^{(0)}$ 
have to be included for the NLO approximation of the photon's 
structure functions $F_{i,\rm ns}^{\,\g}$, $i = 1, 2$. Correspondingly, 
the NNLO evolution additionally requires, besides $c_{i,\rm ns}^{(2)}$,
the functions $P_{\rm ns}^{(2)}$, $c_{i,\g}^{(2),\rm ns}$ and 
$P_{\rm ns\g}^{(2)}$. In general, the terms up to the orders 
$a_{\rm em}^{} a_{\rm s}^k$ ($a_{\rm s}^{k+1}$) in the first (second) 
part of Eq.~(\ref{Pexp}) and $a_{\rm em}^{} a_{\rm s}^{k-1}$ 
($a_{\rm s}^k$) in the photonic (hadronic) coefficient functions,
respectively, contribute to the N$^{k^{}}$LO approximation together
with the coefficients $\beta_{l \leq k}$ in Eq.~(\ref{bQCD}).

We now turn to the NNLO solution of the evolution equation 
(\ref{qevol}). Here we confine ourselves to the (sufficiently general)
case of a constant ratio $\exp L_R = \m_f^2 /\m_r^2$. In this case the 
`{\it U}-matrix' technique, developed for the NLO hadronic evolution in 
ref.~\cite{FP82} and generalized to higher orders in refs.~\cite
{EKL,BV98}, can be applied. In what follows we use the abbreviations 
\beq
  a_{\rm s} \: = \: a_{\rm s}(\m_r^2 \!=\! e^{-L_R}\m_f^2) 
  \:\: , \quad\quad
  a^{}_{\rm 0} \: = \: a_{\rm s}(\m_{r,0}^2 \!=\! e^{-L_R}\m_{f,0}^2)
\eeq
and suppress all references to $x$ or the Mellin variable $N$. It is
understood that in $x$-space all products of quantities depending on 
$x$ have to be read as Mellin convolutions (\ref{conv}).

The general solution of Eq.~(\ref{qevol}) can be decomposed as 
\beq
  \qV^{\g}(\m_f^2) \: = \: \qV^{\,\g}_{\rm inhom}(\m_f^2) 
  \, + \, \qV^{\,\g}_{\rm hom}(\m_f^2) \:\: 
\eeq
with the boundary conditions
\beq
\label{qmu0}
  \qV^{\,\g}_{\rm inhom}(\m_{f,0}^2) \: = \: 0 \:\: , \quad\quad
  \qV^{\,\g}_{\rm hom}(\m_{f,0}^2) = \qV^{\g}(\m_{f,0}^2) \:\: .
\eeq
The homogeneous solution can be written as
\beq
\label{hsol}
  \qV^{\,\g}_{\rm hom}(\m_f^2) \: = \: \UV(a_{\rm s})
  \left( \frac{a_{\rm s}}{a^{}_0} \right)^{{- \footnotesize \RV}_0} 
  \UV^{-1} (a^{}_0) \:\qV^\g(\m_{f,0}^2)
\eeq
with
\beq
\label{Uexp}
  \RV_0   \: = \: \frac{1}{\beta_0} \, \PV^{(0)} \:\: , \quad\quad
  \UV(a)  \: = \: \uV + a\, \UV_1 + a^2\, \UV_2 + \ldots \:\: .
\eeq
The expansion coefficients $\UV_k$ are complicated functions of the 
splitting-function matrices $\PV^{(l\leq k)}$ in Eq.~(\ref{Pexp}) and
the coefficients $\beta_{l\leq k}$ of the $\beta$-function (\ref{bQCD}) 
of QCD. A recursive explicit representation (see ref.~\cite{BV98} for
a derivation) is given by
\beq
\label{Ures}
  \UV_k \: = \: 
  - \frac{1}{k} \Big[ \eV_{-} \widetilde{\RV}_k \eV_{-} +
    \eV_{+} \widetilde{\RV}_k \eV_{+} \Big]
  + \frac{\eV_{+} \widetilde{\RV}_k \eV_{-}}{r_{-} - r_{+_{}} - k}
  + \frac{\eV_{-} \widetilde{\RV}_k \eV_{+}}{r_{+} - r_{-_{}} - k} 
  \:\: .
\eeq
Here $r_{\pm}$ denote the eigenvalues of $\RV_0$ in Eq.~(\ref{Uexp}) 
and $\eV_{\pm}$ the corresponding projectors, viz
\beq
  \RV_0 \: = \: r_{-} \eV_{-} + r_{+} \eV_{+}
\eeq
with
\bea
  r_{\pm}^{} & = & 
  \frac{1}{2 \beta_0} \bigg[ P_{\rm qq}^{(0)} + P_{\rm gg}^{(0)}
  \pm \sqrt{ \Big( P_{\rm qq}^{(0)} - P_{\rm gg}^{(0)}\Big) ^2 +
  4 P_{\rm qg}^{(0)} P_{\rm gq}^{(0)} } \, \bigg]
  \nn \\[1ex] 
  \eV_{\pm}^{} & = & 
  \frac{1}{r_{\pm}-r_{\mp}} \Big[ \RV_0-r_{\mp} \uV \Big] \:\: .
\eea
The matrices $\widetilde{\RV}_k$ in Eq.~(\ref{Ures}) read
$\widetilde{\RV}_1 \: = \: \RV_1$ and 
\beq
  \widetilde{\RV}_{k>1} \: = \: \RV_k + \sum_{l=1}^{k-1}\, 
  \RV_{k-l} \UV_l \:\: , \quad\quad
  \RV_{\, k\geq 1} \: = \: \frac{1}{\beta_0} \widehat{\PV}^{(k)} 
  - \sum_{l=1}^{k} \,\frac{\beta_l}{\beta_0}\, \RV_{k-l} \:\: ,
\eeq
where $\widehat{\PV}^{(k)}$ represents the coefficients of 
$a_{\rm s}^{k+1}$ in the second part of Eq.~(\ref{Pexp}).

The coefficients $\UV_1$ and $\UV_2$ in Eq.~(\ref{Uexp}) are required 
at NNLO. The quantities $\UV_{k > 2}$, on the other hand, receive 
contributions from beyond-NNLO splitting functions, hence these terms 
can be left out. It is understood that then also $\UV^{-1}(a_0^{})$ is
expanded, and that all terms beyond second order in the coupling 
constant are removed in Eq.~(\ref{hsol}). In this manner the spurious
poles in Eq.~(\ref{Ures}) at $N$-values where $r_{-} - r_{+} \pm k$ 
vanishes are completely cancelled in the solution (\ref{hsol}).

The inhomogeneous solution $\qV^{\,\g}_{\rm inhom}$ with the boundary
condition (\ref{qmu0}) is given by
\beq
\label{isol}
  a_{\rm em}^{-1} \: \qV^{\,\g}_{\rm inhom}(a_{\rm s}) \: = \: 
  - \UV(a_{\rm s}) \, a_{\rm s}^{{\footnotesize -\RV}_0} 
  \int_{a^{}_0}^{a_{\rm s}} \! da \, a^{ {\footnotesize \RV}_0 - 2} 
  \,\UV^{-1}(a) \,\RV^{\g}(a) 
\eeq
with
\beq
\label{Rgexp}
  \RV^{\g}(a_{\rm s}) \: = \: \sum_{k=0} \, a_{\rm s}^k \, \RV^{\g}_k 
  \: \equiv \: \frac{1}{\beta_0} \PV^{(0)}_\g + \sum_{k=1} a_{\rm s}^k
  \left( \frac{1}{\beta_0} \widehat{\PV}^{(k)}_\g - \sum_{l=1}^{k} \,
  \frac{\beta_{l}} {\beta_0}\, \RV^{\g}_{k-l} \right) \:\: ,
\eeq
where $\widehat{\PV}^{(k)}$ stands for the coefficients of 
$a_{\rm em}^{} a_{\rm s}^{k}$ in the first part of Eq.~(\ref{Pexp}).
Using Eqs.~(\ref{Uexp}) and (\ref{Rgexp}), the expansion of 
Eq.~(\ref{isol}) leads to
\bea
\label{iexp}
  a_{\rm em}^{-1}\: \qV_{\rm inhom}^{\,\g}(\m_f^2) 
  & \! = \! & a_{\rm s}^{-1}
    \left( \uV + a_{\rm s}\,\UV_1 + a_{\rm s}^{2}\, \UV_2 \right) 
    \left\{ \uV - \left( \frac{a_{\rm s}}{a^{}_0} \right)
    ^{ {\footnotesize \uV} - {\footnotesize \RV}_0} \right\} 
    \left( 1-\RV_0 \right)^{-1} \RV^{\,\g}_0 
  \nn \\[0.5ex] & & \mbox{\hspace{-5mm}} - \, 
    \left( \uV + a_{\rm s} \UV_1 \right) 
    \left\{ \uV - \left( \frac{a_{\rm s}}{a^{}_0} \right)^{- 
    {\footnotesize \RV}_0} \right\} 
    \RV_0^{-1} \left( \RV^{\,\g}_1 - \UV_1 \RV^{\,\g}_0 \right) 
  \\[0.5ex] & & \mbox{\hspace{-5mm}} - a_{\rm s}
    \left\{ \uV - \left( \frac{a_{\rm s}}{a^{}_0} \right)^ 
    {- {\footnotesize \uV} - {\footnotesize \RV}_0} \right\} 
    \left( 1+\RV_0 \right)^{-1}
    \left( \RV^{\,\g}_2 - \UV_1 \RV^{\,\g}_1 - [\, \UV_2 - \UV_1^2 \, ]
    \,\RV^{\,\g}_0 \right) 
  \, + \, \ldots \:\: , \:\: \nn
\eea
where we have again only written out those terms which can be 
consistently \cite{RoDr} included at NNLO.
The non-singlet solutions can be obtained from Eqs.~(\ref{hsol}) and 
(\ref{iexp}) be replacing all quantities by their scalar non-singlet 
analogues. In particular Eq.~(\ref{Ures}) is replaced by 
$\, U_{k,\rm ns} = - \frac{1}{k} \widetilde{R}_{k,\rm ns}$.
%
%
\setcounter{equation}{0}
\section{NNLO quantities in {\boldmath $x$}-space}
%
%
We now proceed to the explicit $x$-space results for the photonic 
coefficient functions and the photon-parton splitting functions 
required for the NNLO evolution. As far as our results are exact, we 
shall write them in terms of the harmonic polylogarithms 
$H_{m_1,...,m_w}(x)$, $m_j = 0,\pm 1$, introduced in ref.~\cite{RVhp} 
to which the reader is referred for a detailed discussion.
The lowest-weight ($w = 1$) functions $H_m(x)$ are given by
\beq
  H_0(x)       \: = \: \ln x \:\: , \quad\quad
  H_{\pm 1}(x) \: = \: \mp \, \ln (1 \mp x) \:\: .
\eeq
The higher-weight ($w \geq 2$) functions are recursively defined as
\beq 
  H_{m_1,...,m_w}(x) \: = \:
    \left\{ \begin{array}{cl} 
    \displaystyle{ \frac{1}{w!}\,\ln^w x \:\: ,} 
       & \quad {\rm if} \:\:\: m^{}_1,...,m^{}_w = 0,\ldots ,0 \\[2ex]
    \displaystyle{ \int_0^x \! dz\: f_{m_1}(z) \, H_{m_2,...,m_w}(z) 
       \:\: , } & \quad {\rm else}
    \end{array} \right.
\eeq
with
\beq
  f_0(x)       \: = \: \frac{1}{x} \:\: , \quad\quad
  f_{\pm 1}(x) \: = \: \frac{1}{1 \mp x} \:\: .
\eeq
In analogy to the notation for harmonic sums \cite{hsum}, a useful 
short-hand notation is 
\beq
  H_{{\footnotesize \underbrace{0,\ldots ,0}_{\scriptstyle m} },\,
  \pm 1,\, {\footnotesize \underbrace{0,\ldots ,0}_{\scriptstyle n} },
  \, \pm 1,\, \ldots}(x) \: = \: H_{\pm (m+1),\,\pm (n+1),\, \ldots}(x) 
  \:\: . 
\eeq
For $w \leq 3$  the harmonic polylogarithms can be expressed in 
terms of standard polylogarithms~\cite{Lewin}; a complete list can be 
found in appendix A of ref.~\cite{MV1}. A {\sc Fortran} program for the 
functions up to weight four has recently been presented~\cite{GRhp}. 

For the convenience of the reader we recall, in this notation, the 
\MSb\ splitting functions and coefficient functions obtained
previously \cite{BaBu,GR83,FP92,GRVg1}. Henceforth suppressing the 
argument `$x$' of the harmonic polylogarithms, the first and 
second-order splitting functions in Eq.~(\ref{Pexp}) read
\beq
\label{p0}
  \d_{\rm p}^{-1}\, P^{(0)}_{\rm p\g}(x) \: = \: 4\, p_{\rm qg}(x) 
  \:\: , \quad\quad
  p_{\rm qg}(x) \: \equiv \: x^2 + (1 - x)^2 \:\: ,
\eeq
and
\bea
  \d_{\rm p}^{-1}\, P^{(1)}_{\rm p\g}(x) & \! = \! & \! C_F \, \Big\{ 
    56 - 116\, x + 80\, x^2 - 4 \left( 1 - 4 x \right) H_{0}
    - 8 \left( 1 - 2 x \right) H_{0,0} - 16\, H_{1} 
  \\ & & \left. \quad \mbox{} 
    + 16\, p_{\rm qg}(x) \Big( - \zeta_2 + H_{0} + H_{1} + 
      H_{0,0} + H_{1,0} + H_{1,1} + H_{2}\Big) \right\}
  \nn \\[1ex] 
  \d_{\rm s}^{-1}\, P^{(1)}_{\rm g\g}(x) & \! = \! & \! C_F \left\{ 
    \frac{16}{3 x} - 64 + 32\, x + \frac{80}{3} x^2 
    - \left( 24 + 40\, x \right ) H_{0}
    - 16 \left( 1 + x \right) H_{0,0} \right\} \:\: , 
\eea
where p = ns, q. The corresponding first-order coefficient functions 
in Eqs.~(\ref{Cpexp}) and (\ref{cpdec}) are
\bea 
  \d_{\rm s}^{-1} \, c_{2,\g}^{(1),\rm s}(x)  & \! = \! &
  \d_{\rm ns}^{-1}\, c_{2,\g}^{(1),\rm ns}(x) \: = \:
  - 4\, p_{\rm qg}(x) \,\Big( 4 + H_0 + H_1 \Big) 
  \nn \\[1ex]
  \d_{\rm s}^{-1} \, c_{L,\g}^{(1),\rm s}(x)  & \! = \! &
  \d_{\rm ns}^{-1}\, c_{L,\g}^{(1),\rm ns}(x) \: = \:
  16\, x (1-x) \:\: . 
\eea
All these results can be directly read off (by replacing $C_A \ra 0$ 
and $1/2\: \Nf \ra 1$) from their hadronic counterparts 
$c_{2,g}^{(1)}(x)\,$ \cite{BBDM,FP82}, $P_{\rm qg}^{(0,1)}(x)$ and
$P_{\rm gg}^{(1)}(x)\,$ \cite{FP80}, in the latter case additionally
keeping in mind that the off-diagonal quantity $P_{\rm g\g}^{(1)}(x)$ 
does not contain a $\d (1-x)$ contribution \cite{FP92,GRVg1}.

The same holds for the NNLO photonic coefficient functions not written
down before. These functions can thus be inferred from the results of 
refs.~\cite{ZvN,MV1}, yielding
\bea
\label{c2p2}
  \lefteqn{ \d_{\rm s}^{-1} \, c_{2,\g}^{(2),\rm s}(x)  \: = \: 
    \d_{\rm ns}^{-1}\, c_{2,\g}^{(1),\rm ns}(x) \: = \: } 
  \nn \\[1ex] & & 
  C_F  \: \left\{ \frac{16}{15 x} - \frac{1294}{15} + \frac{478}{5} x
    - \frac{72}{5} x^2
    + 32 \left( 1 - \frac{13}{6} x + \frac{9}{2} x^2 + \frac{6}{5} 
     x^3\right) \zeta_2 \right.
  \nn \\[0.5ex] & & \mbox{}
    + 64 \left(1 + \frac{9}{4} x^2 \right) \zeta_3 
    - \frac{16}{15} \left( \frac{1}{x} + \frac{59}{2} - \frac{339}{8} x
      + 81\, x^2 \right) H_{0} 
    - 28 \left(1 - \frac{20}{7} x + \frac{12}{7} x^2\right) H_{1}
  \nn \\[0.5ex] & & \mbox{}
    + \frac{16}{15} \left( \frac{1}{x^2} + 90 + 40\, x + 36\, x^3 
      \right) H_{-1,0} 
    - 6 \left(1 - \frac{44}{9} x + 24\, x^2 + \frac{32}{5} x^3 \right) 
      H_{0,0}
  \nn \\[0.5ex] & & \mbox{}
    - 52 \left(1 - \frac{40}{13} x + \frac{36}{13} x^2\right) 
      \Bigl(H_{1,0} + H_{1,1}\Bigr)
    - 32 \left(1 - \frac{7}{2} x + \frac{9}{2} x^2\right) H_{2}
  \nn \\[0.5ex] & & \mbox{}
   + 32\, x^2 \left( \zeta_2 \Big[ H_{-1} + H_{0} + 
       H_{1} \Big] + 2 H_{-1,-1,0} - H_{-1,0,0} 
     - \frac{5}{4} H_{0,0,0} - H_{1,0,0} \right.
  \nn \\[0.5ex] & & \left. \mbox{}
   - \frac{1}{2} H_{2,0} - \frac{1}{2} H_{2,1} - H_{3} \right)
   - 32\, p_{\rm qg}(- x) \Big( \zeta_2\, H_{-1} + 2 H_{-1,-1,0}
     - H_{-1,0,0} \Big) 
  \nn \\[0.5ex] & & \mbox{}
   + 32\, p_{\rm qg}(x) \left( \zeta_2\, H_{0} 
     - \frac{5}{8} H_{0,0,0} + \frac{1}{2}\zeta_2\, H_{1} 
     - \frac{1}{4} H_{1,0,0} - H_{1,1,0} - \frac{3}{2} H_{1,2}
     - \frac{5}{4} H_{1,1,1} 
 \right. 
  \nn \\[0.5ex] & & \left. \left. \mbox{}
     - \frac{3}{4} H_{2,0} - H_{2,1} - H_{3} \right) 
   + 64 \left( 1 + x^2 \right) H_{-2,0}
   \right\}
  \\[2ex] 
\label{cLp2}
  \lefteqn{ \d_{\rm s}^{-1} \, c_{L,\g}^{(2),\rm s}(x)  \: = \: 
    \d_{\rm ns}^{-1}\, c_{L,\g}^{(1),\rm ns}(x) \: = \: } 
  \nn \\[1ex] & & 
    C_F \, \left\{ \frac{64}{15 x} - \frac{256}{15} - \frac{608}{5} x
    + \frac{672}{5} x^2
    + \frac{32}{3} \left(x + \frac{12}{5} x^3\right) \zeta_2
    - \frac{64}{15} \left( \frac{1}{x} +\frac{13}{4} + \frac{39}{2} x 
      - 9 x^2 \right) H_{0}
  \right. \nn \\[0.5ex] & & \mbox{}
    - 16 \left( 1 + 3 x - 4 x^2 \right) H_{1} 
  \left.
    + \frac{64}{15} \left( \frac{1}{x^2} - 5x + 6x^3\right) H_{-1,0}
    - \frac{128}{3} \left(x + \frac{3}{5} x^3 \right) H_{0,0}
    - 32\, x H_{2} \right\} \:\: . 
  \nn \\[0.5ex] & & 
\eea
Analogous to Eqs.~(3.2)--(3.7) of ref.~\cite{NV1} and Eqs.~(3.3)--(3.6)
of ref.~\cite{NV2} for the hadronic second-order coefficient functions
\cite{ZvN}, we also provide compact approximate representations of 
these results in terms of logarithms,
\beq
\label{logs}
  L_0 \: = \: \ln x \:\: , \quad\quad L_1 \: = \: \ln (1-x) \:\: .
\eeq
With deviations up to a few permille, Eqs.~(\ref{c2p2}) and 
(\ref{cLp2}) can be parametrized by (p = ns, s)
\bea
 \d_{\rm p}^{-1} \, c_{2,\g}^{(2),\rm p}(x) & \! = \! &
    \frac{1}{3}\, \Big\{ (26.67 -317.5\, (1-x) ) L_1^3 - 72.00\, L_1^2 
    - ( 1287\, x^{-1} - 908.6) L_1 + 1598\, L_1 L_0
 \nn \\ & & \quad \mbox{} 
    - 13.86  L_0^3 - 27.74\, L_0^2 - 67.33\, L_0 - 1576 - 2727\, x
    + 3823\, x^2 - 0.59\, \delta (1\! -\! x) \Big\}\:\: , 
 \nn \\ [1.5ex] 
 \d_{\rm p}^{-1} \, c_{L,\g}^{(2),\rm p}(x) & \! = \! &
   \frac{1}{3}\, \Big\{ (- 187.2 + 193.7 \,x) (1-x) L_1^2 
    + 737.0\, (1-x) L_1 + 452.2\, L_1 L_0
 \nn \\ & & \quad \mbox{}
    - 84.87\, x L_0^2 - 64.25\, (1-x) L_0 - 66.44\, (1 - x) \Big\} 
    \:\: .
\eea
In the \MSb\ scheme the small $\d (1\! -\! x)$  contribution in the 
first relation (of course absent in the exact expression, but useful 
for obtaining high-accuracy moments and convolutions) is not relevant 
in $x$-space at the present accuracy in the electromagnetic coupling. 

The ${\cal O}(a_{\rm em}^{} a_{\rm s}^2)$ NNLO photon-parton splitting
functions $P_{\rm p\g}^{(2)}(x)$, ${\rm p = ns, ps, g}$ --- which are 
not related to their hadronic analogues by any simple substitutions ---
are not completely fixed by our results in Sect.~3. Therefore we 
resort, for the time being, to approximations analogous to those 
derived in refs.~\cite{NV1,NV2,NV3} for the hadronic parton-parton 
splitting functions. 
In the \MSb\ scheme adopted for our calculations, the leading-$\Nf$ 
term of the NNLO photon-quark splitting function, for example, can be 
cast in the form 
\beq
\label{Pform} 
  \d_{\rm ns}^{-1} P_{\rm ns\g,1}^{(2)}(x) \: = \: 
  \d_{\rm s}^{-1}  P_{\rm q\g,1}^{(2)}(x)  \: = \: 
  \sum_{m=1}^{4}\! A_m \, L_1^m + f_{\rm smooth}(x) + 
  \sum_{n=1}^{4} B_n\, L_0^n \:\: .
\eeq
Here $L_1$ and $L_0$ are the end-point logarithms defined in 
Eq.~(\ref{logs}), and $f_{\rm smooth}$ is finite for $0 \leq x \leq 1$.
This regular function constitutes the mathematically complicated part 
of Eq.~(\ref{Pform}). See Eqs.~(21)--(25) of ref.~\cite{RVhp} for a 
systematic procedure (not applied in Eqs.~(\ref{c2p2}) and (\ref{cLp2})
above) to extract the end-point terms of the harmonic polylogarithms 
entering the exact expression.

At NLO the leading large-$x$ soft-gluon contribution, $\ln^2 (1\!-\!x)$,
to $P_{\rm ns\g}^{(1)}$ cancels in the physical kernel (\ref{dFinh}). 
Anticipating the same cancellation at NNLO, the coefficient $A_4$ in 
Eq.~(\ref{Pform}) can be inferred from the known splitting functions
and coefficient functions. Besides this term we keep two of the
remaining three large-$x$ logarithms and two of the small-$x$ 
logarithms in Eq.~(\ref{Pform}), and choose a two-parameter ansatz 
(a low-order polynomial in $x$) for $f_{\rm smooth}$. 
These parameters and the coefficients of the selected logarithms are 
then determined from the six coefficients of $a_{\rm s}^2 \Nf^0$ on the 
right-hand-sides of Eqs.~(\ref{kqmom}). Varying all these choices we 
obtain about 50 approximations. The two representatives spanning the 
error band for most of the $x$-range, marked by `A' and `B' below, are 
finally selected as our estimates for $P_{\rm ns\g,1}^{(2)}$ and its 
residual uncertainty.

Analogous procedures are applied for the ($\Nf^1$) pure-singlet 
photon-quark splitting function and the $\Nf^0$ term of the NNLO 
photon-gluon splitting function. The non-singlet and gluon $\Nf^1$
pieces are smaller in absolute size and uncertainty than the 
$\Nf^0$ terms, hence for them it suffices to select just one central 
representative. The resulting approximations are displayed graphically 
in Fig.~1 and Fig.~2. For the non-singlet case, the selected 
parametrizations (shown as solid lines) read
\bea
  \d_{\rm ns}^{-1} P^{(2)}_{\rm ns\g, A}(x)\! & = &
   128/27\: L_1^4 + 3.8636\: L_1^3 + 97.512\: L_1^2 - 1319.749\: x^2 
  \nn \\ & & \mbox{}
   + 511.199\: x + 84.808\: L_0 - 22.878\: L_0^3
   \:\: + \:\: \d_{\rm ns}^{-1} P^{(2)}_{\rm ns\g,2}(x)
  \nn \\[1ex]
\label{qp1}
  \d_{\rm ns}^{-1} P^{(2)}_{\rm ns\g, B}(x)\! &=&
   128/27\: L_1^4 - 5.4658\: L_1^3 - 295.331\: L_1 - 1803.989\: x
  \nn \\ & & \mbox{}
   + 740.532 - 259.036\: L_0^2 + 27.110\, L_0^4 \}
   \:\: + \:\: \d_{\rm ns}^{-1} P^{(2)}_{\rm ns\g,2}(x)
\eea
with
\bea
\label{qp2}
  \d_{\rm ns}^{-1} P^{(2)}_{\rm ns\g,2}(x)\! & = &
  \Nf\: \left\{ - 0.2949\: L_1^3 + 34.854\: L_1 + 157.995\: x 
  - 73.672 
  \right. \nn \\ & & \quad\quad \left. \mbox{}
  - 33.059\: L_0 + 2.887\: L_0^3  \right\} \:\: .
\eea
The corresponding approximations chosen for the pure-singlet splitting 
functions are given by
\bea
  \d_{\rm s}^{-1} P^{(2)}_{\rm ps\g, A}(x)\! & = &
  \Nf \left\{ (1 - x) \left( - 110.167\: x^2 + 876.629\: x 
  + 23.605\: x^{-1} \right) 
  \right.  \nn \\ & & \quad\quad \left. \mbox{}
  + 668.725\: x L_0 + 387.125\: x L_0^2 + 121.403\: L_0 \right\} 
  \nn \\[1ex]
\label{ps2}
  \d_{\rm s}^{-1} P^{(2)}_{\rm ps\g, B}(x)\! & = &
  \Nf \left\{ (1 - x) \left( - 34.853\: x + 305.244 
  + 50.950\: x^{-1} \right) 
  \right. \nn \\ & & \quad\quad \left. \mbox{} 
  + 101.246\: x L_0 + 220.083\: L_0 - 5.0738\: L_0^4 \right\} \:\: .
\eea
Finally the NNLO photon-gluon splitting function and its present
uncertainty are parametrized by
\bea
  \d_{\rm s}^{-1} P^{(2)}_{\rm g\g, A}(x)\! & = &
   (1 - x) \left( 769.616 \: L_1 + 1329.961\: x^2 - 391.569\: x \right)
  \nn \\ & & \mbox{}
   + 317.267\: L_0^2 + 60.519\: L_0^3 + 15.018\: x^{-1} L_0
   \:\: + \:\: \d_{\rm s}^{-1} P^{(2)}_{\rm g\g,2}(x)
  \nn \\[1ex]
\label{gp1}
  \d_{\rm s}^{-1} P^{(2)}_{\rm g\g, B}(x)\! & = &
   (1 - x) \left( - 105.632\: L_1^2 - 415.549\: x^2 - 429.907 \right)
  \nn \\ & & \mbox{}
   - 357.604\: L_0 - 146.286\: L_0^3 + 64.666\: x^{-1} L_0
   \:\: + \:\: \d_{\rm s}^{-1} P^{(2)}_{\rm g\g,2}(x) 
\eea
with
\bea
\label{gp2}
   \d_{\rm s}^{-1} P^{(2)}_{\rm g\g,2}(x)\! & = &
   \Nf \left\{ (1 - x) \left( 43.748\: L_1 + 61.028\: x^2 
   - 70.910 - 33.914\: x^{-1} \right)
  \right.  \nn \\ & & \quad\quad \left. \mbox{}
   - 105.172\: L_0 + 1.3972\: L_0^2  \right\} \:\: .
\eea
In all cases the averages $1/2\: [{\rm A} + {\rm B}]$ represent our
central results.

Denoting the coefficients of $a_{\rm s}^l$ on the right-hand-side of 
Eq.~(\ref{Pdsg}) by $\Delta_{\rm DIS_\g}^{(l)}$, the additional NLO
($l=1$) contributions to the DIS$_\g$ splitting functions are given by
\cite{GRVg1}
\bea
\label{dPnlo}
  \d_{\rm p}^{-1} \Delta_{\rm DIS_{\g_{}}}^{(1),\rm p}(x) 
  & \! = \! & 4 C_F\, \Big\{ 
     7 - 10 x
   + \left( 1 - 16 x + 32 x^2 \right) H_{0}
   + \left( 6 - 12 x + 16 x^2 \right) ( H_2 - \zeta_2 )
  \nn \\ & & \mbox{}
   + \left( 5 - 36 x + 32 x^2 \right) H_{1} 
   + \left( 2 - 4 x + 8 x^2 \right) H_{0,0} 
   + 4\, p_{\rm qg}(x) ( H_{1,0} + 2 H_{1,1} ) 
  \Big\} \quad\quad
  \nn \\[1ex]
  \d_{\rm s}^{-1} \Delta_{\rm DIS_{\g_{}}}^{(1),\rm g}(x) 
  & \! = \! & 8/3\: C_F\, \Big\{ 
   2x^{-1} - 20 + 2x + 16\, x^2 
   - \left(3 + 15\, x - 4 x^2\right) H_{0}
  \nn \\ & & \mbox{}
   - \left( 3 + 4x^{-1} - 3x - 4\, x^2 \right) H_{1}
   + 6 (1 + x) ( \zeta_2 - H_{0,0} - H_{2} ) \Big\}
\eea
with p = ns, s. The corresponding NNLO terms can be parametrized as
\bea
\label{dPqnn}
  \lefteqn{ \d_{\rm p}^{-1} \Delta_{\rm DIS_{\g_{}}}^{(2),\rm p}(x) 
  \:\, = \:\, 9.482\, L_1^4 + ( 33.37 + 2585\, (1-x)\, )\, L_1^3 
         + 122.6\, L_1^2 + ( 5598\, + 7949\, (1-x)\, )\, L_1 } 
  \nn \\ & & \mbox{} - 9825\, L_0 L_1
  - 2.963\, L_0^4 + 7.407\, L_0^3 - ( 176.0 - 2616\, x )\, L_0^2
  - 828.6\, L_0 - 1851 + 30120\, x
  \nn \\[0.2ex] & & \mbox{} 
  - 7595\, x^2 - 0.65 \: \d (1\! -\! x)
  \nn \\ & & \mbox{}
  + \Nf\, \Big\{
  - ( 1.044 + 32.57\, (1-x)\, )\, L_1^3 + 26.75\, L_1^2
  - ( 0.266 - 615.1\, (1-x)\, )\, L_1 + 4.557\, L_0 L_1 
  \nn \\ & & \mbox{} 
  - 0.529\, L_0^3 + ( 12.23 - 38.59\, x )\, L_0^2
  + 41.91\, L_0 + 75.74 + 733.7\, x - 1003\, x^2 
  + 0.05\, \d (1\! -\! x) \Big\}
  \nn \\ & & \mbox{} 
  + \Nf f_{\rm ps}\, \Big\{ 
  ( 2.083\, L_1^2 - 68.96\, L_1) (1-x)^3 - 86.40\, (1-x)^2 L_0 L_1 
  + 1.778\, L_0^4 + 3.278\, L_0^3
  \\ & & \mbox{}
  + (41.86 + 105.6\,x ) (1-x) L_0^2 + 1.241\, (1-x)^2 L_0
  - (15.80\, x^{-1} + 94.39 - 8.281\, x) (1-x)^3 \Big\} \quad\quad
  \nn
\eea
and
\bea
\label{dPgnn}
  \d_{\rm s}^{-1} \Delta_{\rm DIS_{\g_{}}}^{(2),\rm g}(x)
  &\! = \! & - ( 25.77\, L_1^3 + 766.7\, L_1) (1-x) - 1337\, L_0 L_1 
  -4.741\, L_0^4 + 4.741\, L_0^3  
  \nn \\ & & \mbox{}
  + (83.11 - 464.9\,x ) L_0^2 + 443.1\, L_0 - (60.36\, x^{-1} - 1772 
  - 650.3\, x) (1-x)
  \nn \\ & & \mbox{}
  + \Nf \, \Big\{ - ( 0.737\, L_1^3 - 216\, L_1) (1-x)
  + 310.6\, L_0 L_1 - (40.99 - 36.66 \,x ) L_0^2 
  \quad \nn \\ & & \mbox{} - 113.7\, L_0
  + (15.80\, x^{-1} - 331.3 - 48.88\, x) (1-x) \Big\} \:\: ,
\eea
where $f_{\rm ps}$ has been defined in Eq.~(\ref{fps}). 
Eqs.~(\ref{dPqnn}) and (\ref{dPgnn}) deviate by a few permille or less
from the (somewhat lengthy) exact expressions deferred to Appendix B.

\begin{figure}[p]
\vspace*{1mm}
\centerline{\epsfig{file=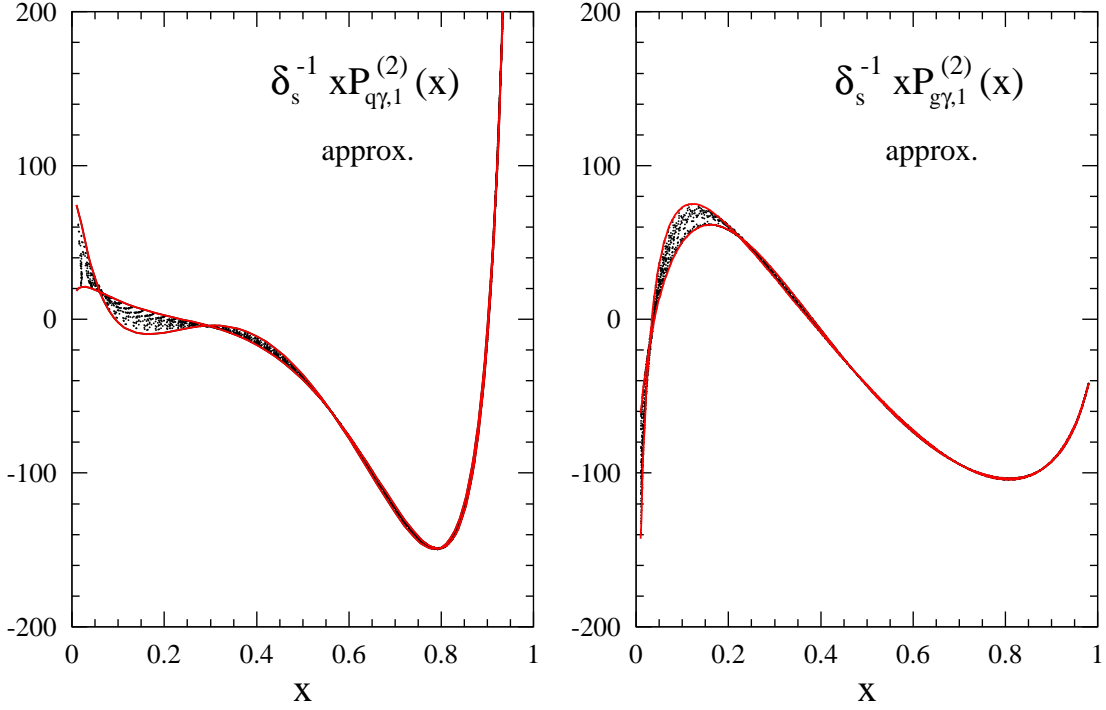,width=15cm,angle=0}}
\vspace{-1.5mm}
\caption{Approximations of the $\Nf^0$ parts $P_{\rm p\g,1}^{(2)}(x)$ 
 of the NNLO photon-quark (p = q) and photon-gluon (p = g) splitting 
 functions, as  obtained from the six moments calculated in Sect.~3. 
 The selected representatives (\ref{qp1}) and (\ref{gp1}) are shown
 by the full curves.} 
\vspace{1mm}
\end{figure}
\begin{figure}[p]
\vspace{1mm}
\centerline{\epsfig{file=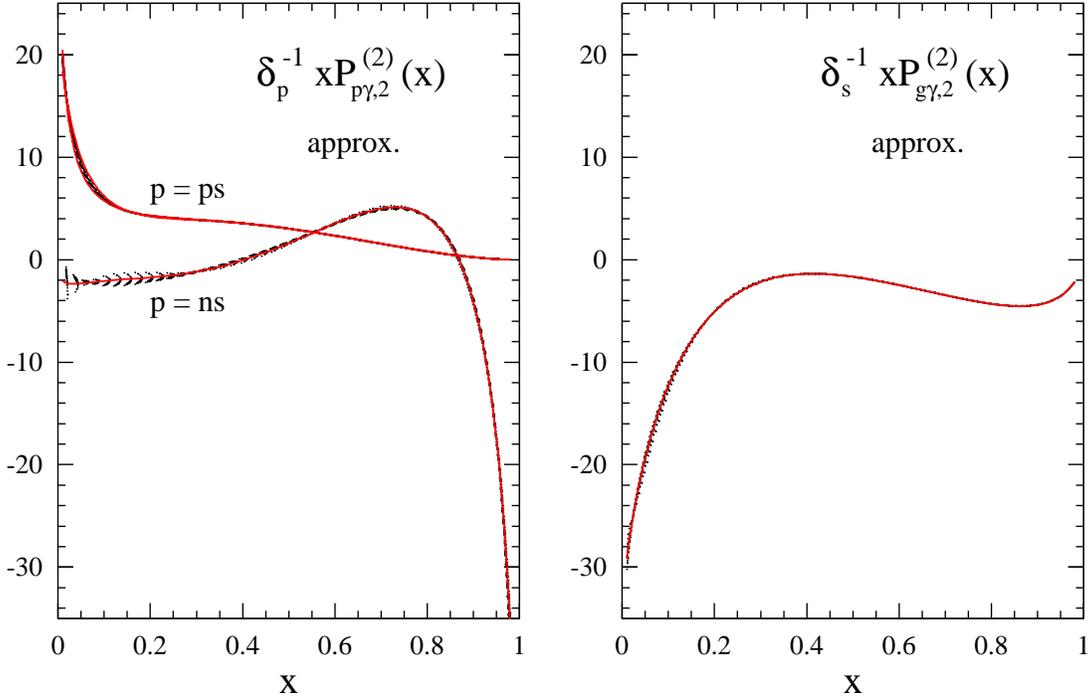,width=15cm,angle=0}}
\vspace{-1.5mm}
\caption{As Fig.~1, but for the $\Nf^1$ contributions 
 $P_{\rm p\g,2}^{(2)}(x)$, with p = ns, ps (left) and p = g (right). 
 The full curves represent the parametrizations in Eqs.~(\ref{qp2}), 
 (\ref{ps2}) and (\ref{gp2}).}  
\end{figure}
%
%
\setcounter{equation}{0}
\section{Numerical results in {\boldmath $x$}-space}
%
%
In this section we finally present the numerical impact of the NNLO
corrections on the evolution of the photon's parton distributions and
on the structure function $F_2^{\,\g}$. Concerning the evolution 
kernels we confine ourselves to the respective inhomogeneous 
contributions $\PV^{\g}$ and ${\cal I}_2^{\,\g}$ to  
Eqs.~(\ref{qevol}) and (\ref{Fevol}). The higher-order homogeneous 
(hadronic) non-singlet and singlet kernels have been discussed in 
detail in refs.~\cite{NV1,NV4} and refs.~\cite{NV2}, respectively; see 
also ref.~\cite{NVrgb} for a recent brief summary using the updated 
NNLO singlet splitting functions~\cite{NV3}. Also our subsequent 
illustrations of the solution of the evolution equations are restricted 
to the photon-specific inhomogeneous piece (\ref{iexp}) and its 
contribution to $F_2^{\,\g}$. For brevity the results are shown only 
with all scales identified, i.e., at $\m_r = \m_f$ for the parton 
evolution and at $\m_r\, (= \m_f) = Q$ for $F_2^{\,\g}$. The 
experimentally elusive longitudinal structure function $F_L^{\,\g}$ 
will not be addressed here.

In Figs.~3 and 4 the singlet photon-quark splitting functions 
$P_{\rm q\g}$ are displayed in the \MSb\ and the DIS$_{\g}$ 
factorization schemes. The NLO and NNLO curves are shown for 
$\a_{\rm s}= 0.2$ which corresponds to a scale between about 20 GeV$^2$
and 50 GeV$^2$, depending on the precise value of $\a_{\rm s}(M_Z^2)$. 
After removing the charge factors $\d_{\rm ns}$ and $\d_{\rm s}$ 
defined in Eq.~(\ref{delta}) only the NNLO terms depend on the number 
of flavours; the curves presented refer to $\Nf = 4$. 
In the \MSb\ scheme (Fig.~3) the NNLO corrections are small except for 
small and very large values of $x$, amounting to less than $2\%\,$ for 
$0.1\leq x \leq 0.95$. Larger corrections for $x\! \ra\! 1$ are obvious 
since the NNLO splitting function $P^{(2)} \sim \ln^4 (1\! -\! x)$ is 
more singular in this limit than its NLO analogue, $P^{(1)} \sim \ln^2 
(1\! - \! x)$. Likewise large NNLO effects are expected for small $x$ 
due to the first non-vanishing pure-singlet term $\sim 1/x$. 

The higher-order corrections to $P_{\rm q\g}$ are somewhat larger in the
DIS$_\g$ scheme (Fig.~4) due to the absorption of the large coefficient 
function $C_{2,\g}$ (which, in particular, reverses the sign of the 
leading large-$x$ contribution). Here the NNLO corrections, under the 
conditions specified above, reach +6\% at $x \simeq 0.6$ and exceed 
$-6\%$ for $x > 0.9$. In the right parts of both figures the relative
NNLO effects are also displayed (dotted curves) for the non-singlet 
splitting functions $P_{\rm ns \g}$, thus the pure-singlet effect can 
be directly read off from these figures. Also this effect is larger
in the DIS$_\g$ scheme where it exceeds 1\% at $x < 0.3$, instead of
only at $x < 0.15$ in the \MSb\ scheme.

The corresponding results for the photon-gluon splitting functions
$xP_{\rm g\g}$ are shown in Fig.~5. The relative NNLO corrections (not
shown separately for these quantities) are larger than in the 
photon-quark cases. However, the absolute size of $P_{\rm g\g}$ is much
smaller than that of $P_{\rm q\g}$ except at small $x$. In the \MSb\
scheme $P_{\rm g\g}$ remains negative at $x \geq 0.2$ at NNLO.
In the DIS$_\g$ scheme, on the other hand, $P_{\rm g\g}$ is positive at 
large $x$, but seems to turn negative at NNLO for $x < 0.05$.  

Fig.~6 depicts, again for $\alpha_{\rm s} = 0.2$ and $\Nf = 4$, the 
inhomogeneous contribution $I_2^{\,\g}$ to the physical non-singlet 
evolution kernel (\ref{Fevol}). Here the NNLO corrections are 
particularly small, about 1\% or less for $0.05 \leq x < 0.95$. As in 
Figs.~3 -- 5, the present uncertainties arising from the residual error 
band for the ${\cal O}(a_{\rm em}^{}a_{\rm s}^{2})$ photon-parton 
splitting functions are estimated by the NNLO$_{\rm A}$ and 
NNLO$_{\rm B}$ curves which derive from the upper and lower 
approximations, respectively, in Eqs.~(\ref{qp1}), (\ref{ps2}) and 
(\ref{gp1}) together with Eqs.~(\ref{qp2}) and (\ref{gp2}). These 
uncertainties are virtually negligible at $x \geq 0.25$ and remain 
perfectly tolerable for $x \geq 0.05$, where they amount to less than 
$\pm 1\%$ for $P_{\rm q\g}$ and $I_2^{\,\g}$ with respect to the 
central results $\frac{1}{2}\, ({\rm NNLO_A}+{\rm NNLO_B})$ not shown 
in the figures. This accuracy rapidly deteriorates towards small values 
of $x$. 

\begin{figure}[p]
\vspace*{1mm}
\centerline{\epsfig{file=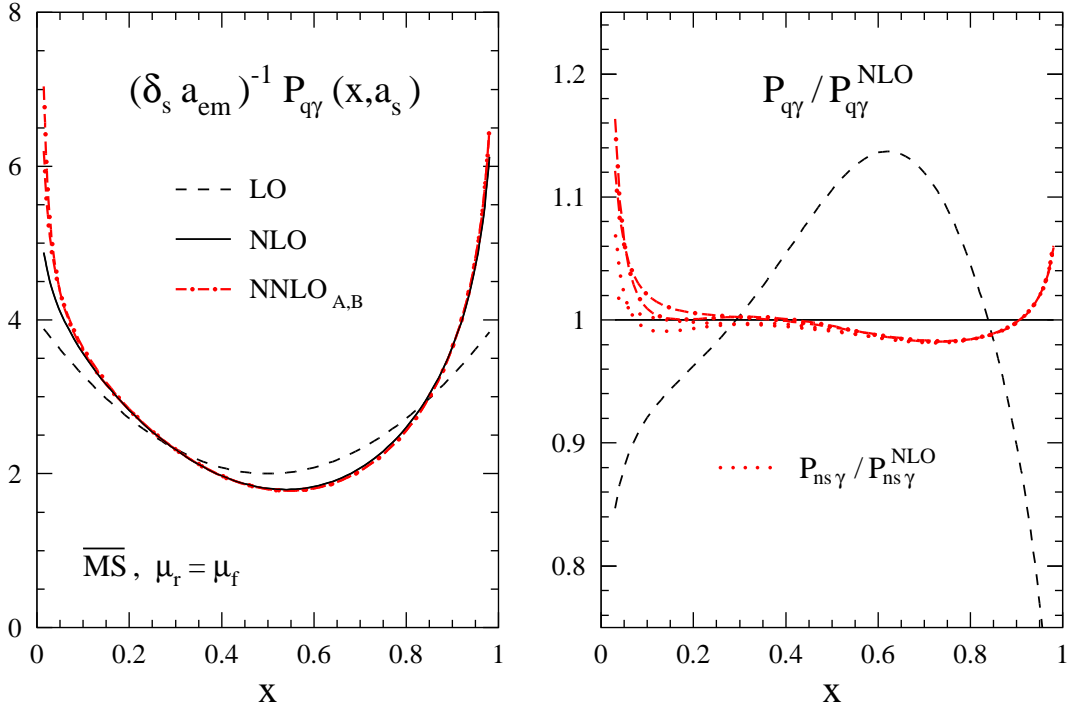,width=15cm,angle=0}}
\vspace{-1.5mm}
\caption{The perturbative expansion (\ref{Pexp}) of the photon-quark
 splitting function $P_{\rm q\g}(x, a_{\rm s})$ in the \MSb\ scheme for 
 $\alpha_{\rm s} = 0.2$ and $\Nf = 4$. The relative NNLO corrections in 
 the right part are also shown for the non-singlet splitting function
 $P_{\rm ns\g}$. Here and in the following three figures the subscripts
 `A' and `B' indicate the respective approximations in Eqs.~(\ref{qp1}),
 (\ref{ps2}) and (\ref{gp2}).}
\vspace{1mm}
\end{figure}
\begin{figure}[p]
\vspace{1mm}
\centerline{\epsfig{file=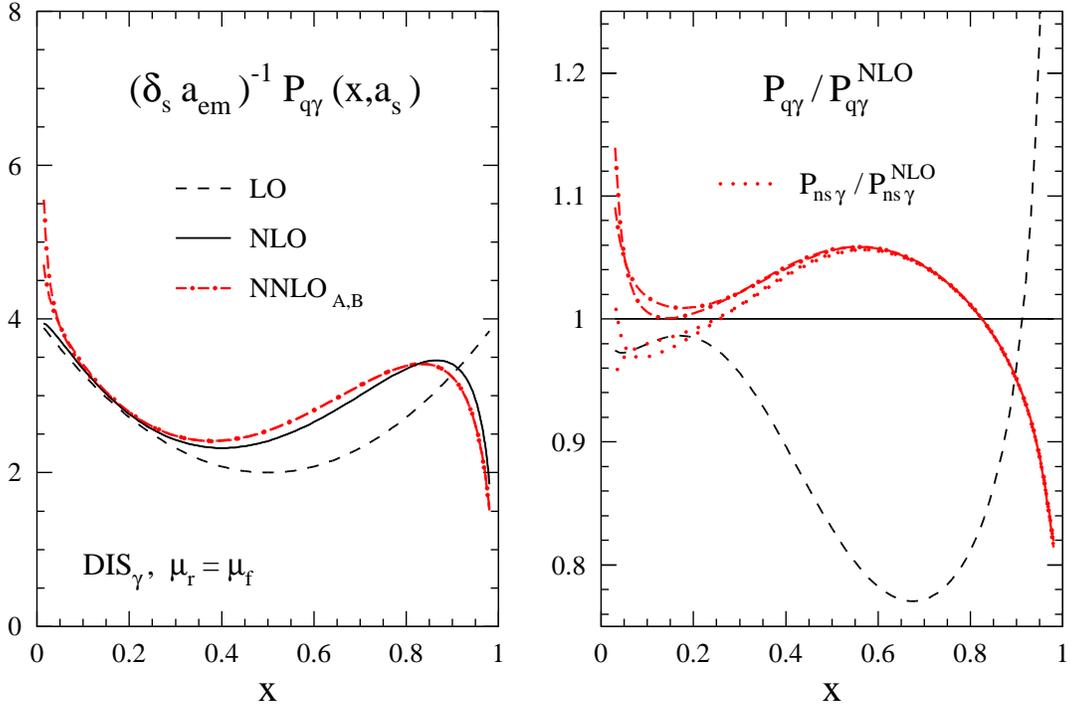,width=15cm,angle=0}}
\vspace{-1.5mm}
\caption{As Fig.~3, but for the photon-quark splitting functions 
 (\ref{Pdsg}) in the DIS$_\g$ scheme.}
\end{figure}

\begin{figure}[p]
\vspace*{1mm}
\centerline{\epsfig{file=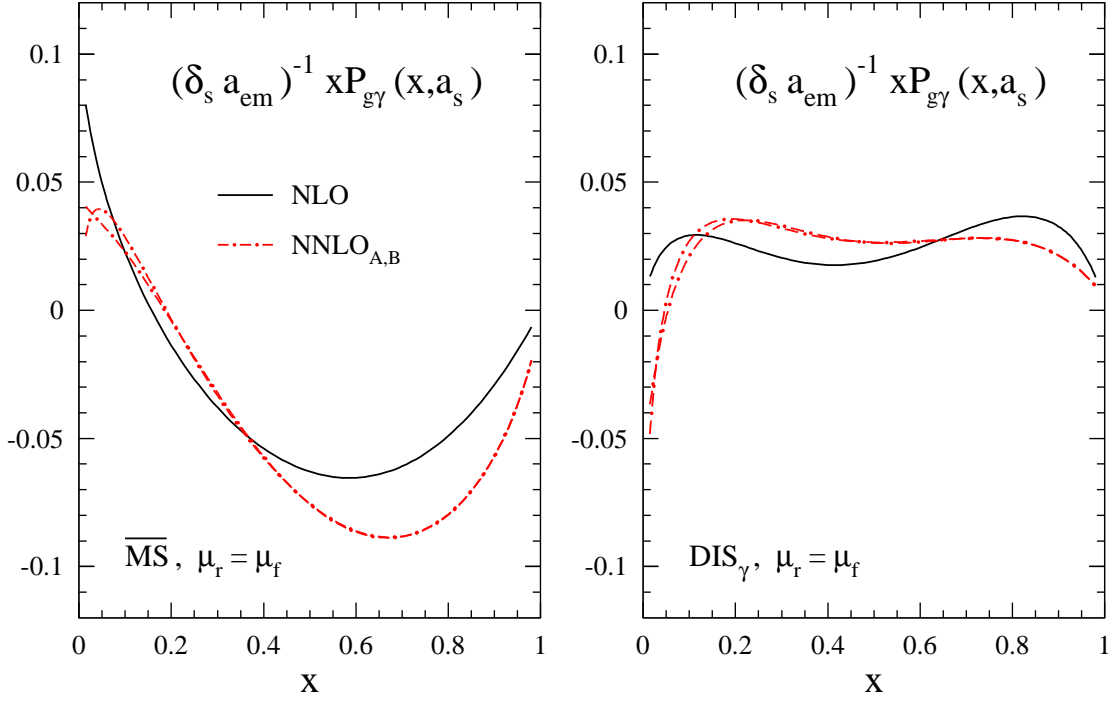,width=15cm,angle=0}}
\vspace{-1.5mm}
\caption{The NLO and NNLO approximations (\ref{Pexp}) for the 
 photon-gluon splitting function $P_{\rm g\g}(x, a_{\rm s})$ in the 
 \MSb\ scheme (left) and the DIS$_\g$ scheme (right) for 
 $\alpha_{\rm s} = 0.2$ and $\Nf = 4$. Note that $x\!\cdot\! P$ is 
 displayed here, unlike in Fig.~3 and Fig.~4.}
\vspace{1mm}
\end{figure}
\begin{figure}[p]
\vspace{1mm}
\centerline{\epsfig{file=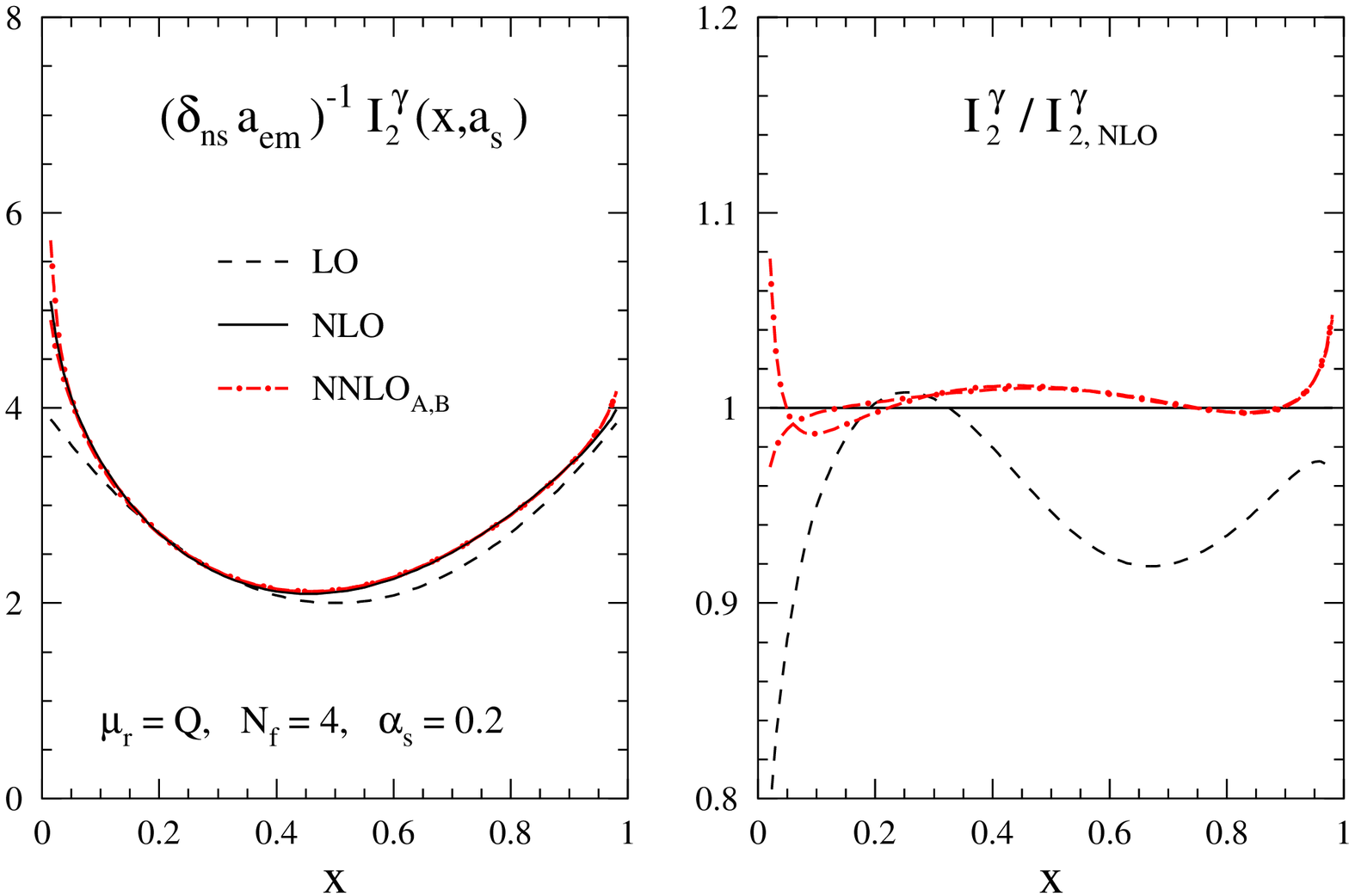,width=15cm,angle=0}}
\vspace{-1.5mm}
\caption{The perturbative expansion (\ref{dFinh}) of the inhomogeneous
 physical kernel $I_{2}^{\,\g}(x,a_{\rm s})$ for the evolution 
 (\ref{Fevol}) of the structure function ${\cal F}_{2,\rm ns}^{\,\g} 
 = 1/x\: F_{2,\rm ns}^{\,\g}$ at the `default' value of the scale
 $\mu_r$.}
\end{figure}

In Fig.~7 and Fig.~8 we present the inhomogeneous solutions 
$\alpha_{\rm em}^{-1}\, xf_{\rm inhom}^{\,\g}$, $f = \S,\, g,$ for the 
singlet-quark and gluon distributions. The NLO approximation derived 
already in  ref.~\cite{GRVg1} is obtained from the NNLO expression 
(\ref{iexp}) by removing the third line, the $\UV_1$ term in the second 
line and the $\UV_2$ contribution to the first line; for the LO result 
also the rest of the second line and the $\UV_1$ term in the first line 
have to be ignored. 
Like the NLO illustrations in refs.~\cite{FP92,GRVg1} the results are 
shown at $\m_f^2 = 50 \mbox{ GeV}^2$ for $\m_{f,0}^2 = 1 \mbox{ GeV}^2$ 
and $\Nf = 3$. For the strong coupling constants we use the realistic 
values $\alpha_{\rm s}(m_{f,0}^2)\equiv 4\pi\, a_0^{} = 0.45$ at LO and 
0.42 at NNLO and NNLO, and solve Eqs.~(\ref{arun}) directly at $D = 4$ 
with the appropriate number of terms in Eq.~(\ref{bQCD}), i.e., 
including the coefficients $\beta_{k\leq m}$ \cite{beta2} for the 
N$^m$LO evolution.
Here and in Fig.~9 the (barely visible) differences of the NNLO%
$_{\rm A}$ and NNLO$_{\rm B}$ curves include the present uncertainties 
\cite{NV4} of the hadronic NNLO splitting functions. The pattern of the
NNLO corrections for the solutions roughly follows that of the 
corresponding splitting functions in Figs.~3 -- 6. At this order 
$\m_f^2 = 50 \mbox{ GeV}^2$ is about the lowest scale where the 
inhomogeneous \MSb\ gluon density is positive over the full $x$-range.

Finally the structure functions $F_{2,\rm inhom}^{\,\g}(x, Q^2\! = \!
50 \mbox{ GeV}^2)$ resulting from these parton densities (and their
non-singlet analogues) are shown in Fig.~9 for the \MSb\ and the 
DIS$_\g$ schemes. In both cases we have taken care to avoid spurious
higher-order contributions which would arise from a simple convolution
of Eq.~(\ref{iexp}) with the corresponding expansion of the hadronic
coefficient functions, see Fig.~3 of ref.~\cite{Lund} for a NLO 
illustration. 
Note that the boundary conditions for $F_{2,\rm inhom}^{\,\g}$ are 
different in the two schemes: in the \MSb\ scheme this quantity is 
given by the corresponding photonic coefficient function (\ref{Cpexp})
at $Q^2 = \m_{f,0}^2$, where it vanishes in the DIS$_\g$ scheme 
according to Eq.~(\ref{F2dsg}). Thus, besides a `physical' input 
describing the $F_2^{\,\g}$ at $Q^2 = \m_{f,0}^2$, a large additional 
`technical' contribution to $\qV^\g(\m_{f,0}^2)$ in Eq.~(\ref{qmu0}) is 
required in the \MSb\ case. 
As mentioned above Eq.~(\ref{Fevol}), the complete structure functions
evolve approximately like non-singlet quantities at large $x$. In 
fact, under the conditions of Fig.~9, the complete DIS$_\g$ result for 
$F_{2,\rm inhom}^{\,\g}$ and this non-singlet approximation differ by 
more than about 1\% only at $x < 0.2$, 0.25 and 0.3 at LO, NLO and 
NNLO, respectively.
  
\begin{figure}[h]
\centerline{\epsfig{file=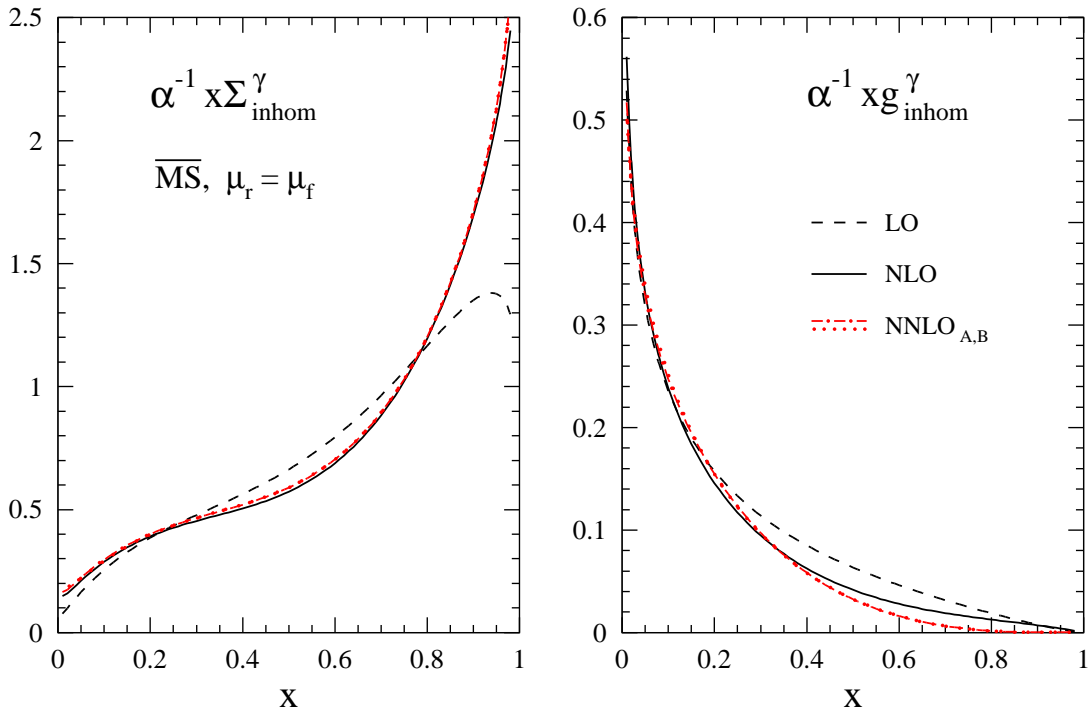,width=15cm,angle=0}}
\vspace{-1.5mm}
\caption{The inhomogeneous LO, NLO and NNLO contributions to the 
 photon's singlet-quark and gluon distributions in the \MSb\ scheme at 
 $\m_f^2 = 50 \mbox{ GeV}^2$, as obtained from Eq.~(\ref{iexp}) for 
 $\Nf = 3$ and $\m_{f,0}^2 = 1 \mbox{ GeV}^2$ with $\alpha_{\rm s} 
 (\m_{f,0}^2)$ = 0.45 at LO and 0.42 at NLO and NNLO.}
\end{figure}
 
\begin{figure}[p]
\vspace*{1mm}
\centerline{\epsfig{file=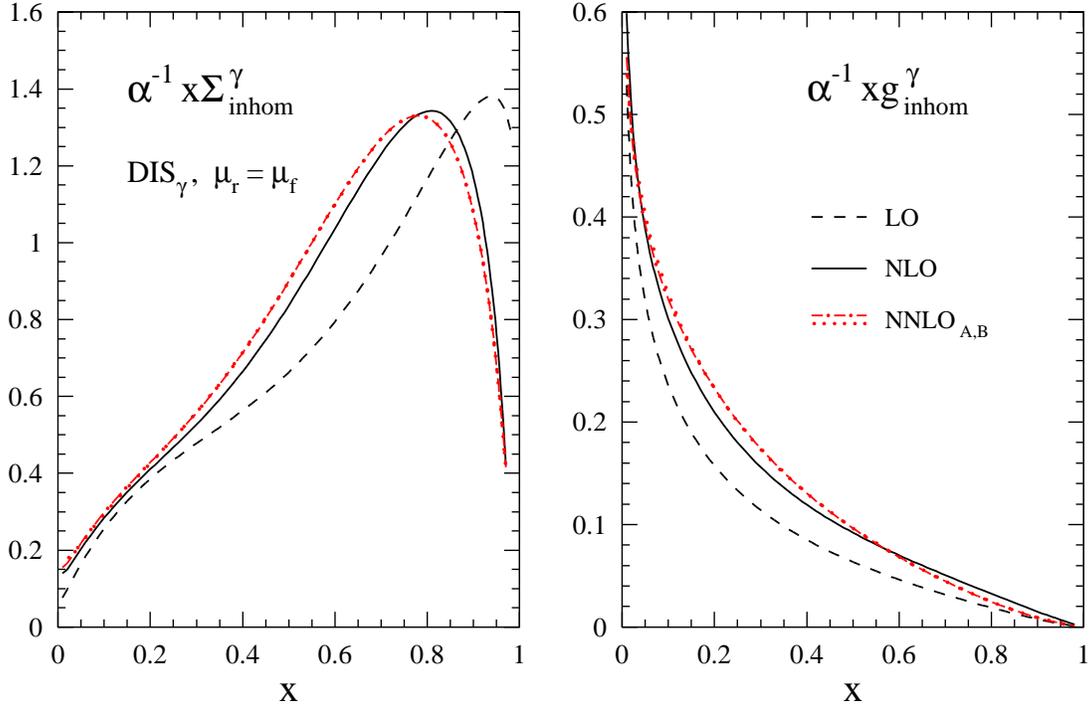,width=15cm,angle=0}}
\vspace{-1.5mm}
\caption{As Fig.~7, but for the inhomogeneous solutions in the DIS$_\g$ 
 factorization scheme (\ref{Pdsg}).} 
\vspace{1mm}
\end{figure}
\begin{figure}[p]
\vspace{1mm}
\centerline{\epsfig{file=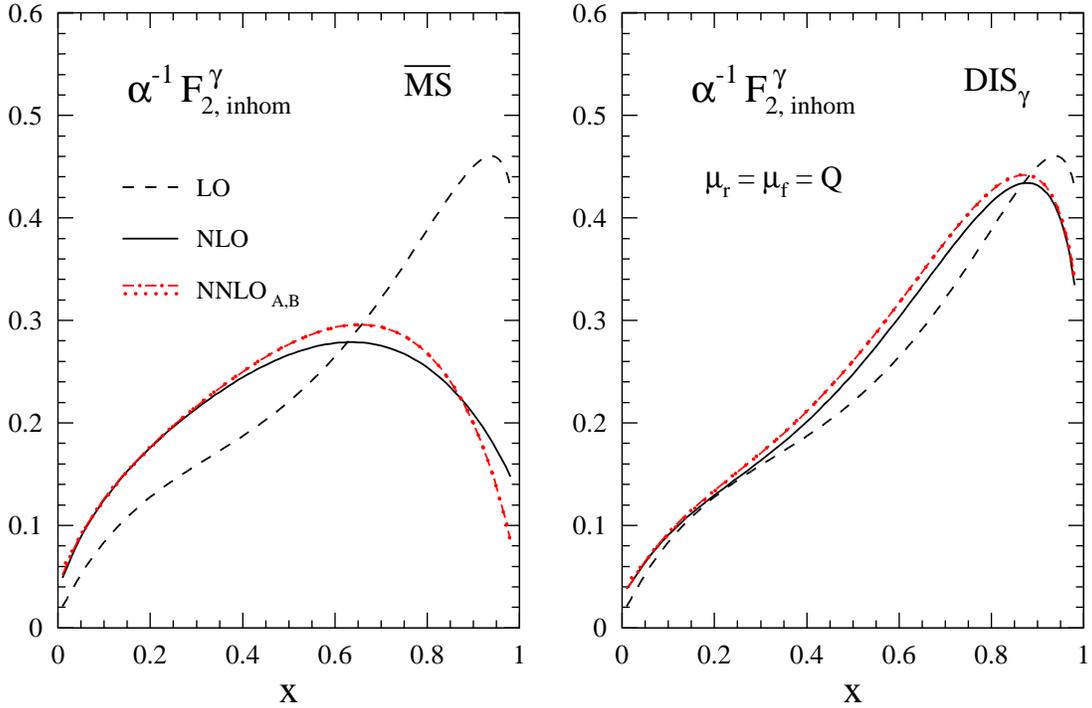,width=15cm,angle=0}}
\vspace{-1.5mm}
\caption{The perturbative expansion of the structure function 
 $F_{2,\rm inhom}^{\,\g}$ at $Q^2 = 50 \mbox{ GeV}^2$, as derived from
 the results shown in the previous two figures and their non-singlet
 counterparts by means of Eq.~(\ref{F12x}) in the \MSb\ scheme (left
 part) and Eq.~(\ref{F2dsg}) in the DIS$_\g$ scheme (right part).}
\end{figure}
%
%
\section{Conclusion}
%
%
We have calculated the next-to-next-to-leading order QCD corrections to 
electron-photon DIS and to the evolution of the photon's quark and 
gluon distributions.
Our exact results for the corresponding photon-parton splitting 
functions are presently confined to the first six even-integer moments. 
Thus the practical applicability of the NNLO evolution is, for the time 
being, restricted to not too small values of the Bjorken variable~$x$. 
This restriction seems to be more serious here than in lepton-hadron 
DIS, as the photon-parton splitting functions enter the evolution 
equations directly, not via smoothening convolutions with 
non-perturbative initial distributions. Consequently the effect of 
these functions is accurately known over a considerably smaller range 
of $x$ than that of their hadronic counterparts; a residual uncertainty 
of about $\pm 1\%$ or less is found for the total photon-quark 
splitting functions at NNLO only at $x \gsim 0.05$.
However, only the perturbative component of the photon's parton 
densities is affected by this uncertainty, and while this component 
dominates at large $x$, it represents only a small correction to the 
hadronic (homogeneous) contribution at small $x$. We thus expect that 
our results are sufficient for extending NLO analyses like those of 
refs.~[51$\,$--$\,$54] 
to NNLO for the full $x$-range covered by the measurements at LEP.

Our present calculations are limited to effectively massless quarks,
hence they do not apply to the charm and bottom contributions to the
structure functions at small and intermediate scales. These
contributions have been computed at NLO in refs.~\cite{LRSV1,LRSV2};
corrected figures for $F_{2,\rm charm}^{\,\g}$ have been presented in
ref.~\cite{LR95}. The NLO effects are found to be fairly small for
this quantity, indicating that the NNLO corrections may be rather 
negligible~\cite{LR95}. While a full massive \mbox{NNLO} calculation 
does not seem feasible at present, these corrections could be estimated 
using the threshold resummation as done for the lepton-nucleon case in 
ref.~\cite{LM99}.

By confining ourselves to the lowest order in the electromagnetic 
coupling we have assumed, as usual also in QCD analyses of 
lepton-hadron DIS, that the QED radiative corrections are treated
elsewhere. The corresponding formalism has been set up in 
ref.~\cite{NVe1} for measurements of the photon structure via the 
process $e^+ e^- \!\ra\! e^+ e^- + X$. The non-factorizable corrections 
due to photon exchange between to two electron lines have been shown to 
be negligible in ref.~\cite{NVe2}, using a pseudoscalar as a simple 
model for the final state $X$. The dominant corrections arising from 
photon emissions of the `tagged' electron line have been investigated 
for realistic measurements of the photon structure in refs.~\cite{LSch}.
Neither of these studies has addressed the QED corrections to the 
subprocess $\g^{\ast} \g \!\ra\! X$. We expect these corrections to be 
rather large, especially at the high scales which, hopefully, will 
become accessible to precise measurements in $e\g$ collisions at the 
future linear collider. Such contributions can be included in our 
present formalism by extending the analysis to higher order of 
electromagnetism. This extension is left to a future publication. 

{\sc Fortran} subroutines of our NNLO coefficient functions and of the
approximations of the NNLO splitting functions can be found at
{\tt http://www.nikhef.nl/$\sim$avogt.}
\subsection*{Acknowledgments}
We thank T. Gehrmann for providing the {\sc Fortran} routine 
\cite{GRhp} for the harmonic polylogarithms.    
The work of S.M. has been supported by the German Federal Ministry for
Research (BMBF) under grant BMBF-05HT9VKA and by the German Research
Society (DFG) under contract No.~FOR~264/2-1. The work of J.V. is part 
of the research program of the Dutch Foundation for Fundamental 
Research of Matter (FOM). The work of A.V. has been supported by the 
European Community TMR network `QCD and Particle Structure' under 
contract No.~FMRX--CT98--0194. 
%
%
\setcounter{equation}{0}
\renewcommand{\theequation}{A.\arabic{equation}}
\section*{Appendix A}
%
%
Here we present the analytic expressions for the anomalous dimensions 
and coefficient functions up to order $a_{\rm em}^{}\, a_{\rm s}^{2}$ 
at the even-integer values $N = 2,\, \ldots ,\, 12$. The notation is as
in Sect.~3; in addition $C_A$ and $C_F$ are the standard QCD colour
factors, $C_A \equiv N_c = 3$ and $C_F = (N_c^2 -1)/(2N_c) = 4/3$, and 
$\z_i$ stands for Riemann's $\z$-function. The photon-quark anomalous 
dimensions are given by
\bea
(\d_{\rm p}\, a_{\rm em})^{-1}\, k_{\rm p}^{N=2} & = &    
       - \frac{4}{3}
       + a_{\rm s} C_F   \left(
          - \frac{148}{27}
          \right)\nn\\[0.5ex]
&   & \mbox{}
       + a_{\rm s}^2 f_{\rm ps} C_F \Nf   \left(
          - \frac{614}{243}
          \right)\nn\\[0.5ex]
&   & \mbox{}
       + a_{\rm s}^2 C_F C_A   \left(
          - \frac{3022}{243}
          - \frac{32}{3} \zeta_3
          \right)\nn\\[0.5ex]
&   & \mbox{}
       + a_{\rm s}^2 C_F^2   \left(
          - \frac{4310}{243}
          + \frac{64}{3} \zeta_3
          \right)\nn\\[0.5ex]
&   & \mbox{}
       + a_{\rm s}^2 C_F \Nf   \left(
            \frac{268}{243}
          \right)\\[2ex]
(\d_{\rm p}\, a_{\rm em})^{-1}\, k_{\rm p}^{N=4} & = &    
       - \frac{11}{15}
       + a_{\rm s} C_F   \left(
          - \frac{56317}{9000}
          \right)\nn\\[0.5ex]
&   & \mbox{}
       + a_{\rm s}^2 f_{\rm ps} C_F \Nf   \left(
          - \frac{3848699}{12150000}
          \right)\nn\\[0.51ex]
&   & \mbox{}
       + a_{\rm s}^2 C_F C_A   \left(
          - \frac{101727401}{9720000}
          - \frac{261}{25} \zeta_3
          \right)\nn\\[0.5ex]
&   & \mbox{}
       + a_{\rm s}^2 C_F^2   \left(
          - \frac{757117001}{24300000}
          + \frac{522}{25} \zeta_3
          \right)\nn\\[0.5ex]
&   & \mbox{}
       + a_{\rm s}^2 C_F \Nf   \left(
            \frac{673127}{607500}
          \right)\\[2ex]
(\d_{\rm p}\, a_{\rm em})^{-1}\, k_{\rm p}^{N=6} & = &    
       - \frac{11}{21}
       + a_{\rm s} C_F   \left(
          - \frac{296083}{46305}
          \right)\nn\\[0.5ex]
&   & \mbox{}
       + a_{\rm s}^2 f_{\rm ps} C_F \Nf   \left(
          - \frac{36356399}{326728080}
          \right)\nn\\[0.5ex]
&   & \mbox{}
       + a_{\rm s}^2 C_F C_A   \left(
          - \frac{4774226099}{466754400}
          - \frac{1240}{147} \zeta_3
          \right)\nn\\[0.5ex]
&   & \mbox{}
       + a_{\rm s}^2 C_F^2   \left(
          - \frac{2933980223981}{81682020000}
          + \frac{2480}{147} \zeta_3
          \right)\nn\\[0.5ex]
&   & \mbox{}
       + a_{\rm s}^2 C_F \Nf   \left(
            \frac{1449736439}{1166886000}
          \right)\\[2ex]
(\d_{\rm p}\, a_{\rm em})^{-1}\, k_{\rm p}^{N=8} & = &    
       - \frac{37}{90}
       + a_{\rm s} C_F   \left(
          - \frac{51090517}{8164800}
          \right)\nn\\[0.5ex]
&   & \mbox{}
       + a_{\rm s}^2 f_{\rm ps} C_F \Nf   \left(
          - \frac{986499379}{18895680000}
          \right)\nn\\[0.5ex]
&   & \mbox{}
       + a_{\rm s}^2 C_F C_A   \left(
          - \frac{16806083022067}{1728324864000}
          - \frac{749}{108} \zeta_3
          \right)\nn\\[0.5ex]
&   & \mbox{}
       + a_{\rm s}^2 C_F^2   \left(
          - \frac{4374484944665803}{113421319200000}
          + \frac{749}{54} \zeta_3
          \right)\nn\\[0.5ex]
&   & \mbox{}
       + a_{\rm s}^2 C_F \Nf   \left(
            \frac{1373611188913}{1080203040000}
          \right)\\[2ex]
(\d_{\rm p}\, a_{\rm em})^{-1}\, k_{\rm p}^{N=10} & = &    
       - \frac{56}{165}
       + a_{\rm s} C_F   \left(
          - \frac{379479917}{62889750}
          \right)\nn\\[0.5ex]
&   & \mbox{}
       + a_{\rm s}^2 f_{\rm ps} C_F \Nf   \left(
          - \frac{10520389912}{366894309375}
          \right)\nn\\[0.5ex]
&   & \mbox{}
       + a_{\rm s}^2 C_F C_A   \left(
          - \frac{6245528269597}{684642974400}
          - \frac{17712}{3025} \zeta_3
          \right)\nn\\[0.5ex]
&   & \mbox{}
       + a_{\rm s}^2 C_F^2   \left(
          - \frac{1091980048536213833}{27182465592975000}
          + \frac{35424}{3025} \zeta_3
          \right)\nn\\[0.5ex]
&   & \mbox{}
       + a_{\rm s}^2 C_F \Nf   \left(
            \frac{29487804752071}{23534602245000}
          \right)\\[2ex]
(\d_{\rm p}\, a_{\rm em})^{-1}\, k_{\rm p}^{N=12} & = &    
       - \frac{79}{273}
       + a_{\rm s} C_F   \left(
          - \frac{9256843807}{1598647050}
          \right)\nn\\[0.5ex]
&   & \mbox{}
       + a_{\rm s}^2 f_{\rm ps} C_F \Nf   \left(
          - \frac{60434951117}{3466052828784}
          \right)\nn\\[0.5ex]
&   & \mbox{}
       + a_{\rm s}^2 C_F C_A   \left(
          - \frac{816463582581304612861}{95815031160008160000}
          - \frac{2563}{507} \zeta_3
          \right)\nn\\[0.5ex]
&   & \mbox{}
       + a_{\rm s}^2 C_F^2   \left(
          - \frac{2960118366121154186145047}{71933134643376126120000}
          + \frac{5126}{507} \zeta_3
          \right)\nn \\[0.5ex]
&   & \mbox{}
       + a_{\rm s}^2 C_F \Nf   \left(
            \frac{11634907579340558837}{9581503116000816000}
          \right) \:\: .
\eea  
The photon-gluon anomalous dimensions read
\bea  
(\d_{\rm s}\, a_{\rm em})^{-1}\, k_{\rm g}^{N=2} & = &
       \,\,\, a_{\rm s} C_F   \left(
            \frac{40}{27}
          \right)\nn \\[0.5ex]
&   & \mbox{}
       + a_{\rm s}^2 C_F C_A   \left(
          - \frac{569}{243}
          + \frac{32}{3} \zeta_3
          \right)\nn\\[0.5ex]
&   & \mbox{}
       + a_{\rm s}^2 C_F^2   \left(
            \frac{4796}{243}
          - \frac{64}{3} \zeta_3
          \right)\nn\\[0.5ex]
&   & \mbox{}
       + a_{\rm s}^2 C_F \Nf   \left(
            \frac{940}{243}
          \right)\\[1ex]
(\d_{\rm s}\, a_{\rm em})^{-1}\, k_{\rm g}^{N=4} & = &
       \,\,\, a_{\rm s} C_F   \left(
            \frac{2951}{4500}
          \right)\nn \\[0.5ex]
&   & \mbox{}
       + a_{\rm s}^2 C_F C_A   \left(
            \frac{97099523}{24300000}
          + \frac{66}{25} \zeta_3
          \right)\nn\\[0.5ex]
&   & \mbox{}
       + a_{\rm s}^2 C_F^2   \left(
            \frac{5594651}{1518750}
          - \frac{132}{25} \zeta_3
          \right)\nn\\[0.5ex]
&   & \mbox{}
       + a_{\rm s}^2 C_F \Nf   \left(
            \frac{252613}{303750}
          \right)\\[1ex]
(\d_{\rm s}\, a_{\rm em})^{-1}\, k_{\rm g}^{N=6} & = &
       \,\,\, a_{\rm s} C_F   \left(
            \frac{15418}{46305}
          \right)\nn \\[0.5ex]
&   & \mbox{}
       + a_{\rm s}^2 C_F C_A   \left(
            \frac{25354637939}{8168202000}
          + \frac{176}{147} \zeta_3
          \right)\nn\\[0.5ex]
&   & \mbox{}
       + a_{\rm s}^2 C_F^2   \left(
            \frac{29816260849}{20420505000}
          - \frac{352}{147} \zeta_3
          \right)\nn\\[0.5ex]
&   & \mbox{}
       + a_{\rm s}^2 C_F \Nf   \left(
            \frac{76134169}{145860750}
          \right)\\[1ex]
(\d_{\rm s}\, a_{\rm em})^{-1}\, k_{\rm g}^{N=8} & = &
       \,\,\, a_{\rm s} C_F   \left(
            \frac{32699}{163296}
          \right)\nn \\[0.5ex]
&   & \mbox{}
       + a_{\rm s}^2 C_F C_A   \left(
            \frac{29969084304431}{12962436480000}
          + \frac{37}{54} \zeta_3
          \right)\nn\\[0.5ex]
&   & \mbox{}
       + a_{\rm s}^2 C_F^2   \left(
            \frac{8625845880277}{11342131920000}
          - \frac{37}{27} \zeta_3
          \right)\nn\\[0.5ex]
&   & \mbox{}
       + a_{\rm s}^2 C_F \Nf   \left(
            \frac{315324047}{843908625}
          \right)\\[1ex]
(\d_{\rm s}\, a_{\rm em})^{-1}\, k_{\rm g}^{N=10} & = &
       \,\,\, a_{\rm s} C_F   \left(
            \frac{599864}{4492125}
          \right)\nn \\[0.5ex]
&   & \mbox{}
       + a_{\rm s}^2 C_F C_A   \left(
            \frac{32269440044009551}{18121643728650000}
          + \frac{1344}{3025} \zeta_3
          \right)\nn\\[0.5ex]
&   & \mbox{}
       + a_{\rm s}^2 C_F^2   \left(
            \frac{873823240941547}{1941604685212500}
          - \frac{2688}{3025} \zeta_3
          \right)\nn\\[0.5ex]
&   & \mbox{}
       + a_{\rm s}^2 C_F \Nf   \left(
            \frac{119478173264}{420260754375}
          \right)\\[1ex]
(\d_{\rm s}\, a_{\rm em})^{-1}\, k_{\rm g}^{N=12} & = &
       \,\,\, a_{\rm s} C_F   \left(
            \frac{1742495}{18270252}
          \right)\nn \\[0.5ex]
&   & \mbox{}
       + a_{\rm s}^2 C_F C_A   \left(
            \frac{117808573254816607171}{83039693672007072000}
          + \frac{158}{507} \zeta_3
          \right)\nn\\[0.5ex]
&   & \mbox{}
       + a_{\rm s}^2 C_F^2   \left(
            \frac{781495667102322375851}{2740309891176233376000}
          - \frac{316}{507} \zeta_3
          \right)\nn\\[0.5ex]
&   & \mbox{}
       + a_{\rm s}^2 C_F \Nf   \left(
            \frac{5137635507236447}{22813102657144800}
          \right) \:\: .
\eea
The photonic coefficient functions for the structure function $F_2$ at 
$\mu_r = \mu_f = Q$ are given by
\begin{eqnarray}  
(\d_{\rm p}\, a_{\rm em})^{-1}\, c_{2,\g}^{\,{\rm p},N=2} & = &
       - 1
       + a_{\rm s} C_F   \left(
          - \frac{4799}{405}
          + \frac{32}{5} \zeta_3
          \right)\nn\\[0.5ex]
&   & \mbox{}
       + a_{\rm s}^2 f_{\rm ps} C_F \Nf   \left(
          - \frac{18007}{1458}
          + \frac{10144}{405} \zeta_3
          - \frac{64}{3} \zeta_5
          \right)\nn\\[0.5ex]
&   & \mbox{}
       + a_{\rm s}^2 C_F C_A   \left(
          - \frac{440774}{3645}
          + \frac{26408}{405} \zeta_3
          - \frac{16}{3} \zeta_4
          + \frac{160}{3} \zeta_5
          \right)\nn\\[0.5ex]
&   & \mbox{}
       + a_{\rm s}^2 C_F^2   \left(
            \frac{28403}{2430}
          + \frac{8296}{405} \zeta_3
          + \frac{32}{3} \zeta_4
          - \frac{320}{3} \zeta_5
          \right)\nn\\[0.5ex]
&   & \mbox{}
       + a_{\rm s}^2 C_F \Nf   \left(
            \frac{65374}{3645}
          - \frac{3008}{405} \zeta_3
          \right)\\[2ex]
(\d_{\rm p}\, a_{\rm em})^{-1}\, c_{2,\g}^{\,{\rm p},N=4} & = &
       - \frac{133}{90}
       + a_{\rm s} C_F   \left(
          - \frac{6410867}{360000}
          + \frac{36}{5} \zeta_3
          \right)\nn\\[0.5ex]
&   & \mbox{}
       + a_{\rm s}^2 f_{\rm ps} C_F \Nf   \left(
          - \frac{207634473797}{4374000000}
          - \frac{2379889}{10125} \zeta_3
          + \frac{928}{3} \zeta_5
          \right)\nn\\[0.5ex]
&   & \mbox{}
       + a_{\rm s}^2 C_F C_A   \left(
          - \frac{425129854859}{2449440000}
          + \frac{11978567}{141750} \zeta_3
          - \frac{261}{50} \zeta_4
          + \frac{164}{3} \zeta_5
          \right)\nn\\[0.5ex]
&   & \mbox{}
       + a_{\rm s}^2 C_F^2   \left(
          - \frac{4566936087251}{61236000000}
          + \frac{18867971}{141750} \zeta_3
          + \frac{261}{25} \zeta_4
          - \frac{584}{3} \zeta_5
          \right)\nn\\[0.5ex]
&   & \mbox{}
       + a_{\rm s}^2 C_F \Nf   \left(
            \frac{4247321033}{153090000}
          - \frac{123812}{14175} \zeta_3
          \right)\\[2ex]
(\d_{\rm p}\, a_{\rm em})^{-1}\, c_{2,\g}^{\,{\rm p},N=6} & = &
       - \frac{1777}{1260}
       + a_{\rm s} C_F   \left(
          - \frac{12660217}{648270}
          + \frac{40}{7} \zeta_3
          \right)\nn\\[0.5ex]
&   & \mbox{}
       + a_{\rm s}^2 f_{\rm ps} C_F \Nf   \left(
          - \frac{506384656389713}{3430644840000}
          - \frac{77458783}{138915} \zeta_3
          + \frac{16400}{21} \zeta_5
          \right)\nn\\[0.5ex]
&   & \mbox{}
       + a_{\rm s}^2 C_F C_A   \left(
          - \frac{22283947715467}{117622108800}
          + \frac{68649953}{694575} \zeta_3
          - \frac{620}{147} \zeta_4
          + \frac{80}{3} \zeta_5
          \right)\nn\\[0.5ex]
&   & \mbox{}
       + a_{\rm s}^2 C_F^2   \left(
          - \frac{986797608696253}{6861289680000}
          + \frac{65099719}{694575} \zeta_3
          + \frac{1240}{147} \zeta_4
          - \frac{880}{7} \zeta_5
          \right)\nn\\[0.5ex]
&   & \mbox{}
       + a_{\rm s}^2 C_F \Nf   \left(
            \frac{3142023113107}{98018424000}
          - \frac{140188}{19845} \zeta_3
          \right)\\[2ex]
(\d_{\rm p}\, a_{\rm em})^{-1}\, c_{2,\g}^{\,{\rm p},N=8} & = &
       - \frac{16231}{12600}
       + a_{\rm s} C_F   \left(
          - \frac{23994462871}{1175731200}
          + \frac{14}{3} \zeta_3
          \right)\nn\\[0.5ex]
&   & \mbox{\hspace*{-3.0cm}}
       + a_{\rm s}^2 f_{\rm ps} C_F \Nf   \left(
          - \frac{11841145531674938821}{32665339929600000}
          - \frac{3563866619}{3061800} \zeta_3
          + \frac{15232}{9} \zeta_5
          \right)\nn\\ [0.5ex]
&   & \mbox{\hspace*{-3.0cm}}
       + a_{\rm s}^2 C_F C_A   \left(
          - \frac{28717038027651600293}{143727495690240000}
          + \frac{79435698877}{785862000} \zeta_3
          - \frac{749}{216} \zeta_4
          + \frac{118}{9} \zeta_5
          \right)\nn\\[0.5ex] 
&   & \mbox{\hspace*{-3.0cm}}
       + a_{\rm s}^2 C_F^2   \left(
          - \frac{954577609184015777449}{5030462349158400000}
          + \frac{154197212633}{2357586000} \zeta_3
          + \frac{749}{108} \zeta_4
          - \frac{260}{3} \zeta_5
          \right)\nn\\[0.5ex] 
&   & \mbox{\hspace*{-3.0cm}}
       + a_{\rm s}^2 C_F \Nf   \left(
            \frac{622760912995104859}{17965936961280000}
          - \frac{1122137}{200475} \zeta_3
          \right)\\[3ex]
(\d_{\rm p}\, a_{\rm em})^{-1}\, c_{2,\g}^{\,{\rm p},N=10} & = &
       - \frac{8704}{7425}
       + a_{\rm s} C_F   \left(
          - \frac{72533010722807}{3486607740000}
          + \frac{216}{55} \zeta_3
          \right)\nn\\[0.5ex]
&   & \mbox{\hspace*{-3.0cm}}
       + a_{\rm s}^2 f_{\rm ps} C_F \Nf   \left(
          - \frac{209277550337208434702179}{287046836661816000000}
          - \frac{4286102504597}{1981027125} \zeta_3
          + \frac{35264}{11} \zeta_5
          \right)\nn\\[0.5ex]
&   & \mbox{\hspace*{-3.0cm}}
       + a_{\rm s}^2 C_F C_A   \left(
          - \frac{9568565675783568703229}{46259614172772000000}
          + \frac{366130385968}{3679050375} \zeta_3
          - \frac{8856}{3025} \zeta_4
          + \frac{200}{33} \zeta_5
          \right)\nn\\[0.5ex]
&   & \mbox{\hspace*{-3.0cm}}
       + a_{\rm s}^2 C_F^2   \left(
          - \frac{26284376777719892724358177}{117545679613013652000000}
          + \frac{3565449892804}{77260057875} \zeta_3
          + \frac{17712}{3025} \zeta_4
          - \frac{2096}{33} \zeta_5
          \right)\nn\\[0.5ex]
&   & \mbox{\hspace*{-3.0cm}}
       + a_{\rm s}^2 C_F \Nf   \left(
            \frac{462262990271149761829}{12721393897512300000}
          - \frac{302180374}{66891825} \zeta_3
          \right)\\[3ex]
(\d_{\rm p}\, a_{\rm em})^{-1}\, c_{2,\g}^{\,{\rm p},N=12} & = &
       - \frac{8110049}{7567560}
       + a_{\rm s} C_F   \left(
          - \frac{6258789011950819}{299266727760000}
          + \frac{44}{13} \zeta_3
          \right)\nn\\[0.5ex]
&   & \mbox{\hspace*{-3.5cm}}
       + a_{\rm s}^2 f_{\rm ps} C_F \Nf   \left(
          - \frac{80246164545568080582270787397}
                 {62342050024259309304000000}
          - \frac{3109586855303}{847767375} \zeta_3
          + \frac{499488}{91} \zeta_5
          \right)\nn\\[0.5ex]
&   & \mbox{\hspace*{-3.5cm}}
       + a_{\rm s}^2 C_F C_A   \left(
          - \frac{109517569276335628551465332167}
                 {517918569432308108064000000}
          + \frac{59504875891511}{615479114250} \zeta_3
          - \frac{2563}{1014} \zeta_4
          + \frac{580}{273} \zeta_5
          \right)\nn\\[0.5ex]
&   & \mbox{\hspace*{-3.5cm}}
       + a_{\rm s}^2 C_F^2   \left(
          - \frac{194382513719914205037949320835327}
                 {777654732002610624258096000000}
          + \frac{542538591728921}{16617936084750} \zeta_3
          + \frac{2563}{507} \zeta_4 
          - \frac{4440}{91}  \zeta_5 
          \right)\nn\\[0.5ex]
&   & \mbox{\hspace*{-3.5cm}}
       + a_{\rm s}^2 C_F \Nf   \left(
            \frac{1937462197313198708954658827}
                 {51791856943230810806400000}
          - \frac{294410659}{79053975} \zeta_3
          \right) \:\: ,
\eea
and the corresponding results for $F_L$ read
\bea    
(\d_{\rm p}\, a_{\rm em})^{-1}\, c_{L,\g}^{\,{\rm p},N=2} & = &
         \frac{4}{3}
       + a_{\rm s} C_F   \left(
          - \frac{232}{135}
          - \frac{32}{5} \zeta_3
          \right)\nn\\[0.5ex]
&   & \mbox{}
       + a_{\rm s}^2 f_{\rm ps} C_F \Nf   \left(
            \frac{2686}{243}
          + \frac{3808}{15} \zeta_3
          - \frac{896}{3} \zeta_5
          \right)\nn\\[0.5ex]
&   & \mbox{}
       + a_{\rm s}^2 C_F C_A   \left(
          - \frac{11788}{135}
          - \frac{6512}{45} \zeta_3
          + 160 \zeta_5
          \right)\nn\\[0.5ex]
&   & \mbox{}
       + a_{\rm s}^2 C_F^2   \left(
            \frac{102566}{1215}
          + \frac{1856}{45} \zeta_3
          - \frac{320}{3} \zeta_5
          \right)\nn\\[0.5ex]
&   & \mbox{}
       + a_{\rm s}^2 C_F \Nf   \left(
            \frac{2624}{405}
          + \frac{512}{45} \zeta_3
          \right)\\[2ex]
(\d_{\rm p}\, a_{\rm em})^{-1}\, c_{L,\g}^{\,{\rm p},N=4} & = &
         \frac{8}{15}
       + a_{\rm s} C_F   \left(
          - \frac{5528}{1125}
          \right)\nn\\[0.5ex]
&   & \mbox{}
       + a_{\rm s}^2 f_{\rm ps} C_F \Nf   \left(
          - \frac{48227851}{1215000}
          + \frac{4384}{135} \zeta_3
          \right)\nn\\[0.5ex]
&   & \mbox{}
       + a_{\rm s}^2 C_F C_A   \left(
          - \frac{51480649}{354375}
          - \frac{174284}{4725} \zeta_3
          + 128 \zeta_5
          \right)\nn\\[0.5ex]
&   & \mbox{}
       + a_{\rm s}^2 C_F^2   \left(
            \frac{912595079}{5670000}
          + \frac{451816}{4725} \zeta_3
          - 256 \zeta_5
          \right)\nn\\[0.5ex]
&   & \mbox{}
       + a_{\rm s}^2 C_F \Nf   \left(
            \frac{520202}{39375}
          - \frac{64}{105} \zeta_3
          \right)\\[2ex]
(\d_{\rm p}\, a_{\rm em})^{-1}\, c_{L,\g}^{\,{\rm p},N=6} & = &
         \frac{2}{7}
       + a_{\rm s} C_F   \left(
          - \frac{137761}{46305}
          \right)\nn\\[0.5ex]
&   & \mbox{}
       + a_{\rm s}^2 f_{\rm ps} C_F \Nf   \left(
            \frac{97510126879}{4084101000}
          + \frac{61778}{315} \zeta_3
          - \frac{1760}{7} \zeta_5
          \right)\nn\\[0.5ex]
&   & \mbox{}
       + a_{\rm s}^2 C_F C_A   \left(
          - \frac{43056771637}{437582250}
          - \frac{17446}{2205} \zeta_3
          + \frac{480}{7} \zeta_5
          \right)\nn\\[0.5ex]
&   & \mbox{}
       + a_{\rm s}^2 C_F^2   \left(
            \frac{101548438631}{1021025250}
          + \frac{87148}{2205} \zeta_3
          - \frac{960}{7} \zeta_5
          \right)\nn\\[0.5ex]
&   & \mbox{}
       + a_{\rm s}^2 C_F \Nf   \left(
            \frac{826872491}{87516450}
          - \frac{64}{63} \zeta_3
          \right)\\[2ex]
(\d_{\rm p}\, a_{\rm em})^{-1}\, c_{L,\g}^{\,{\rm p},N=8} & = &
         \frac{8}{45}
       + a_{\rm s} C_F   \left(
          - \frac{51097}{25515}
          \right)\nn\\[0.5ex]
&   & \mbox{}
       + a_{\rm s}^2 f_{\rm ps} C_F \Nf   \left(
            \frac{166525045327877}{1620304560000}
          + \frac{617672}{1575} \zeta_3
          - \frac{1664}{3} \zeta_5
          \right)\nn\\[0.5ex]
&   & \mbox{}
       + a_{\rm s}^2 C_F C_A   \left(
          - \frac{103132581580249}{1485279180000}
          - \frac{393671}{779625} \zeta_3
          + \frac{128}{3} \zeta_5
          \right)\nn\\[0.5ex]
&   & \mbox{}
       + a_{\rm s}^2 C_F^2   \left(
            \frac{2384408424295187}{35646700320000}
          + \frac{15446822}{779625} \zeta_3
          - \frac{256}{3} \zeta_5
          \right)\nn\\[0.5ex]
&   & \mbox{}
       + a_{\rm s}^2 C_F \Nf   \left(
            \frac{36203618923}{5304568500}
          - \frac{1216}{1485} \zeta_3
          \right)\\[2ex]
(\d_{\rm p}\, a_{\rm em})^{-1}\, c_{L,\g}^{\,{\rm p},N=10} & = &
         \frac{4}{33}
       + a_{\rm s} C_F   \left(
          - \frac{509195549}{352182600}
          \right)\nn\\[0.5ex]
&   & \mbox{}
       + a_{\rm s}^2 f_{\rm ps} C_F \Nf   \left(
            \frac{551115536481640147}{2761393330080000}
          + \frac{694547002}{1091475} \zeta_3
          - \frac{10240}{11} \zeta_5
          \right)\nn\\[0.5ex]
&   & \mbox{}
       + a_{\rm s}^2 C_F C_A   \left(
          - \frac{321273125948555419}{6230251067040000}
          + \frac{290422414}{156080925} \zeta_3
          + \frac{320}{11} \zeta_5
          \right)\nn\\[0.5ex]
&   & \mbox{}
       + a_{\rm s}^2 C_F^2   \left(
            \frac{9053411269935853949}{188465094777960000}
          + \frac{1721766952}{156080925} \zeta_3
          - \frac{640}{11} \zeta_5
          \right)\nn\\[0.5ex]
&   & \mbox{}
       + a_{\rm s}^2 C_F \Nf   \left(
            \frac{195749323625753}{38073756520800}
          - \frac{272}{429} \zeta_3
          \right)\\[2ex]
(\d_{\rm p}\, a_{\rm em})^{-1}\, c_{L,\g}^{\,{\rm p},N=12} & = &
         \frac{8}{91}
       + a_{\rm s} C_F   \left(
          - \frac{197186301194}{179847793125}
          \right)\nn\\[0.5ex]
&   & \mbox{\hspace*{-1cm}}
       + a_{\rm s}^2 f_{\rm ps} C_F \Nf   \left(
            \frac{4911117022004232715153043}{15569942563501326000000}
          + \frac{463564050788}{496621125} \zeta_3
          - \frac{126208}{91} \zeta_5
          \right)\nn\\[0.5ex]
&   & \mbox{\hspace*{-1cm}}
       + a_{\rm s}^2 C_F C_A   \left(
          - \frac{47756006182490429534639}{1197687889500102000000}
          + \frac{5631384334}{2152024875} \zeta_3
          + \frac{1920}{91} \zeta_5
          \right)\nn\\[0.5ex]
&   & \mbox{\hspace*{-1cm}}
       + a_{\rm s}^2 C_F^2   \left(
            \frac{281711596115081884853551}{7784971281750663000000}
          + \frac{4695483724}{717341625} \zeta_3
          - \frac{3840}{91} \zeta_5
          \right)\nn\\[0.5ex]
&   & \mbox{\hspace*{-1cm}}
       + a_{\rm s}^2 C_F \Nf   \left(
            \frac{780777857066540089}{194429852191575000}
          - \frac{3392}{6825} \zeta_3
          \right) \:\: .
\eea
%
%
\setcounter{equation}{0}
\renewcommand{\theequation}{B.\arabic{equation}}
\section*{Appendix B}
%
%
The exact expression for the ${\cal O}(a_{\rm em}^{} a_{\rm s}^{2})$
contribution (\ref{dPqnn}) to the transformation (\ref{Pdsg}) of the 
photon-quark splitting function to the DIS$_\g$ scheme is given by
\bea 
\label{dPex1}   
\lefteqn{ \d_{\rm p}^{-1} \Delta_{\rm DIS_{\g_{}}}^{(2),\rm p}(x)
\: = \: } \nn \\[1ex]
& &\hspace*{-0.5cm} 
16 C_F^2 \left\{ 
            - \left(  56 + \frac{2}{5} \frac{1}{x^2} + 82 x + 50 x^2 
                + \frac{72}{5} x^3\right) H_{-1} \zeta_2
          - \left(  \frac{431}{30} + \frac{4}{15} \frac{1}{x} 
                   + \frac{6}{5} x - \frac{102}{5} x^2\right) H_{2}
\right. 
\nn \\
&  &  
\left. \mbox{}
          - \left(   48 + \frac{4}{15} \frac{1}{x^2} + \frac{308}{3} x 
                + 76 x^2 + \frac{48}{5} x^3\right) H_{-1,-1,0}
          - 4 (   9 + 10 x + 16 x^2) H_{-2,-1,0}
\right. 
\nn \\
&  & 
\left. \mbox{}
          - \left(   21 + \frac{2}{15} \frac{1}{x^2} - \frac{130}{3} x 
                + 34 x^2 - \frac{24}{5} x^3\right) H_{1} \zeta_2
          - \left(   \frac{1883}{120} - \frac{4}{15} \frac{1}{x} 
                - \frac{1683}{20} x + \frac{373}{5} x^2\right) H_{1}
\right. 
\nn \\
&  & 
\left. \mbox{}
          - 2 \left(   7 - 14 x + 6 x^2 \right) H_{1,1} \zeta_2
          - 4 \left(   3 -  6 x + 7 x^2 \right) H_{3,1}
          - 2 \left(   5 -  2 x + 6 x^2 \right) H_{2} \zeta_2
\right. 
\nn \\
&  & 
\left. \mbox{}
          - 2 \left(   5 - 10 x - 4 x^2 \right) H_{1} \zeta_3
          - 4 \left(   2 - 2 x + 9 x^2 \right) H_{4}
          - \frac{1679}{180} - \frac{4}{45} \frac{1}{x} 
                   - \frac{4549}{120} x + \frac{276}{5} x^2 
\right. 
\nn \\
&  & 
\left. \mbox{}
          - \left(   \frac{701}{72} - \frac{4}{45} \frac{1}{x} 
                   + \frac{129}{40} x + \frac{357}{5} x^2\right) H_{0}
          - 8 \left(   1 - 2 x + 3 x^2 \right) H_{3,0}
          - 4 \left(  2 - 4 x + 5 x^2 \right) H_{2,2}
\right. 
\nn \\
&  & 
\left. \mbox{}
          - \left(   \frac{809}{120} + \frac{4}{15} \frac{1}{x} 
                   + \frac{15521}{180} x + \frac{213}{5} x^2
                   + \frac{32}{5} x^3\right) H_{0,0}
          - \left(  \frac{5}{2} + 3 x + 20 x^2\right) H_{0,0,0,0}
\right. 
\nn \\
&  & 
\left. \mbox{}
          - 2 \left(   3 - 5 x + 2 x^2 \right) H_{2,0}
          - 2 \left(   3 - 6 x + 10 x^2 \right) H_{1,3}
          - 2 \left(   3 - 6 x - 2 x^2 \right) H_{1,0} \zeta_2
\right. 
\nn \\
&  & 
\left. \mbox{}
          - \left(   \frac{11}{2} + 17 x - 60 x^2\right) H_{2,1}
          - \left(   \frac{9}{2} + \frac{47}{3} x + 16 x^2 
                   + \frac{48}{5} x^3\right) H_{3}
          - \left(   \frac{9}{2} + 11 x - 15 x^2\right) H_{1,0}
\right. 
\nn \\
&  & 
\left. \mbox{}
          -  \left( 3 - 8 x + 4 x^2 \right) H_{1,2}
          - 2\left( 1 - 2 x + 6 x^2 \right) H_{1,1,0,0}
          -  \left( 1 - 2 x + 12 x^2 \right ) H_{2,0,0}
\right. 
\nn \\
&  & 
\left. \mbox{}
          - \left(   \frac{1}{2} + \frac{287}{6} x + 36 x^2 
          + \frac{48}{5} x^3\right) H_{0,0,0}
          - \left(   \frac{1}{2} + 2 x - 8 x^2\right) H_{1,0,0}
          + \left(\frac{6}{5} + \frac{4}{5} x + \frac{66}{5} x^2\right) 
            \zeta_2^2
\right. 
\nn \\
&  & 
\left. \mbox{}
          + \left(\frac{9}{2} + 73 x + 16 x^2 + \frac{96}{5} x^3\right) 
            H_{0} \zeta_2
          + \left(\frac{27}{4} - \frac{95}{2} x + 30 x^2\right) H_{1,1}
          + (8 + 36 x^2) H_{0,0} \zeta_2
\right. 
\nn \\
&  & 
\left. \mbox{}
          + 4\left(3 + 2 x + 8 x^2 \right) H_{-3,0}
          + 2\left(1 - 2 x - 2 x^2 \right) H_{1,0,0,0}
          +  \left(1 - 2 x + 16 x^2 \right) H_{2,1,1}
\right. 
\nn \\
&  & 
\left. \mbox{}
          + 2 \left(3 + 6 x + 4 x^2 \right) ( 2H_{-1,0,0,0} + 2H_{-1,3} 
                     + 8 H_{-1,-1} \zeta_2 + 8 H_{-1,-1,-1,0} 
                     - 4 H_{-1,-2,0} 
\right. 
\nn \\
&  & 
\left. \mbox{}
                     - 4 H_{-1,-1,2} -4 H_{-1,0} \zeta_2 
                     - 7 H_{-1} \zeta_3 - 6 H_{-1,-1,0,0} )
          -  2 \left(   19 + 14 x + 32 x^2 \right) H_{-2} \zeta_2
\right. 
\nn \\
&  & 
\left. \mbox{}
          + 8\left(2 +   x + 8 x^2 \right) H_{0} \zeta_3
          + 4\left(5 + 2 x + 8 x^2 \right) H_{-2,2}
          + 4\left(7 + 6 x + 12 x^2 \right) H_{-2,0,0}
\right. 
\nn \\
&  & 
\left. \mbox{}
          + \left(13 + \frac{308}{3} x + 62 x^2 + 24 x^3\right)
          \zeta_3
          + \left(\frac{431}{30} + \frac{8}{15} \frac{1}{x} 
                + \frac{4328}{45} x 
                - \frac{102}{5} x^2 + \frac{32}{5} x^3\right) \zeta_2
\right. 
\nn \\
&  & 
\left. \mbox{}
          + \left(\frac{31}{2} - 74 x + 60 x^2\right) H_{1,1,1}
          + \left(40 + \frac{4}{15} \frac{1}{x^2} + \frac{200}{3} x 
              + 44 x^2 + \frac{48}{5} x^3\right) H_{-1,0,0}
\right. 
\nn \\
&  & 
\left. \mbox{}
          + \left(32 + \frac{4}{15} \frac{1}{x^2} + \frac{92}{3} x 
              + 12 x^2 + \frac{48}{5} x^3\right) H_{-1,2}
          + \left(16 + \frac{172}{3} x + 76 x^2 
                + \frac{48}{5} x^3\right) H_{-2,0}
\right. 
\nn \\
&  & 
\left. \mbox{}
          + \Big( 8 H_{1,1,1,1} - 2 H_{1,1,0} - 6 H_{1,1,1,0} 
                 - 10 H_{1,1,2} - 12 H_{1,2,0} - 14 H_{1,2,1} 
                 - 6 H_{2,1,0} \Big) \,  p_{\rm qg}(x)
\right. 
\nn \\
&  & 
\left. \mbox{}
          + \left(\frac{613}{15} + \frac{8}{45} \frac{1}{x^2} 
               + \frac{4}{15} \frac{1}{x} + \frac{4274}{45} x 
               + \frac{288}{5} x^2 + \frac{32}{5} x^3\right) H_{-1,0}
          \right\}
\nn
\\[1ex]
& &\hspace*{-0.5cm}  \mbox{}
+ 16 C_A C_F
\left\{  
          -  \left(  32 + \frac{11}{45} \frac{1}{x^2} + \frac{394}{9} x 
                   + 24 x^2 + \frac{44}{5} x^3\right) H_{-1,0}
          - \frac{233}{10} - \frac{11}{45} \frac{1}{x}  +
                   \frac{12283}{360} x - \frac{7}{10} x^2 
\right. 
\nn \\
&  & 
\left.
          - \left(   \frac{15}{2} - 27 x + \frac{59}{3} x^2\right) 
                   H_{1,0,0}
          - \left(   \frac{1825}{72} - \frac{467}{4} x 
                   + \frac{842}{9} x^2\right) H_{1}
          - \left(   \frac{157}{6} + 16 x + 96 x^2\right) \zeta_3
\right. 
\nn \\
&  & 
\left.
          - \left(   \frac{167}{18} - \frac{464}{9} x 
                   + \frac{464}{9} x^2\right) H_{1,0}
          - \left(   \frac{571}{72} - \frac{542}{9} x 
                   + \frac{248}{9} x^2 - \frac{44}{5} x^3\right) H_{0,0}
\right. 
\nn \\
&  & 
\left.
          - \left(   \frac{701}{36} - \frac{466}{9} x 
                   + \frac{433}{9} x^2\right) H_{1,1}
          - \left(   \frac{56}{3} + 22 x + \frac{158}{3} x^2\right) 
                   H_{-2,0}
          - 2 \left(   2 + 5 x + 3 x^2 \right) H_{-1,2}
\right. 
\nn \\
&  & 
\left.
          - \left(   \frac{95}{6} - \frac{191}{6} x 
                   + \frac{433}{9} x^2\right) H_{2}
          - \left(   \frac{40}{3} + \frac{122}{3} x 
                   + \frac{88}{3} x^2\right) H_{-1,0,0}
          - 4\left(   1 + 2 x + 3 x^2 \right) H_{-2,0,0}
\right. 
\nn \\
&  & 
\left.
          - 3\left(   1 + 2 x + 4 x^2 \right) 
                   ( 3 H_{0} \zeta_3 + 2 H_{-3,0} + 2 H_{-2,2} )
          - \left(   \frac{1283}{120} - \frac{11}{45} \frac{1}{x}  
                   - \frac{21301}{360} x + \frac{3814}{45} x^2\right) 
                   H_{0}
\right. 
\nn \\
&  & 
\left.
          - \left(   \frac{11}{6} + \frac{85}{3} x + 6 x^2\right) 
                   H_{0} \zeta_2
          - \left(   \frac{1}{10} - \frac{3}{5} x 
                   + \frac{22}{5} x^2\right) \zeta_2^2
          + \left(\frac{11}{12} + \frac{121}{6} x 
                   + \frac{29}{3} x^2\right) H_{0,0,0}
\right. 
\nn \\
&  & 
\left.
          + \left(3 - 6 x + 8 x^2 \right) H_{2} \zeta_2
          + \left(5 + 10 x + 16 x^2 \right) H_{-2} \zeta_2
          + \left(6 + 12 x + 16 x^2 \right) H_{-2,-1,0}
\right. 
\nn \\
&  & 
\left.
          + \left(\frac{11}{6} + \frac{19}{3} x + 6 x^2\right) H_{3}
          + \left(\frac{34}{3} - \frac{107}{3} x 
                   + \frac{79}{3} x^2\right) H_{1} \zeta_2
          + \left(\frac{46}{3} + \frac{137}{3} x 
                   + \frac{97}{3} x^2\right) H_{-1} \zeta_2
\right. 
\nn \\
&  & 
\left.
          + \left(\frac{95}{6} - \frac{1361}{18} x + \frac{433}{9} x^2 
                - \frac{44}{5} x^3\right) \zeta_2
          + \left(\frac{68}{3} +\frac{214}{3} x 
                + \frac{158}{3} x^2\right) H_{-1,-1,0}
          - \frac{11}{3}  x^2  H_{2,1}
\right. 
\nn \\
&  & 
\left.
          + \Big( 7H_{-1} \zeta_3 + 4 H_{-1,-2,0} - 8 H_{-1,-1} \zeta_2 
                - 8 H_{-1,-1,-1,0} + 6 H_{-1,-1,0,0} +
                4 H_{-1,-1,2} + 4 H_{-1,0} \zeta_2 
\right. 
\nn \\
&  & 
\left.
                - 2 H_{-1,0,0,0} - 2 H_{-1,3} \Big) \, p_{\rm qg}( - x)
          + 4 x \left( H_{0,0,0,0} + H_{4} - 2 H_{0,0} \zeta_2 \right)
          + \bigg( H_{1} \zeta_3 + 4 H_{1,0} \zeta_2 
\right. 
\nn \\
&  & 
\left. 
               - 2 H_{1,0,0,0} + 4 H_{1,1} \zeta_2 
               - 2 H_{1,1,0,0} - 2 H_{1,3} - 2 H_{2,0,0} 
               - \frac{11}{3} H_{1,1,0} - \frac{11}{3} H_{2,0} 
               - \frac{11}{6} H_{1,1,1} \bigg)\, p_{\rm qg}(x)
          \right\}
\nn
\\[1ex]
& &\hspace*{-0.5cm} \mbox{}
 + \frac{16}{3} N_F C_F 
\left\{  
            \left(1 - 2 x \right) ( H_{0} \zeta_2 - H_{3} )
          + \left(3 - 6 x + 2 x^2 \right) H_{1,0,0}
          - \left( \frac{1}{2} - x + 2 x^2\right) H_{0,0,0}
\right. 
\nn \\
&  & 
\left. \mbox{}
          + \left(4 + 8 x + 4 x^2) ( H_{-1,0,0} - H_{-1} \zeta_2 - 2
                     H_{-1,-1,0}  \right)
          - \left(4 - 8 x + 4 x^2 \right) H_{1} \zeta_2
\right. 
\nn \\
&  & 
\left. \mbox{}
          + \left(\frac{41}{12} - \frac{44}{3} x + \frac{80}{3} x^2 
                  - \frac{24}{5} x^3\right) H_{0,0}
          - \left(   7 - \frac{55}{3} x + \frac{70}{3} x^2 -
                   \frac{24}{5} x^3\right) \zeta_2
\right. 
\nn \\
&  & 
\left. \mbox{}
          + \left(\frac{83}{20} - \frac{2}{15} \frac{1}{x}  
              - \frac{1091}{60} x + \frac{553}{15} x^2\right) H_{0}
          + \left(\frac{13}{3} - \frac{80}{3} x 
              + \frac{80}{3} x^2\right) H_{1,0}
\right. 
\nn \\
&  & 
\left. \mbox{}
          + \left(7 - 13 x + \frac{70}{3} x^2\right) H_{2}
          + \frac{83}{10} + \frac{2}{15} \frac{1}{x} + \frac{7}{60} x
                - \frac{69}{5} x^2
          + \left(\frac{55}{6} - \frac{76}{3} x 
                + \frac{70}{3} x^2\right) H_{1,1}
\right. 
\nn \\
&  & 
\left. \mbox{}
          + \left(11 - 6 x + 18 x^2 \right) \zeta_3
          + 2 x^2 H_{2,1}
          + \left(2 H_{1,1,0} + H_{1,1,1} + 2 H_{2,0} \right) 
                p_{\rm qg}(x) 
          + 8 \left(1 + x^2 \right) H_{-2,0}
\right. 
\nn \\
&  & 
\left. \mbox{}
          + \left(12 + \frac{2}{15} \frac{1}{x^2}  + \frac{16}{3} x 
                     + \frac{24}{5} x^3\right) H_{-1,0}
          + \left(\frac{137}{12} - \frac{105}{2} x 
                     + \frac{125}{3} x^2\right) H_{1}
          \right\}
\nn
\\[1ex]
& &\hspace*{-0.5cm} \mbox{}
 + 16 f_{\rm ps} N_F C_F \left\{
         2 (1 - 2 x)
           \left(  H_{4} - H_{0} \zeta_3 - H_{0,0} \zeta_2 +
           H_{0,0,0,0} - \frac{2}{5} \zeta_2^2 \right)
       + \left(1 - 4 x - \frac{16}{3} x^2\right) H_{0,0,0}
\right.
\nn \\
&  &
\left. \mbox{}
       - \frac{175}{27} - \frac{20}{27} \frac{1}{x} + \frac{4327}{27} x 
              - \frac{4132}{27} x^2
       - \left( 3 - \frac{40}{27} \frac{1}{x} - 33 x 
              + \frac{850}{27} x^2\right) H_{1}
\right.
\nn \\
&  &
\left. \mbox{}
       + \left(1 + 71 x + \frac{1124}{27} x^2\right) H_{0}
       + \left(7 + 33 x - \frac{44}{9} x^2\right) H_{0,0}
       - \left(4 + 11 x + \frac{148}{9} x^2\right)
           ( \zeta_2 - H_{2})
\right.
\nn \\
&  &
\left. \mbox{}
       - \left(1 + 8 x - \frac{16}{3} x^2\right)
           ( \zeta_3 + H_{0} \zeta_2 - H_{3})
          \right\} \:\: ,
\eea
where the function $p_{\rm qg}(x)$ has been defined in Eq.~(\ref{p0}).
The corresponding result for the NNLO transformation of the photon-%
gluon splitting function, parametrized in Eq.~(\ref{dPgnn}), reads
\bea 
\label{dPex2}
\lefteqn{ \d_{\rm s}^{-1} \Delta_{\rm DIS_{\g_{}}}^{(2),\rm g}(x)
\: = \: } \nn \\[1ex]
& &\hspace*{-0.5cm}
32 C_F^2 
          \left\{ 
            \left( - \frac{67}{3} - \frac{4}{45} \frac{1}{x^2} 
                   - \frac{209}{9} x + \frac{4}{5} x^3\right) H_{-1,0}
          - \left(   8 + \frac{4}{3} x - \frac{8}{3} x^2\right) H_{-2,0}
          - \left( \frac{36}{5} + \frac{26}{5} x\right) \zeta_2^2
\right. 
\nn \\
&  & 
\left. \mbox{}
          + \left( - \frac{85}{12} + \frac{13}{3} \frac{1}{x} 
                   + \frac{25}{12} x + \frac{2}{3} x^2\right) H_{1,1}
          - \left(   5 + \frac{22}{3} x\right) H_{0} \zeta_2
          + \left(\frac{1}{2} + \frac{29}{6} x - 2 x^2\right) H_{0,0,0}
\right. 
\nn \\
&  & 
\left. \mbox{}
          + \left( - \frac{5}{2} - \frac{4}{3} \frac{1}{x} 
                     - \frac{13}{2} x + 2 x^2\right) \zeta_3
          + \left( - \frac{5}{2} + \frac{2}{3} \frac{1}{x} +
                     \frac{5}{2} x - \frac{2}{3} x^2\right) H_{1,0,0}
\right. 
\nn \\
&  & 
\left. \mbox{}
          + \left( - \frac{9}{4} - \frac{326}{9} x - \frac{2}{3} x^2 
                + \frac{4}{5} x^3\right) \zeta_2
          + \left(\frac{1}{3} + \frac{11}{3} \frac{1}{x} 
                - \frac{4}{3} x - \frac{8}{3} x^2\right) H_{1,0}
          - (7 - x)\, H_{0} \zeta_3
\right. 
\nn \\
&  & 
\left. \mbox{}
          + 4 (1 - x) \left( H_{-2} \zeta_2 + 2 H_{-2,-1,0} - H_{-2,0,0}
                    + \frac{3}{4} H_{1} \zeta_2 \right)
          + (1 + x) \bigg(  3 H_{-1} \zeta_2 - 3 H_{-1,0,0} 
\right. 
\nn \\
&  & 
\left. \mbox{}
          + 6 H_{-1,-1,0} 
                     + 4 H_{2} \zeta_2 - 3 H_{0,0} \zeta_2  - H_{2,0,0}
                     - 2 H_{2,1,0} - 7 H_{2,1,1} - 2 H_{3,1} 
                     + \frac{5}{2} H_{0,0,0,0} + 3 H_{4} \bigg)
\right. 
\nn \\
&  & 
\left. \mbox{}
          + \left(\frac{9}{4} + 13 x + \frac{2}{3} x^2\right) H_{2}
          + \left(\frac{41}{12} + \frac{13}{12} \frac{1}{x} +
                  \frac{15}{2} x - 12 x^2\right) H_{1}
          + \left(6 + 5 x + \frac{4}{3} x^2\right) H_{2,0}
\right. 
\nn \\
&  & 
\left. \mbox{}
          + \left(\frac{39}{8} + \frac{2113}{72} x - \frac{8}{3} x^2 
          - \frac{4}{5} x^3\right) H_{0,0}
          + (5 + 6 x) H_{3}
          + \left(\frac{13}{2} + \frac{7}{2} x + \frac{14}{3} x^2\right)
            H_{2,1}
\right. 
\nn \\
&  & 
\left. \mbox{}
          + \Big( H_{-1} \zeta_2 -  H_{-1,0,0} + 2 H_{-1,-1,0} \Big) 
            p_{\rm g\gamma}(-x)
          +\left( H_{1} \zeta_2 - H_{1,1,0} - \frac{7}{2} H_{1,1,1} 
            \right) p_{g\gamma}(x) 
\right. 
\nn \\
&  & 
\left. \mbox{}
          + \left(\frac{3389}{360} + \frac{4}{45} \frac{1}{x} 
          + \frac{311}{40} x - \frac{64}{5} x^2\right) H_{0}
          + \frac{2179}{180} + \frac{149}{180} \frac{1}{x} -
                 \frac{77}{15} x - \frac{39}{5} x^2 
          - 8 H_{-3,0}
          \right\}
\nn
\\[1ex]
& &\hspace*{-0.5cm} \mbox{}
+ 32 C_A C_F 
          \left\{
          - \left(   \frac{20}{3}  + \frac{2}{3} \frac{1}{x} 
                   + \frac{2}{3} x   - \frac{16}{3} x^2\right) H_{-1,0}
          - \left(   \frac{19}{3}  - \frac{8}{3} \frac{1}{x} +
                     \frac{25}{3} x  + 4 x^2\right) H_{2,1}
\right. 
\nn \\
&  & 
\left. \mbox{}
          - \left(   \frac{11}{3}  + \frac{14}{3} x  
                   + \frac{4}{3} x^2\right) H_{3}
          - \left(   \frac{31}{9}  - \frac{2}{3} \frac{1}{x} +
                     \frac{128}{9} x - 17 x^2\right) H_{1}
\right. 
\nn \\
&  & 
\left. \mbox{}
          - \left(   \frac{8}{3}   - \frac{4}{3} \frac{1}{x} +
                     \frac{5}{3} x   + \frac{8}{3} x^2\right) H_{2,0}
          - \left(   \frac{37}{18} + \frac{67}{18} x  
                   + \frac{35}{9} x^2\right) H_{0,0}
\right. 
\nn \\
&  & 
\left. \mbox{}
          - \left(   1     - 5 x      - \frac{4}{3} x^2\right) H_{-2,0}
          + \left(\frac{1}{3} - \frac{8}{3} \frac{1}{x} + \frac{5}{3} x 
                  + \frac{2}{3} x^2\right) H_{1,1}
          + \left(\frac{7}{9} - \frac{2}{3} \frac{1}{x} 
                  - \frac{95}{9} x - \frac{56}{9} x^2\right) H_{2}
\right. 
\nn \\
&  & 
\left. \mbox{}
          - \left(\frac{7}{9} - \frac{89}{9} x - \frac{56}{9} x^2\right)
               \zeta_2
          +  \left( H_{-1,-1,0} - H_{-1,0,0} - H_{-1,2}  
               + \frac{3}{2} H_{-1} \zeta_2 \right) p_{\rm g\gamma}(-x) 
\right. 
\nn \\
&  & 
\left. \mbox{}
          + \left( H_{1,0,0} + 2 H_{1,1,0} + 2 H_{1,2} + 3 H_{1,1,1} 
               - \frac{3}{2} H_{1} \zeta_2 \right) p_{\rm g\gamma}(x) 
          + \left(\frac{11}{3} + \frac{29}{3} x + \frac{4}{3} x^2\right)
               H_{0} \zeta_2
\right. 
\nn \\
&  & 
\left. \mbox{}
          + (1 - x) \, \Big(   3 H_{-2} \zeta_2  - 2 H_{-3,0}
                      + 2 H_{-2,-1,0} - 2 H_{-2,0,0} - 2 H_{-2,2} 
                      - H_{0} \zeta_3 - 2 H_{0,0,0,0} \Big)
\right. 
\nn \\
&  & 
\left. \mbox{}
          + (1 + x) \, \Big(   2 H_{2,0,0} - 3 H_{2} \zeta_2  
                      + 4 H_{2,1,0} + 6 H_{2,1,1} + 4 H_{2,2} 
                      + 4 H_{3,0} + 6 H_{3,1} \Big)
\right. 
\nn \\
&  & 
\left. \mbox{}
          + \left(\frac{21}{10} + \frac{27}{10} x\right) \zeta_2^2
          + \left(\frac{19}{6} + \frac{103}{6} x + 2 x^2\right) \zeta_3
          + \left(\frac{7}{2} - \frac{5}{3} \frac{1}{x} + \frac{1}{2} x 
                  - \frac{7}{3} x^2\right) H_{1,0}
\right. 
\nn \\
&  & 
\left. \mbox{}
          + \left(\frac{97}{18} + \frac{67}{9} x 
                 - \frac{128}{9} x^2\right) H_{0}
          + \frac{317}{18} - \frac{1}{6} \frac{1}{x} - \frac{295}{6} x 
          + \frac{571}{18} x^2 
          + 2 x \, ( H_{0,0,0} - H_{0,0} \zeta_2 + 2 H_{4} )
          \right\}
\nn
\\[1ex]
& &\hspace*{-0.5cm} \mbox{}
+ \frac{32}{3} N_F C_F 
          \left\{
            \left(\frac{11}{3} - \frac{26}{9} \frac{1}{x} 
          + \frac{1}{3} x - \frac{10}{9} x^2\right) H_{1}
          - \frac{127}{9} +\frac{10}{9} \frac{1}{x} - \frac{35}{9} x 
          + \frac{152}{9} x^2 
\right. 
\nn \\
&  & 
\left. \mbox{}
          - \left( 3 + 21 x + \frac{10}{9} x^2\right) H_{0}
          + \left(\frac{13}{3} + \frac{7}{3} x + \frac{8}{3} x^2\right) 
             ( \zeta_2 - H_{2} )
          - \left(\frac{16}{3} + \frac{22}{3} x + \frac{4}{3} x^2\right)
             H_{0,0}
\right. 
\nn \\
&  & 
\left. \mbox{}
          + \Big( H_{1,0} + 2 H_{1,1} \Big) p_{\rm g\gamma}(x)
          + 2 (1 + x) \,  \Big( H_{3} - \zeta_3 - H_{0} \zeta_2 
          + H_{2,0} + 2 H_{2,1} \Big)
          \right\} \:\: .
\eea
Here we have used the abbreviation
\[
   p_{\rm g\gamma}(x) \: = \: 
   \frac{4}{3x} + 1 - x - \frac{4}{3} x^2 \:\: .
\]
For the numerical evaluation of the harmonic polylogarithms up to 
weight four entering Eqs.~(\ref{dPex1}) and (\ref{dPex2}) we have 
employed the program of ref.~\cite{GRhp}.
%
%

\end{document}